\RequirePackage{fix-cm}
\documentclass[twocolumn]{revtex4}

\usepackage{soul}
\RequirePackage{graphicx}
\RequirePackage{amsmath}
\RequirePackage{amssymb}
\RequirePackage{bm}
\RequirePackage{color}
\RequirePackage[pdfencoding=auto, psdextra]{hyperref}
\RequirePackage{verbatim}
\newcommand{\ba}{\begin{eqnarray}}
\newcommand{\ea}{\end{eqnarray}}
\begin{document}

\title{Strong decay widths and mass spectra of charmed baryons}

\author{H. Garc{\'i}a-Tecocoatzi}

\affiliation{Center for High Energy Physics, Kyungpook National University, 80 Daehak-ro, Daegu 41566, Korea}
\affiliation{INFN, Sezione di Genova, Via Dodecaneso 33, 16146 Genova, Italy}

\author{A. Giachino} 
\affiliation{INFN, Sezione di Genova, Via Dodecaneso 33, 16146 Genova, Italy}
\affiliation{Institute of Nuclear Physics Polish Academy of Sciences Radzikowskiego 152, 31-342 Cracow, Poland}

\author{J. Li} 
\affiliation{Center for High Energy Physics, Kyungpook National University, 80 Daehak-ro, Daegu 41566, Korea}

\author{A. Ramirez-Morales}
\affiliation{Center for High Energy Physics, Kyungpook National University, 80 Daehak-ro, Daegu 41566, Korea}

\author{E. Santopinto}\email[]{elena.santopinto@ge.infn.it}
\affiliation{INFN, Sezione di Genova, Via Dodecaneso 33, 16146 Genova, Italy}

%\author{H. Garc{\'i}a-Tecocoatzi\thanksref{addr4,addr5}\and  A. Giachino\thanksref{addr1,addr2} \and J. Li\thanksref{addr5} \and A. Ramirez-Morales\thanksref{addr5} \and E. Santopinto\thanksref{e1,addr1}
 %etc.
%}

%\thankstext{e1}{e-mail:santopinto@ge.infn.it}

%\institute{Department of Physics, University of La Plata (UNLP), 49 y 115 cc. 67, 1900 La Plata, Argentina \label{addr4} \and Center of High Energy Physics, Kyungpook National University, 80 Daehak-ro, Daegu 41566, Korea \label{addr5} \and
%          Dipartimento di Fisica dell'Universitá di Genova,
%via Dodecaneso 33, 16146 Genova, Italy \label{addr2} \and INFN, Sezione di Genova, Via Dodecaneso 33, 16146 Genova, Italy \label{addr1}
           %\emph{Present Address:} if needed\label{addr3}
%}

\begin{abstract}
 The total decay widths of the charmed baryons are calculated by means of the $^3P_0$ model. Our calculations consider in the final states: the charmed baryon-(vector/pseudoscalar) meson pairs and the (octet/ decuplet) baryon-(pseudoscalar/vector) charmed meson pairs, within a constituent quark model. Furthermore, we calculate the masses of the charmed baryon ground  states and their excitations up to the $D$-wave in a constituent quark model both in the three-quark and in the quark-diquark schemes, utilizing a Hamiltonian model based on a harmonic oscillator potential plus a mass splitting term that encodes the spin, \mbox{spin-orbit}, isospin, and flavor interactions. 
The parameters of the Hamiltonian model are fitted to the experimental data of the charmed baryon masses and decay widths.  As the experimental uncertainties of the data affect the fitted model parameters, we have thoroughly propagated these uncertainties into our predicted charmed baryon masses and decay widths via a Monte Carlo bootstrap approach, which is often absent in other theoretical studies on this subject. 
Our quantum number assignments and predictions of the masses and strong partial decay widths are in reasonable agreement with the available data. Thus, our results show the ability to guide future measurements in LHCb, Belle and Belle II experiments.
 Finally, the  appendices provide some details of our calculations, in which we include the flavor coupling coefficients, which are useful for further theoretical investigations.

\end{abstract}

\maketitle

%%%%%%%%%%%%%%%%%%%%%%%%%%%%%%%%%%%%%%%%%%%%%%%%%%%%%
\section{Introduction}
The discovery of new baryon resonances in high-energy physics experiments always enriches our knowledge of the hadron zoo, and provides essential information to explain the fundamental forces that govern nature. 
 In particular, the hadron mass patterns carry information regarding the way the quarks interact with one another, and provide further insight into the fundamental binding mechanism of matter at an elementary level.

The number of observed charmed baryons has increased owing to the LHCb and Belle experiments. 
In 2017, the LHCb collaboration announced the observation of five narrow $\Omega_{ c}$ states in the $\Xi _{c}^{+}K^{-}$  decay channel \cite{PhysRevLett.118.182001}.
Later,  Belle  observed five resonant states in the $\Xi_c^{+} K^{-}$ invariant mass distribution and unambiguously confirmed four of the states announced by LHCb, $\Omega_{ c}(3000)$, $\Omega_{ c}(3050)$, $\Omega_{ c}(3066)$ and $\Omega_{ c}(3090)$, although no signal was found for the $\Omega_{ c}(3119)$ state \cite{PhysRevD.97.051102}.
Belle also measured a signal excess at 3188 MeV, corresponding to the $\Omega_{ c}(3188)$ state reported by LHCb 
\cite{PhysRevD.97.051102}. In 2020, the LHCb collaboration observed three new states, $\Xi^0_{c}(2923)$, $\Xi^0_{c}(2939)$ and $\Xi^0_{c}(2965)$ \cite{Aaij:2020yyt}; however, their $J^P$ quantum numbers were not reported.  These results reported by LHCb implied that the
  $\Xi^0_{c}(2930)$ broad state  observed by Belle \cite{Belle:2017jrt} and BaBar \cite{BaBar:2007xtc} resolves into two narrower states, $\Xi^0_{c}(2923)$ and $\Xi^0_{c}(2939)$. Nevertheless, 
 a puzzle emerges in the experimental data, since 
Ref. \cite{Aaij:2020yyt} reported a narrow state with a central mass of about 2965 MeV, which is close to a resonance seen by the Belle collaboration at 2970 MeV \cite{Yelton:2016fqw,Belle:2006edu}, and confirmed by the BaBar collaboration \cite{BaBar:2007zjt}; hence, further studies are required in order to determine whether these observations correspond to different baryons or to the same one.
Moreover, the available charm baryon data are limited, especially for the $\Sigma_c$ resonances; indeed, only three states  are  reported by the PDG \cite{Zyla:2020zbs}, $\Sigma_c(2455)$, $\Sigma_c(2520)$ and $\Sigma_c(2800)$, while new analyses are being carried out in this sector \cite{Belle:2021qip}. More recently, in 2021, the Belle collaboration measured the spin and parity of the $\Xi_c(2970)$ state to be $J^P=1/2^ +$ \cite{PhysRevD.103.L111101}, under an assumption that the lowest partial wave dominates the decay.

The application of the non-relativistic quark model to the light baryon spectrum  owes its origins to the pioneering investigations by Isgur and Karl \cite{ISGUR1977109,Isgur:1978xj}, which were further extended in \cite{Copley:1979wj} to the $\Lambda_c$ and $\Sigma_c$ baryons and to the 
$\Lambda_b$ and $\Sigma_b$ baryons in \cite{Capstick:1986bm}.
 Over the last few years, the interest in heavy-light baryon spectroscopy has grown once more. 
Examples of the recent ample literature on theoretical investigations into the heavy baryon spectroscopy: the reports of the QCD-motivated relativistic quark-diquark model based on the quasi-potential approach \cite{Ebert:2011kk,Ebert:2007nw}, the non-relativistic quark model \cite{Copley:1979wj,Roberts:2007ni,Yoshida:2015tia,Chen:2016iyi},  the QCD sum rules in the
framework of the Heavy Quark Effective Theory (HQET) \cite{Chen:2015kpa,Chen:2016phw,Bagan:1992tp,Yang:2021lce} and the symmetry-preserving Schwinger-Dyson equation approach \cite{Gutierrez-Guerrero:2019uwa}.
Alternative discussions employing other models can be found in Refs. \cite{Garcilazo:2007eh,Hasenfratz:1980ka,Savage:1995dw,Kim:2020imk,Kim:2021ywp}, and lattice QCD studies in Refs.  \cite{Vijande:2014uma,Liu:2009jc,Briceno:2012wt,Bahtiyar:2020uuj}.
For extra references, see the review articles \cite{Korner:1994nh,Chen:2016spr,Crede:2013kia,Amhis:2019ckw,Cheng:2015iom}. 
 The spin-parity quantum numbers for most of the charmed baryon states which are reported by the PDG \cite{Zyla:2020zbs} are not measured yet, but they have been extracted from quark model predictions.
%Indeed, only few data determine the quantum numbers for the {\color{red}charmed baryon} states; some of these kinds of assignments, are yet unmeasured, have been extracted from quark model predictions from the PDG \cite{Zyla:2020zbs}. 
Furthermore, it is unclear whether the heavy baryons behave as quark-diquark or three-quark systems. Thus, a full understanding of the internal structure of the charmed baryons still requires thorough theoretical and experimental studies. 

Numerous studies have been conducted on the heavy baryon decay widths. Nevertheless, a complete calculation of all charmed baryon partial strong  decay widths for ground and excited states up to the $D$-wave shell within the same model has never been performed.
For example, within the framework of the chiral quark models in Ref.~\cite{Zhong:2007gp}, only the open-flavor strong decay widths $\Lambda_c\to \Sigma_c \pi, D^0 p$  and $\Sigma_c\to \Lambda_c \pi,  \Sigma_c \pi, D^0 p$ were calculated. Additionally, in Ref.~\cite{Liu:2012sj} the $\Xi_c^{'}$ strong decays were considered up to the $D$-wave shell, while no predictions of the other charmed baryon decays were made. In Refs.~\cite{Wang:2017kfr,Wang:2017hej} the authors calculated the $S$- and $P$-wave  heavy baryon decay widths; however, their analysis was limited to baryons decaying only into ground-state charmed baryons plus pseudoscalar mesons. Moreover, no $D$-wave or radial excitations were reported. In the framework of the heavy hadron chiral perturbation theory in Ref.~\cite{Cheng:2006dk}, certain decays of  $\Lambda_c$, $\Sigma_c$ and $\Xi_c^{'}$ baryons were computed although these calculations did not include the charmed baryon-vector meson channels and did not give predictions for the $\Omega_c$ states. In Ref.~\cite{Cheng:2015naa} the calculations were performed only for the $S$- and $P$-wave $\Lambda_c$, $\Sigma_c$ and $\Xi_c^{'}$ states that decay into a ground-state charmed baryon plus a pion.
Adopting a non-relativistic quark model, in Ref.~\cite{Nagahiro:2016nsx}  only the decay widths of the charmed baryons 
$\Lambda_{c}^{*}(2595)$, $\Lambda_{c}^{*}(2625)$, $\Lambda_{c}^{*}(2765)$, $\Lambda_{c}^{*}(2880)$ and $\Lambda_{c}^{*}(2940)$ into $\Sigma_{c}(2455) \pi$ and $\Sigma_{c}^{*}(2520) \pi$, and of $\Sigma_{c}(2455)$ and $\Sigma_{c}^{*}(2520)$ into $\Lambda_{c} \pi$, were evaluated. In a more recent work \cite{Arifi:2021orx}, the same decay widths were calculated by adding relativistic corrections, and the previous analysis was extended to the decay widths of bottom baryons. In the context of the elementary emission model \cite{Yao:2018jmc}, the strong and radiative decays of  charmed and bottom baryons were investigated. However, the study was restricted to the low-lying $\lambda$-mode $D$-wave excitations and the charmed baryon-vector meson channels or the charmed meson-octet/decuplet baryon channels were not included. In the framework of QCD sum rules in Ref. \cite{Zhu:2000py}, the author studied only the $P$-wave $\Lambda_c \to \Sigma_c + \pi$ decays and the $P$-wave $\Lambda_c$ electromagnetic decays, while in \cite{Chen:2017sci} the authors calculated the $P$-wave charmed baryon decays into ground-state charmed baryons accompanied by a pseudoscalar meson. In \cite{PhysRevD.75.094017}, the $^3P_0$ model was applied  to calculate the strong decays of $\Lambda_c$, $\Sigma_c$, and $\Xi_c$ excited states up to the $D$-wave shell. Nevertheless the decay widths into charmed baryon-vector mesons were not calculated, nor was the $\Omega_c$ sector considered. The $^3P_0$ model was also applied in \cite{Guo:2019ytq,Gong:2021jkb,Lu:2018utx}. In these references, however only the $\Lambda_c$ decays were studied. In \cite{Chen:2016iyi}, the Eichten, Hill and Quigg formula, in combination with the $^3P_0$ model, was applied in order to calculate the 1$P$ and 2$S$
$\Lambda_c$, $\Sigma_c$ and $\Xi_c$ decays into charmed baryon and pseudoscalar mesons.
%   $^3P_0$ model  and eem to do \cite{}
%  $^3P_0$ model for $\Omega_c$ states \cite{Santopinto2019},
%It is also worth to mention that the LHCb collaboration has just announced the observation of a new bottom baryon, $\Xi_b(6227)^-$, in both $\Lambda_b^0 K^-$ and $\Xi^0_b \pi$ decay modes \citep{PhysRevLett.121.072002}, and of two bottom resonances, $\Sigma_b(6097)^\pm$, in the $\Lambda_b^0 \pi^\pm$ channels \citep{PhysRevLett.122.012001}. 
{{}
\\
In Ref.~\cite{Santopinto2019}, prompted by the observation of the five $\Omega_c$ by LHCb \cite{PhysRevLett.118.182001}, we calculated the $\Omega_c$ decay widths in the $\Xi_c^{+} K^{-}$ and $\Xi_c^{'+} K^{-}$ channels within the $^3P_0$ model.  In that study, we also calculated the $\Omega_b$ decay widths in the $\Xi_b^{+} K^{-}$ and $\Xi_b^{'+} K^{-}$ channels and gave predictions for the mass spectra of both $\Omega_c$ and $\Omega_b$ ground states and $P$-wave excitations.
Subsequently, in Ref.~\cite{Bijker:2020tns}, we extended our model to the $\Xi_c^{'}$ and the $\Xi_b^{'}$ states and calculated the mass spectra and  the strong partial decay widths of the  $\Xi_c^{'}$-ground states  and $P$-wave excitations into $^2\Sigma_c \bar{K}$, $^2\Xi_c^{'} \pi$, $^4\Sigma_c \bar{K}$, $^4\Xi_c^{'} \pi$, $\Lambda_c^{} \bar{K}$, $\Xi_c^{} \pi$ and $\Xi_c^{} \eta$  and of the $\Xi_b^{'}$-ground states  and $P$-wave excitations into $^2\Sigma_b \bar{K}$, $^2\Xi_b^{'} \pi$, $^4\Sigma_b \bar{K}$, $^4\Xi_b^{'} \pi$, $\Lambda_b^{} \bar{K}$, $\Xi_b^{} \pi$ and $\Xi_b^{} \eta$, within both the Elementary Emission Model (EEM) and the $^3P_0$ model. %We also calculated the electromagnetic decay widths for %$\Xi'_{c/b}$ and $\Xi_{c/b}$ radiative decays.
In the present article we further extend our model to the whole charmed baryon states ($cqq,cqs$ and $css$ systems) by employing the same mass formula originally introduced in Ref. \cite{Santopinto2019}. Additionally, in the present paper, the parameters of the model are fitted in order to globally reproduce all the available charmed baryon experimental states.
The experimental uncertainties are also propagated to the model parameters by means of the Monte Carlo bootstrap method \cite{Efron1994}, which is an advanced method used to properly estimate the error propagation by taking into account the correlation between the fitted parameters. 
In this way, we perform a global fit of a single model, in which the same set of parameters predicts the charmed baryon masses and strong partial decay widths.
Moreover, considering the well-established observation by Isgur and Karl in Ref.~\cite{ISGUR1977109} that the harmonic oscillator wave functions are a good approximation of the eigenfunctions of low-lying states% of a system bound by Coulomb-plus-linear potentials
, and also taking into account that the calculations of the strong decay widths are barely sensitive to the specific model used \cite{PhysRevD.47.1994}, our strong partial decay width predictions are the most complete calculations in the charmed baryon sector up to date.
% which is similar to the equal spacing mass rule observed in $SU_f(3)$, ground states in  the Gell-Mann Okubo
%and G\"ursey and Radicati mass formulas \cite{Gell-Mann:1962yej,Okubo:1961jc,Gursey:1964htz}, but now it is generalized for the charm sector,   and the $^3P_0$ model for the decay widths.
}
The paper is organized as follows: in Sec. \ref{methodology}, we introduce the details of the methodology used to construct the charmed baryon states and to calculate the mass spectra and decay widths. The theoretical details for the calculation of the charmed baryon mass spectra include contributions due to \mbox{spin-orbit-}, \mbox{spin-}, \mbox{isospin-} and \mbox{flavor-dependent} interactions. Thus, we develop a formalism for obtaining the $S$-, $P$- and $D$-wave charm baryon mass spectrum. We also describe the calculation of the total decay widths of the charmed baryons via the $^3P_0$ model. In Sec.~\ref{Parameter determanation}, we carefully study the parameters of the mass formula presented in Ref.~\cite{Santopinto2019} and perform a global fit to the data on the well-established charmed baryons and their uncertainties, which have been propagated by means of the bootstrap method. In Sec.~\ref{Discussion}, we present the masses and widths of all charmed baryons up to \mbox{$D$-wave} and discuss our assignments for all the available experimental data.  In section \ref{comparison}, we discuss 
why the presence or absence of the $\rho$-mode excitations in the experimental spectrum is the key to distinguishing between the quark-diquark and three-quark behaviours \cite{Santopinto2019}. Finally, in Sec.~\ref{conclusion}, we state our conclusions.

%%%%%%%%%%%%%%%%%%%%%%%%%%%%%%%%%%%%%%%%%%%%%%%%%%%%%%%
\section{Methodology}
\label{methodology}
\subsection{Mass spectra of charmed baryons}
\label{secIIB}
The masses of the charmed baryon states are calculated as the eigenvalues of the Hamiltonian   of Ref. \cite{Santopinto2019}, which is modeled as:
\begin{eqnarray}
\label{eq:mass}
	H = H_{\rm h.o.}+P_s\; {\bf S }^2 + P_{sl} \; {\bf S} \cdot {\bf L} +P_I\;  \bm{I}^2+P_f \; {\bf C_2}(\mbox{SU(3)}_{\rm f}),%\mathtt  
	\nonumber \\
	\label{MassFormula}
\end{eqnarray}
%here
${\bf S}, {\bf L}, {\bm I}$ and ${\bf C_2}(\mbox{SU(3)}_{\rm f})$ are the spin, orbital momentum, isospin and Casimir operators, respectively. These terms are weighted with the model parameters $P_s,P_{sl},P_I$ and $P_f$, as indicated in Eq.~\ref{MassFormula}. Notice that our mass formula in Eq. \ref{MassFormula} is independent  of $I_z$, the isospin projection; therefore, the charge channels are degenerated in this model.% ${\bf C_2}(\mbox{SU(3)}_{\rm f}) $

For the case in which the baryon is modeled as a three-quark system, the three-dimensional h.o. Hamiltonian reads as,
\begin{eqnarray}
 H_{\rm h.o.} =\sum_{i=1}^3m_i + \frac{\mathbf{p}_{\rho}^2}{2 m_{\rho}} 
+ \frac{\mathbf{p}_{\lambda}^2}{2 m_{\lambda}} 
+\frac{1}{2} m_{\rho} \omega^2_{\rho} \boldsymbol{\rho}^2   
+\frac{1}{2}  m_{\lambda} \omega^2_{\lambda} \boldsymbol{\lambda}^2
	\nonumber \\
\label{eq:Hho}
\end{eqnarray}
written in terms of  Jacobi coordinates, $ \boldsymbol{\rho}$ and $\boldsymbol{\lambda}$, and 
conjugated momenta, $\mathbf{p}_{\rho}$ and $ \mathbf{p}_{\lambda}$.
The $H_{\mathrm{h.o}}$ eigenvalues are 
\begin{eqnarray}
    \label{eq:freq}
    \sum_{i=1}^3m_i + \omega_{\rho} \; n_{\rho} + \omega_{\lambda}\; n_{\lambda};\; \mathrm{with}\;
\omega_{\rho(\lambda)}=\sqrt{\frac{3K_c}{m_{\rho(\lambda)}}},
\end{eqnarray}
where $m_{i}$ are the constituent quark masses,
$m_1$ and $m_2$ correspond to the light quarks and $m_3$ to the charm quark;
$m_\rho$ is defined as $m_\rho=(m_1+m_2)/2$,
%are the masses of the light quarks the charm baryon,
and $m_\lambda=3m_\rho m_3/(2m_\rho+m_3)$.
%is the mass of the charm quark. 
We use the well-known definitions for
$ n_{\rho(\lambda)}= 2 k_{\rho(\lambda)}+l_{\rho(\lambda)}$, 
$k_{\rho(\lambda)}=0,1,...$,   and $l_{\rho(\lambda)}=0,1,...$; here, $l_{\rho(\lambda)}$ are the orbital angular momenta of the $\rho$($\lambda$) oscillators,  and $k_{\rho(\lambda)}$ is the number of nodes (radial excitations) in the $\rho$($\lambda$) oscillators. $K_c$ is the spring constant.

\textcolor{black}{
Additionally, we present a simplification of the three-quark system that utilizes only one  relative coordinate $\boldsymbol{r}$ and momentum $ \mathbf{p}_{r}$, namely, the quark-diquark system. Here, the two light quarks are regarded as a single diquark object}. The quark-diquark Hamiltonian reads as,
\begin{eqnarray}
 H_{\rm h.o.} =m_D + m_c 
+ \frac{\mathbf{p}_{r}^2}{2 \mu }
+\frac{1}{2} \mu \omega^2_{r} \boldsymbol{r}^2,
	%\nonumber \\
\label{eq:Hhodi}
\end{eqnarray}
with $\mathbf{p}_{r}=(m_c\mathbf{p}_{D}-m_D\mathbf{p}_{c})/(m_c +m_D)$. The $H_{\rm h.o}$ eigenvalues are
\begin{eqnarray}
    \label{eq:freq_di}
    m_D + m_c + \omega_{r}\; n_{r};\; \mathrm{with}\;
\omega_{r}=\sqrt{\frac{3K_c}{ \mu}},
\end{eqnarray}
\textcolor{black}{where $m_D$ is the diquark mass, $m_c$ is the charm quark mass; $\mu$ is the reduced mass of the system, and is defined as $\mu=m_c m_D/(m_c +m_D)$; $n_{r}$ and $K_c$ are defined as in the three-quark system.}

\subsection{Charmed baryon states}
\label{secIIA}
In the three-quark model, the baryon states are  thought as a $qqQ$ system, where $Q=c$ and $q=u,d,s$. The three-quark Hamiltonian in Eq. \ref{eq:Hho} is expressed in terms of two coordinates, $\boldsymbol{\rho}$ and $\boldsymbol{\lambda}$~\cite{PhysRevD.18.4187}, that encode the spatial degrees of freedom of the system with associated effective masses, $m_{\rho}$ and $m_{\lambda}$. Note that in heavy-light baryons, for which $m_{\rho} \ll m_{\lambda}$, the two excitation modes can be decoupled from each other as long as the heavy-light quark mass difference is significant.

In the quark-diquark system, the baryon states are  thought as a  $DQ$ system, where $Q=c$ and $D=D_{\Omega}, D_{\Xi}, D_{\Sigma,\Lambda}$ are the diquarks that correspond to the $\Omega_c$, $\Xi_c' (\Xi_c)$, $\Sigma_c$ and $\Lambda_c$ baryons, respectively. The quark-diquark Hamiltonian in Eq.\ref{eq:Hhodi} is expressed in terms of one spatial coordinate $\boldsymbol{r}$ with an associated reduced mass ${\mu}$;  $i.e.$, the quark-diquark system resembles a diatomic molecule.

We construct the ground and excited states in order to establish the quantum numbers of the charmed baryon states. 

We consider that a single quark is described by its spin, flavor, color and spatial degrees of freedom. The baryon states should be in color singlet. Moreover, in our model the light quarks are considered to be identical particles; hence, their wave function should be antisymmetric in order to satisfy the Pauli Principle. Since the two light quarks should be in the  ${\bf \bar{3}_{\rm c}}$, their
 spin-flavor and orbital wave functions should have the same permutation symmetry:  spin-flavor symmetric in the $S$-wave or $D$-wave ${(\bf {\bar{3}}}_{\rm f}$  with spin 0 or ${\bf 6}_{\rm f}$ with spin 1), and  spin-flavor  antisymmetric in the  $P$-wave ${(\bf {\bar{3}}}_{\rm f}$  with spin 1 or ${\bf 6}_{\rm f}$ with spin 0).%%%%%%%%%%%%%%%%%%%%%%%%%%%%%%%%%%%%%%%%%%%%%%%%%%%%%%%%%%
 
Formally, a three-quark (quark-diquark) quantum state, written as $\left| l_{\lambda}(l_{r}),l_{\rho}, k_{\lambda}(k_r),k_{\rho}\right\rangle$, is defined by its total angular momentum ${\bf J_{\rm tot}=} {\bf L}_{\rm tot} + {\bf S}_{\rm tot} $, where ${\bf L_{\rm tot} =l}_{\rho}+{\bf l}_{\lambda}$ \textcolor{black}{(${\bf L_{\rm tot} =l}_{r})$} and ${\bf S}_{\rm tot} = {\bf S}_{\rm lt}+\frac{1}{2}$, ${\bf S}_{\rm lt}$ is the coupled spin of the light quarks and the number of nodes is $k_{\lambda, \rho}(k_r$). In addition, in order to unambiguously define these quantum states, we assign to them the flavor $\mathcal{F}$ and spectroscopy $^{2S+1}L_{J}$ representations. In the following paragraphs, we will construct the possible states for the two different flavor representations available for the charmed baryons, in the energy bands $N=n_\rho+n_\lambda$ $(N=n_r)$ and $N=0,1,2$ in order to find $S$-, $P$-, $D$-wave charmed baryon states.
 
\subsubsection{The symmetric ${\bf {6}}_{\rm f}$-multiplet for the three-quark model}
The $\Omega_c$, $\Xi'_c$, and $\Sigma_c$ baryons form a flavor sextet in the charmed baryon sector.  These charmed baryons have a symmetric flavor-wave function  of the light quarks which,  in combination with their  antisymmetric color-wave function,  produces an antisymmetric wave function. This implies that the product of the spatial and spin-wave functions of the light quarks should be symmetric. In the energy band $N=0$, if $l_\rho=l_\lambda=0$, the spatial wave function of the two light quarks is symmetric. That is, these states have a symmetric-spin wave function of the two light quarks, meaning $\bf S_{\rm lt}=1$. Hence, two ground states, $\bf J_{\rm tot}={\bf S}_{\rm tot}=1/2,3/2$, exist. %written as (borre estos estados por que para los demas no aparece la notacion dirac) $\vert qqc,1/2,1/2,0,0,0,{\bf {6}}_{\rm f} \rangle$ and $\vert qqc,3/2,3/2,0,0,0,{\bf {6}}_{\rm f} \rangle$, respectively.
For the energy band $N=1$, there are two different possibilities. If $l_\rho=0$ and $l_\lambda=1$, we again have spatial-symmetric wave functions under the interchange of light quarks, which must be coupled with two possible spin configurations, $\bf S_{\rm tot}=1/2,3/2$, with $\bf{\rm{\bf L_{\rm tot}}}=1$, yielding five {$P_\lambda$}-wave excitations. If  $l_\rho=1$ and $l_\lambda=0$, the spatial wave function is antisymmetric under the interchange  of light quarks implying that the two light quark spin wave function antisymmetric, meaning $\bf S_{\rm lt}=0$, which yields two $P_\rho$-wave states obtained from $\bf J_{\rm tot}=1+1/2=1/2,3/2$.  In the energy band $N=2$, when $l_\rho=0$ and $l_\lambda=2$, the total spatial wave function is symmetric. Thus, it must be combined with two possible spin configurations, $\bf S_{\rm tot}=1/2,3/2$, obtaining six  $D_\lambda$-wave excitations. Additionally, there are radial excitation modes in this band. For the case $k_\rho=0$ and $k_\lambda=1$, the spatial wave function is symmetric. Thus, the two light quark spin wave function must also be symmetric, yielding two $\lambda$-radial excitations, $\bf J_{\rm tot}={\bf S}_{\rm tot}=1/2,3/2$, since $\bf L_{\rm tot}=0$. The same situation appears when $k_\rho=1$ and $k_\lambda=0$; the spatial wave function is symmetric. Thus, there are two $\rho$-radial excitations, corresponding to $\bf J_{\rm tot}={\bf S}_{\rm tot}=1/2,3/2$. In the case $l_\rho=1$ and $l_\lambda=1$, which yields ${\bf L_{\rm tot}=2, 1, 0}$, the two light quark spatial wave function  antisymmetric, implying that we have to couple it with the light quark  antisymmetric spin configuration,   $\bf S_{\rm lt}=0$, thus obtaining five possible states: two $D$-wave states, two $P$-wave states, and one $S$- wave state. Finally, if $l_\rho=2$ and $l_\lambda=0$, the spatial wave functions are symmetric. Hence, we have to combine them with $\bf S_{\rm tot}=1/2,3/2$, obtaining six {$D_\rho$}-wave excitations.

\subsubsection{The symmetric ${\bf {6}}_{\rm f}$-multiplet for the quark-diquark model} \label{DQ6plet}

When the $\Omega_c$, $\Xi'_c$, and $\Sigma_c$ baryons are seen as quark-diquark systems, the two constituent light quarks of the diquark are considered to be correlated, with no internal spatial excitations ($S$-wave); $i.e$., it is hypothesized that we are within the limit where the diquark internal spatial excitations are higher in energy than the scale of the resonances studied. Since the hadron must be colorless, the diquark transforms as $\bf \bar 3$ under $SU_c(3)$; thus, the product of the spin and flavor wave functions of the diquark configuration should be symmetric. The flavor wave functions of the ${\bf {6}}_{\rm f}$ representation are symmetric. As a result, we can only combine with axial-vector diquarks; that is, with ${\bf S_{\rm lt}=1}$.  For the energy band $N=0$, ${\bf L_{\rm tot}=0}$, and thus
$\bf J_{\rm tot}={\bf S}_{\rm tot}=1/2,3/2$, yielding two ground states. In the next band $N=1$, ${\bf L_{\rm tot}=1}$ has to be coupled with ${\bf S_{\rm tot}=1/2,3/2}$, yielding five $P$-wave excitations. For the band $N=2$, ${\bf L_{\rm tot}=2}$, and we must combine with ${\bf S_{\rm tot}=1/2,3/2}$, to get six $D$-wave states. Moreover, there is a radial degree of freedom $k=1$; with ${\bf L_{\rm tot}=0}$, we have $\bf J_{\rm tot}={\bf S}_{\rm tot}=1/2,3/2$, and hence find two radial excitations. 

\subsubsection{The  antisymmetric ${\bf {\bar{3}}}_{\rm f}$-plet for the three-quark model}
The $\Lambda_c$ and $\Xi_c$ baryons form a flavor-anti-triplet in the charmed baryon sector. These charmed baryons have an  antisymmetric flavor wave function of the light quarks, and which, in combination with the antisymmetric color wave function, produces a symmetric combination. This implies that the product of the spatial and spin wave functions of the light quarks should be  antisymmetric. For the energy band $N=0$, if $l_\rho=l_\lambda=0$, the spatial wave function of the two light quarks is symmetric, thus their spin wave function should be antisymmetric. This corresponds to $\bf{S}_{\rm lt}=0$, producing only one ground state. % $\vert qqc,1/2,1/2,0,0,0,{\bf {3}}_{\rm c} \rangle$.
For the energy band $N=1$, if $l_\rho=0$ and $l_\lambda=1$, we have a light quark symmetric spatial wave function; thus, their spin wave function is antisymmetric. It implies $\bf{S}_{\rm tot}=1/2$ and, in combination with the total $\bf L_{\rm _{tot}}=1$, yields two $P_\lambda$ states. If $l_\rho=1$ and $ l_\lambda=0$, the spatial wave function of the two light quarks is  antisymmetric. Thus, their spin wave function is symmetric, giving two possible configurations: $\bf{S}_{\rm tot}=1/2,3/2$. This, in combination with $\bf L_{\rm tot}=1$, constructs five $P_\rho$ states.
In the energy band $N=2$, in the case of $l_\rho=0$ and $l_\lambda=2$, the total spatial wave function is symmetric; it is therefore combined with the
light quark antisymmetri spin configuration, $\bf S_{\rm lt}=0$, giving two {$D_\lambda$}-wave excitations. The two possible radial excitations, $k_\rho=0$ and $k_\lambda=1$, and $k_\rho=1$ and $k_\lambda=0$, are symmetric in the spatial wave function. They should be combined with the light quark  antisymmetric spin configuration,  $\bf S_{\rm lt}=0$, producing one $\lambda$-radial excitation and one $\rho$-radial excitation. The  antisymmetric spatial wave functions of the configuration $l_\rho=1$ and $l_\lambda=1$ are coupled to ${\bf L_{\rm tot}=0, 1, 2}$, and the angular momenta should be combined with $\bf{S}_{\rm tot}=1/2,3/2$,   coming from the light quark spin configuration $\bf S_{\rm lt}=1$, producing  thirteen mixed excited states: six $D$-wave states, five $P$-wave states, and two $S$- wave states. Finally, the symmetric configuration $l_\rho=2$ and $l_\lambda=0$, combined with the light quark  antisymmetric spin configuration, $\bf S_{\rm lt}=0$, gives two {$D_\rho$}-wave excitations.

\subsubsection{The  antisymmetric  ${\bf {\bar{3}}}_{\rm f}$-plet for the quark-diquark model}
Moreover, $\Lambda_c$ and $\Xi_c$ baryons are described as quark-diquark systems. In this case, as discussed in subsection \ref{DQ6plet}, the diquark presents an $S$-wave configuration, given its lack of internal spatial excitations. Considering that it is 
$\bf \bar 3$ in the color representation $SU_c(3)$, we conclude that the product of the spin and flavor wave functions of the diquark configuration should be symmetric. In the  antisymmetric ${\bf {\bar{3}}}_{\rm f}$-plet, the flavor wave function is  antisymmetric; thus, the spin wave function of the diquark correspond to a scalar configuration, ${\bf S_{\rm lt}=0}$. For the energy band $N=0$, we have ${\bf L=0}$; thus, we only have one ground state $\bf J_{\rm tot}= \bf{S}_{\rm tot}=1/2$. In the next band, $N=1$, we must combine ${\bf L{\rm _{tot}}=1}$ with $\bf{S}_{\rm tot}=1/2$, which yields two $P$-wave states. In the band $N=2$, we have ${\bf L{\rm _{tot}}=1}$, and, on coupling to $\bf{S}_{\rm tot}=1/2$, we get two $D$-wave states. Finally, with a radial excitation $k_r=1$ and ${\bf L{\rm _{tot}}=0}$, there is only one state.

\subsection{Charmed baryon decay widths}
\begin{figure}
    \centering
    a)
    
    \includegraphics[scale=.5]{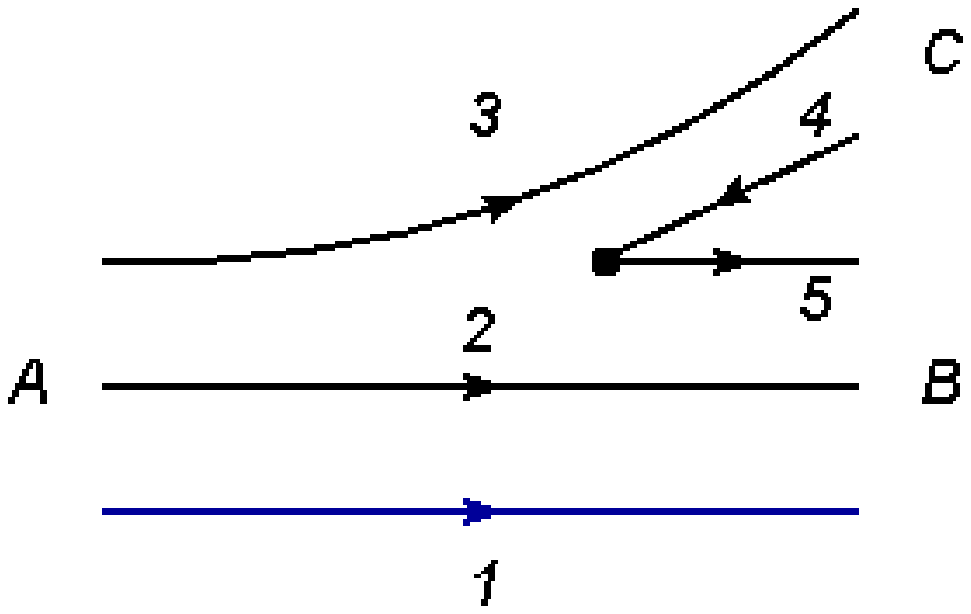}
    
    b)
    
    \includegraphics[scale=.5]{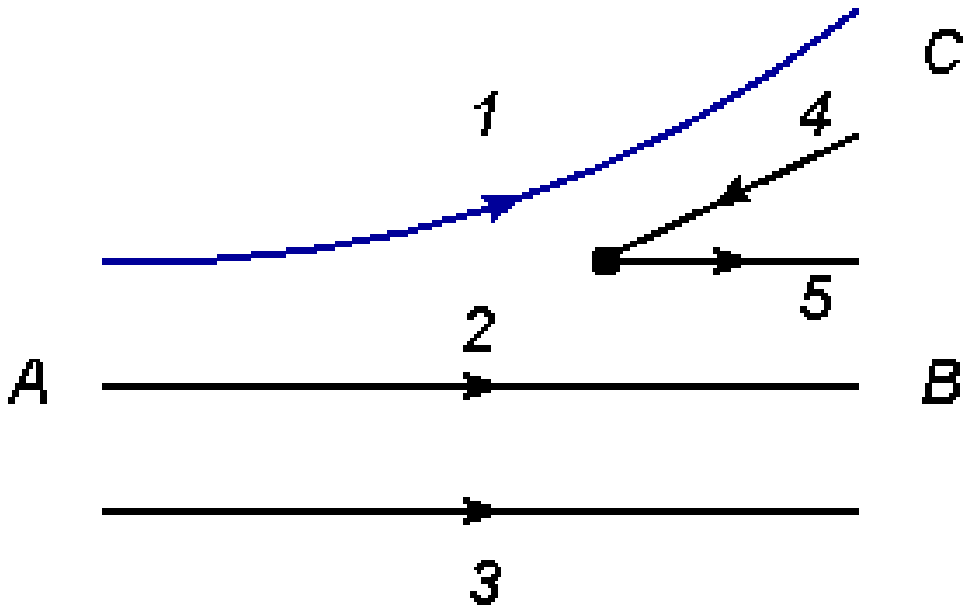}
    \caption{{} The $^3P_0$ pair-creation model (color online). The red line 1 denotes a
charm quark, while the remaining black lines denote light quarks. In diagram  a) the charmed baryon $A$ decays to a charmed baryon $B$ and a light meson $C$. In diagram b) the charmed baryon $A$ decays to a light baryon and a charmed meson $C$.  }
    \label{fig:3p0}
\end{figure}
\label{secIIC}
The open-flavor strong decays of a charmed baryon $A$ to another baryon $B$ plus a meson $C$, $A\rightarrow BC$, have been studied by means of the $^3P_0$ model ~\cite{MICU1969521,PhysRevD.8.2223,PhysRevD.94.074040,PhysRevD.75.094017}. 
 According to this model, a $q\bar q$ pair
%of quarks with $J^{PC}=0^{++}$ 
is created from the vacuum when a $qqc$ baryon decays and  regroups into an outgoing meson and
a baryon via the quark rearrangement process as depicted in Fig. \ref{fig:3p0}.   {{} In the present study, we consider the decay of a charmed baryon $A$ to a charmed baryon $B$ plus a light meson $C$, see Fig \ref{fig:3p0} a), and also the case in which the final state is a light baryon $B$ plus a charmed meson $C$, see Fig.  \ref{fig:3p0} b)}. Within the
non-relativistic limit, the transition operator is written as
\begin{eqnarray}
T^\dagger &=& - 3 \gamma_0 \sum_m\: \langle 1\;m;1\;-m|0\;0 \rangle\,
\int\!{\rm d}^3{\textbf{P}}_4\; {\rm d}^3{\textbf{P}}_5
\delta^3({\textbf{P}}_4+{\textbf{P}}_5)\nonumber\\&\times& {\cal
Y}^m_1\Big({\textbf{P}}_4-{\textbf{P}_5}\Big)\; \chi^{4
5}_{1, -\!m}\; \varphi^{4 5}_0\;\, \omega^{4 5}_0\;
b^\dagger_{4}({\textbf{P}}_4)\; d^\dagger_{5}({\textbf{P}}_5),
\label{tmatrix}
\end{eqnarray}
where $4$ and $5$ are the  indices of the quark and
anti-quark created. $\varphi^{45}_{0}=(u\bar u +d\bar d +s \bar s)/\sqrt
3$ and $\omega_{0}^{45}=(r\bar r +b\bar b +g \bar g)/\sqrt
3$ are the flavor and color
singlet-wave functions, respectively. $\chi_{{1,-m}}^{45}$ is the spin-triplet
state. $\mathcal{Y}_{1}^{m}(\mathbf{k})\equiv
|\mathbf{k}|Y_{1}^{m}(\theta_{k},\phi_{k})$ is a solid harmonic
polynomial corresponding to the $P$-wave quark pair. $\gamma_0$ is a
dimensionless constant related to the strength of the $q\bar q$ pair
creation vertex from the vacuum. $\gamma_0$ is a free parameter of the $^3P_0$ model.    

The total decay width $\Gamma$  is the sum of the partial widths for the open channels $BC$, $\Gamma=\sum_{BC}\Gamma_{A \rightarrow BC}$, where 
%of a initial baryon $A$ decaying to a final baryon $B$ plus a meson $C$,
the partial widths $\Gamma_{A \rightarrow BC}$, are computed as  
\begin{eqnarray}
	\Gamma_{A \rightarrow BC}& = & \frac{2 \pi \gamma_0^2}{2J_{A}+1} \mbox{ } \Phi_{A \rightarrow BC}(q_0) \nonumber \\
	&\times& \sum_{M_{J_A},M_{J_B}} \big|\mathcal{M}^{M_{J_A},M_{J_B}}\big|^2 \mbox{ }.  %\nonumber
        \label{gamma}
\end{eqnarray}
Here, $\mathcal{M}= \langle BC|T^\dagger|A\rangle$ is the $^3P_0$ amplitude written in terms of hadron h.o. wave functions and the sum runs over the projections $M_{J_A,J_B}$ of the total angular momenta $J_{A,B}$ of $A$ and $B$.
$q_0$ is the relative momentum between $
B$ and $C$, and the coefficient $\Phi_{A \rightarrow BC}(q_0)$ is the relativistic phase space factor~\cite{PhysRevD.94.074040},
\begin{eqnarray}
	\label{eqn:rel-PSF}
	\Phi_{A \rightarrow BC}(q_0) = q_0 \frac{E_B(q_0) E_C(q_0)}{m_A}  \mbox{ }, \nonumber\\
	\mbox{with}\;\;E_{B,C} = \sqrt{m_{B,C}^2 + q_0^2}; \nonumber
\end{eqnarray}
where $m_A$ is the initial charmed baryon mass in its rest frame. The masses $m_B$ and $m_C$ and energies $E_B$ and $E_C$ correspond to the final baryon and meson, respectively.

The h.o. wave functions depend on the parameters $\alpha_{\rho(\lambda)}$, see  \ref{BWF}, which, in Ref.~\cite{PhysRevD.94.074040}, are regarded as free parameters. Conversely, in the present study, $\alpha_{\rho(\lambda)}$ are related to the baryon $\rho$- and $\lambda$-mode h.o. frequencies as defined in Eq. \ref{eq:freq}; this relation is established by  $\alpha^2_{\rho(\lambda)}=\omega_{\rho(\lambda)}m_{\rho(\lambda)}$. Therefore, $\alpha_{\rho(\lambda)}$ will depend on the fit parameter $K_{c}$ and constituent quark masses.  The h.o. wave functions and coordinate system conventions used in our decay width calculations are given in \ref{app}. The decay widths are calculated for each charmed baryon type; the available open-flavor channels include all the pseudoscalar and vector mesons. The open-flavor channels share an extra parameter $R$ related to the meson size, which has been discussed extensively in the literature \cite{PhysRevD.32.189,PhysRevD.35.907,PhysRevD.72.094004}; we adopt $R=2.1/$GeV which is taken from Refs. \cite{PhysRevD.75.094017,PhysRevD.53.3700}. The flavor-meson-wave functions are given in \ref{appme}. All the possible flavor couplings, $\mathcal{F}_{A\rightarrow BC}=\langle\phi_B \phi_C|\phi_0 \phi_A  \rangle$ are given in \ref{flavor}. The masses of the decay products are listed in Table~\ref{tab:exp_dec} in \ref{app2}. It is important to mention that the application of the $^3P_0$ model is restricted to the three-quark system.

%%%%%%%%%%%%%%%%%%%%%%%%%%%%%%%%%%%%%%%%%%%%%%%%%%%%%%%%%%%%%%%
\section{Parameter determination and uncertainties}
\label{Parameter determanation}
\begin{table}[htbp!]
\begin{tabular}{c | c c}\hline \hline
            &  three-quark & diquark \\ 
 Parameter  &  Value       & Value    \\ \hline
 $m_{c}$ & $1606^{+58}_{-61}$ MeV & $1563^{+22}_{-24}$ MeV \\ 
 $m_{s}$ & $455^{+29}_{-27}$ MeV & $\dagger$ \\ 
 $m_{u,d}$ & $284^{+30}_{-31}$ MeV & $\dagger$ \\ 
 $m_{D_{\Omega}}$          & $\dagger$ & $947^{+3}_{-3}$ MeV \\ 
 $m_{D_{\Xi}}$             & $\dagger$ & $791^{+18}_{-14}$ MeV \\ 
 $m_{D_{\Sigma,\Lambda}}$ & $\dagger$ & $613^{+20}_{-17}$ MeV \\
 $K_c$   & $0.0290^{+0.0007}_{-0.0008}$ GeV$^{3}$ & $0.0195^{+0.0007}_{-0.0007}$ GeV$^{3}$ \\
 $P_s$     & $23^{+3}_{-3}$ MeV & $24^{+3}_{-3}$ MeV \\ 
 $P_{sl}$     & $18^{+5}_{-5}$ MeV & $17^{+5}_{-5}$ MeV \\ 
 $P_{I}$     & $45^{+8}_{-8}$ MeV & $41^{+9}_{-9}$ MeV \\ 
 $P_{f}$     & $52^{+6}_{-6}$ MeV & $52^{+7}_{-7}$ MeV \\ 
\hline\hline
\end{tabular}
\caption{Fitted parameters for the three-quark model (second column) and  the quark-diquark model (third column). The $\dagger$ indicates that the parameter is absent in that model.}
\label{tab:comb_fit}
\end{table}

\subsection{Mass spectra of charmed baryons}
We fitted a selection of experimentally observed charmed baryon states, $\Omega_{c}$, $\Sigma_{c}$, $\Lambda_{c}$, $\Xi'_{c}$, and $\Xi_{c}$,  to the masses predicted by Eq.~\ref{eq:Hho} and Eq.~\ref{eq:Hhodi} to obtain the constituent quark and diquark masses ($m_{c}$, $m_s$, $m_{u,d}$, $m_{D_{\Omega}}$, $m_{D_{\Xi}}$, and $m_{D_{\Sigma,\Lambda}}$) and the model parameters ($P_s, P_{sl},$ $ P_I, P_f$ and $K_{c}$). The fitted model parameters and masses minimize the sum of the squared differences between the experimental baryon masses and those predicted by the model (least-squares method). %\footnote{$K_{c}$ is defined in Eq.\ref{eq:freq} and is chosen to avoid the $\omega_{\rho(\lambda)}$ dependency on each baryon and h.o. mode.}

The measured baryon masses come with statistical and systematic uncertainties. Furthermore, the models in Eq.~\ref{eq:Hho} and Eq.~\ref{eq:Hhodi} are approximate descriptions of the charm baryons. Thus, to take into account the possible deviations of these models from the experimental observations, we assigned a model uncertainty to each model. The model uncertainty, $\sigma_{mod}$, is calculated in accordance with Ref.~\cite{Zyla:2020zbs} and is such that $\chi^2/NDF\simeq 1$, where 
\begin{equation}
\label{eq:opt_chi}
\chi^2=\sum_{i} \dfrac{(M_{mod,i}-M_{exp,i})^2} {\sigma_{mod}^2 + \sigma_{exp,i}^2},
\end{equation}
 $M_{mod,i}$ are the predicted charm baryon masses, $M_{exp,i}$ are the experimental charm baryon masses included in the fit with uncertainties $\sigma_{exp,i}$, and $NDF$ is the number of degrees of freedom. We obtained $\sigma_{mod}\!=\!15.42$ MeV for the three-quark model and $\sigma_{mod}\!=\!13.63$ MeV for the quark-diquark model.

%The sum in Eq.~\ref{eq:opt_chi} runs over the states considered in the fit.

To integrate the experimental and model uncertainties into our fit, we carried out a statistical simulation of error propagation. To do so, we randomly sampled the experimental masses from a Gaussian shaped distribution with a mean equal to the central mass value and a width equal to the squared sum of the uncertainties. We fitted the model by using a sampled mass corresponding to each experimental observed state included in the fit, and we repeated the procedure $10^4$ times. In this manner, we obtained a Gaussian distribution for each constituent quark mass, model parameter, and the baryon mass itself. Next, we assigned the mean of the distribution as the value of the parameter and used its difference from the distribution quantiles at 68\% confidence level (C.L.), in order to extract the uncertainty. This method is known as the Monte Carlo bootstrap uncertainty propagation \cite{Efron1994,Molina2020}.

The experimental masses and their corresponding uncertainties used in the fit and error propagation are marked with (*) in Tables~\ref{tab:All_mass_Omega}-\ref{tab:All_mass_Xi}. These mass measurements are summarized in the PDG database~\cite{Zyla:2020zbs}. However, the charmed baryon masses predicted by Eq.~\ref{eq:Hho} are degenerated in comparison with different $u$ or $d$ quark configurations, since the model will assign identical masses to baryons with the same number of $u$ or $d$ quark contents. This is a consequence of isospin symmetry.  As these groups of mass states have the same quantum numbers, our quantum number assignments are not affected by the mass degeneracy. In our calculations, to account for this degeneracy, we fitted the arithmetic mean of the measured masses and adopted a conservative approach to the uncertainty by defining it as the standard deviation among the measured masses, plus their highest reported experimental uncertainty. The calculations were carried out by using \texttt{MINUIT} \cite{JAMES1975343} and \texttt{NumPy} \cite{Harris2020}. The results of the fit are shown in Table~\ref{tab:comb_fit}. The constituent quark masses obtained agree with previous theoretical determinations \cite{Isgur:1978xj}. Furthermore, the model parameters used in the present study are in the range of our previous work \cite{Santopinto2019}, where phenomenological considerations were considered to determine them. Tables~\ref{tab:3quark_corr} and \ref{tab:diquark_corr} show the correlation of the fitted parameters in the three-quark and quark-diquark model, respectively. In the three-quark model, the constituent quark masses are highly correlated, indicating that the quark masses exhibit similar behavior inside the baryon. Moreover, the spring constant $K_c$ is also highly correlated with the quark masses, as expected from Eq.~\ref{eq:freq}. In the quark-diquark model, the charm-quark mass is totally uncorrelated with the diquark masses; this is a consequence of the diatomic structure of the modeled baryon. In the same manner, $K_c$ is correlated with the diquark masses, as expected from Eq.~\ref{eq:freq_di}.

\begin{table}[htbp]
\begin{tabular}{c  c  c  c  c  c  c  c  c}\hline \hline
         &  $m_{c}$       &     $m_{s}$    &    $m_{n}$  &      $K_c$    & $P_s$ & $P_{sl}$ & $P_{I}$ & $P_{f}$ \\ \hline
 $m_{c}$ &     1   &   &   &   &    &   &   &  \\ 
 $m_{s}$ & -0.76 &  1   &   &   &   &   &   &  \\ 
 $m_{u,d}$ & -0.82 & 0.76 &  1   &   &   &   &   & \\ 
 $K_{c}$   & -0.77 & 0.7 & 0.69 &  1   &   &   &   &   \\ 
 $P_{s}$     & 0.26 & -0.29 & -0.27 & -0.14 & 1 &   &   & \\ 
 $P_{sl}$     & -0.1 & 0.08 & 0.08 & 0.37 & -0.21 & 1  &   & \\ 
 $P_{I}$     & 0.11 & 0.12 & -0.19 & -0.16 & 0.21 & -0.02 & 1  &  \\ 
 $P_{f}$     & -0.42 & 0.04 & 0.28 & 0.36 & -0.51 & 0.21 & -0.68 & 1 \\ \hline \hline
\end{tabular}
\caption{Correlation between fitted parameters: three-quark system.}
\label{tab:3quark_corr}

\end{table}

\begin{table*}[htbp]
\centering
\begin{tabular}{c | c c c c c c c c c}\hline \hline
         &  $m_{c}$ & $m_{D_{\Omega}}$ & $m_{D_{\Xi}}$ & $m_{D_{\Sigma,\Lambda}}$ &  $K_c$   & $P_s$ & $P_{sl}$ & $P_I$ & $P_f$ \\ \hline
 $m_{c}$ &     1   &   &   &   &    &   &   &  &  \\ 
$m_{D_{\Omega}}$ & 0.0 &  1   &   &   &   &    &   &   &  \\ 
$m_{D_{\Xi}}$ & 0.0 & 0.85 &  1   &   &   &   &    &   & \\ 
$m_{D_{\Sigma,\Lambda}}$ & 0.0 & -0.83 & 0.3 &  1   &   &   &   &    &   \\ 
 $K_c$   & 0.0 & 0.33 & 0.3 & -0.52 & 1   &   &   &   & \\ 
 $P_{s}$     & 0.0 & -0.18 & -0.1 & 0.18 & -0.14 & 1   &   &   & \\ 
 $P_{sl}$     & 0.0 & 0.14 & 0.1 & -0.18 & 0.37 & -0.21 & 1   &   &  \\ 
 $P_{I}$     & 0.0 & -0.72 & -0.78 & 0.63 & -0.16 & 0.21 & -0.02 & 1   & \\ 
 $P_{f}$     & 0.0 & 0.7 & 0.68 & -0.88 & 0.36 & -0.51 & 0.21 & -0.68 & 1 \\ \hline \hline
\end{tabular}
\caption{Correlation between fitted parameters: quark-diquark system.}
\label{tab:diquark_corr}
\end{table*}

\subsection{Charmed baryon decay widths}
The parameter determination and the error propagation for the decay widths were carried out in analogy with the above procedure for the charmed baryon masses. The  pair-creation constant $\gamma_0$ of Eq.~\ref{gamma} was obtained by fitting data of selected charmed baryon decay widths.

To compute the uncertainty of the decay widths, we considered all possible sources of uncertainty. First, the error coming from the baryon mass $m_{A}$ and parameter $K_{c}$ were included by calculating a decay width for all the statistically simulated constituent quark masses, $m_{A}$ and $K_{c}$ and $\gamma_{0}$; each width calculation was then repeated $10^4$ times. Next, we included the experimental uncertainties of the decay products $m_{B}$ and $m_{C}$. These experimental uncertainties, the values of which are shown in Table \ref{tab:exp_dec}, were propagated to the decay widths by means of the same random sampling technique described for the masses. Furthermore, a model uncertainty, $\sigma_{mod}\!=\!4.44$, was included. We set the decay width value as the population mean of the Gaussian distribution obtained, with an error equivalent to the difference between this mean and the distribution quantiles at 68\% C.L. The value and uncertainty obtained are $\gamma_{0}=19.6 \pm 5.1$. These calculations are only performed when the charmed baryons are modeled as three-quark systems.

%%%%%%%%%%%%%%%%%%%%%%%%%%%%%%%%%%%%%%%
\section{Results and Discussion}
\label{Discussion}
%%%%%%%%%%OMEGA_C%STATES%%%%%%%%%%%%%%%
\begin{table*}[htbp]
\begin{tabular}{c  c| c c c c c }\hline \hline
            &  & Three-quark &  Quark-diquark    &               &              &  \\ 
$\Omega_{c}(ssc)$&  & Predicted   &    Predicted   &  Experimental &  Predicted            & Experimental \\ 
$\mathcal{F}={\bf {6}}_{\rm f}$  & $^{2S+1}L_{J}$ & Mass (MeV)  &   Mass (MeV)   &  Mass (MeV)   &  $\Gamma_{tot}$ (MeV) & $\Gamma$ (MeV) \\ \hline
\hline
 $N=0$  &  &  &  &  &  \\ 
$\vert l_{\lambda}\!\!=\!0, l_{\rho}\!\!=\!0, k_{\lambda}\!\!=\!0, k_{\rho}\!\!=\!0 \rangle$ & $^{2}S_{1/2}$ & $2709^{+10}_{-10}$ & $2702^{+9}_{-9}$ & $2695.0\pm 1.7$ (*) & $0$ & $<10^{-7}$ \\ 
$\vert l_{\lambda}\!\!=\!0, l_{\rho}\!\!=\!0, k_{\lambda}\!\!=\!0, k_{\rho}\!\!=\!0 \rangle$ & $^{4}S_{3/2}$ & $2778^{+9}_{-9}$ & $2776^{+9}_{-9}$ & $2765.9\pm 2.0$ (*) & $0$ & $\dagger$ \\ 
\hline
 $N=1$  &  &  &  &  &  \\ 
$\vert l_{\lambda}\!\!=\!1, l_{\rho}\!\!=\!0, k_{\lambda}\!\!=\!0, k_{\rho}\!\!=\!0 \rangle$ & $^{2}P_{1/2}$ & $3008^{+10}_{-10}$ & $3000^{+9}_{-9}$ & $3000.4\pm 0.22$ (*) & $4^{+2}_{-2}$ & $4.5\pm 0.6\pm 0.3$ (*) \\ 
$\vert l_{\lambda}\!\!=\!1, l_{\rho}\!\!=\!0, k_{\lambda}\!\!=\!0, k_{\rho}\!\!=\!0 \rangle$ & $^{4}P_{1/2}$ & $3050^{+15}_{-15}$ & $3048^{+14}_{-14}$ & $3050.2\pm 0.13$ & $8^{+4}_{-4}$ & $0.8\pm 0.2\pm 0.1$ \\ 
$\vert l_{\lambda}\!\!=\!1, l_{\rho}\!\!=\!0, k_{\lambda}\!\!=\!0, k_{\rho}\!\!=\!0 \rangle$ & $^{2}P_{3/2}$ & $3035^{+10}_{-10}$ & $3025^{+10}_{-10}$ & $3065.6\pm 0.28$ & $26^{+13}_{-13}$ & $3.5\pm 0.4\pm 0.2$ \\ 
$\vert l_{\lambda}\!\!=\!1, l_{\rho}\!\!=\!0, k_{\lambda}\!\!=\!0, k_{\rho}\!\!=\!0 \rangle$ & $^{4}P_{3/2}$ & $3077^{+9}_{-9}$ & $3073^{+8}_{-8}$ & $3090.2\pm 0.5$ & $7^{+3}_{-3}$ & $8.7\pm 1.0\pm 0.8$ \\ 
$\vert l_{\lambda}\!\!=\!1, l_{\rho}\!\!=\!0, k_{\lambda}\!\!=\!0, k_{\rho}\!\!=\!0 \rangle$ & $^{4}P_{5/2}$ & $3122^{+12}_{-12}$ & $3115^{+12}_{-11}$ & $3119.1\pm 1.0$  & $50^{+25}_{-24}$ &$<2.6$ \\ 
$\vert l_{\lambda}\!\!=\!0, l_{\rho}\!\!=\!1, k_{\lambda}\!\!=\!0, k_{\rho}\!\!=\!0 \rangle$ & $^{2}P_{1/2}$ & $3129^{+9}_{-9}$ & $\dagger\dagger$ & $\dagger$ & $14^{+2}_{-10}$ & $\dagger$ \\ 
$\vert l_{\lambda}\!\!=\!0, l_{\rho}\!\!=\!1, k_{\lambda}\!\!=\!0, k_{\rho}\!\!=\!0 \rangle$ & $^{2}P_{3/2}$ & $3156^{+10}_{-10}$ & $\dagger\dagger$ & $\dagger$ & $72^{+36}_{-35}$ & $\dagger$ \\ 
\hline
 $N=2$  &  &  &  &  &  \\ 
$\vert l_{\lambda}\!\!=\!2, l_{\rho}\!\!=\!0, k_{\lambda}\!\!=\!0, k_{\rho}\!\!=\!0 \rangle$ & $^{2}D_{3/2}$ & $3315^{+15}_{-14}$ & $3306^{+14}_{-14}$ & $\dagger$ & $11^{+5}_{-5}$ & $\dagger$ \\ 
$\vert l_{\lambda}\!\!=\!2, l_{\rho}\!\!=\!0, k_{\lambda}\!\!=\!0, k_{\rho}\!\!=\!0 \rangle$ & $^{2}D_{5/2}$ & $3360^{+17}_{-16}$ & $3348^{+17}_{-17}$ & $\dagger$ & $24^{+12}_{-12}$ & $\dagger$ \\ 
$\vert l_{\lambda}\!\!=\!2, l_{\rho}\!\!=\!0, k_{\lambda}\!\!=\!0, k_{\rho}\!\!=\!0 \rangle$ & $^{4}D_{1/2}$ & $3330^{+25}_{-25}$ & $3328^{+24}_{-23}$ & $\dagger$ & $16^{+8}_{-8}$ & $\dagger$ \\ 
$\vert l_{\lambda}\!\!=\!2, l_{\rho}\!\!=\!0, k_{\lambda}\!\!=\!0, k_{\rho}\!\!=\!0 \rangle$ & $^{4}D_{3/2}$ & $3357^{+18}_{-19}$ & $3354^{+17}_{-17}$ & $\dagger$ & $30^{+15}_{-15}$ & $\dagger$ \\ 
$\vert l_{\lambda}\!\!=\!2, l_{\rho}\!\!=\!0, k_{\lambda}\!\!=\!0, k_{\rho}\!\!=\!0 \rangle$ & $^{4}D_{5/2}$ & $3402^{+13}_{-13}$ & $3396^{+12}_{-12}$ & $\dagger$ & $62^{+31}_{-31}$ & $\dagger$ \\ 
$\vert l_{\lambda}\!\!=\!2, l_{\rho}\!\!=\!0, k_{\lambda}\!\!=\!0, k_{\rho}\!\!=\!0 \rangle$ & $^{4}D_{7/2}$ & $3466^{+23}_{-23}$ & $3455^{+23}_{-23}$ & $\dagger$ & $123^{+61}_{-62}$ & $\dagger$ \\ 
$\vert l_{\lambda}\!\!=\!0, l_{\rho}\!\!=\!0, k_{\lambda}\!\!=\!1, k_{\rho}\!\!=\!0 \rangle$ & $^{2}S_{1/2}$ & $3342^{+14}_{-14}$ & $3331^{+15}_{-15}$ & $\dagger$ & $2^{+1}_{-1}$ & $\dagger$ \\ 
$\vert l_{\lambda}\!\!=\!0, l_{\rho}\!\!=\!0, k_{\lambda}\!\!=\!1, k_{\rho}\!\!=\!0 \rangle$ & $^{4}S_{3/2}$ & $3411^{+13}_{-13}$ & $3404^{+12}_{-12}$ & $\dagger$ & $3^{+1}_{-1}$ & $\dagger$ \\ 
$\vert l_{\lambda}\!\!=\!0, l_{\rho}\!\!=\!0, k_{\lambda}\!\!=\!0, k_{\rho}\!\!=\!1 \rangle$ & $^{2}S_{1/2}$ & $3585^{+15}_{-15}$ & $\dagger\dagger$ & $\dagger$ & $18^{+9}_{-9}$ & $\dagger$ \\ 
$\vert l_{\lambda}\!\!=\!0, l_{\rho}\!\!=\!0, k_{\lambda}\!\!=\!0, k_{\rho}\!\!=\!1 \rangle$ & $^{4}S_{3/2}$ & $3654^{+16}_{-16}$ & $\dagger\dagger$ & $\dagger$ & $24^{+12}_{-12}$ & $\dagger$ \\ 
$\vert l_{\lambda}\!\!=\!1, l_{\rho}\!\!=\!1, k_{\lambda}\!\!=\!0, k_{\rho}\!\!=\!0 \rangle$ & $^{2}D_{3/2}$ & $3437^{+14}_{-14}$ & $\dagger\dagger$ & $\dagger$ & $198^{+98}_{-98}$ & $\dagger$ \\ 
$\vert l_{\lambda}\!\!=\!1, l_{\rho}\!\!=\!1, k_{\lambda}\!\!=\!0, k_{\rho}\!\!=\!0 \rangle$ & $^{2}D_{5/2}$ & $3482^{+16}_{-16}$ & $\dagger\dagger$ & $\dagger$ & $115^{+57}_{-56}$ & $\dagger$ \\ 
$\vert l_{\lambda}\!\!=\!1, l_{\rho}\!\!=\!1, k_{\lambda}\!\!=\!0, k_{\rho}\!\!=\!0 \rangle$ & $^{2}P_{1/2}$ & $3446^{+13}_{-13}$ & $\dagger\dagger$ & $\dagger$ & $2^{+1}_{-1}$ & $\dagger$ \\ 
$\vert l_{\lambda}\!\!=\!1, l_{\rho}\!\!=\!1, k_{\lambda}\!\!=\!0, k_{\rho}\!\!=\!0 \rangle$ & $^{2}P_{3/2}$ & $3473^{+14}_{-14}$ & $\dagger\dagger$ & $\dagger$ & $3^{+1}_{-1}$ & $\dagger$ \\ 
$\vert l_{\lambda}\!\!=\!1, l_{\rho}\!\!=\!1, k_{\lambda}\!\!=\!0, k_{\rho}\!\!=\!0 \rangle$ & $^{2}S_{1/2}$ & $3464^{+13}_{-13}$ & $\dagger\dagger$ & $\dagger$ & $88^{+43}_{-44}$ & $\dagger$ \\ 
$\vert l_{\lambda}\!\!=\!0, l_{\rho}\!\!=\!2, k_{\lambda}\!\!=\!0, k_{\rho}\!\!=\!0 \rangle$ & $^{2}D_{3/2}$ & $3558^{+15}_{-15}$ & $\dagger\dagger$ & $\dagger$ & $217^{+109}_{-107}$ & $\dagger$ \\ 
$\vert l_{\lambda}\!\!=\!0, l_{\rho}\!\!=\!2, k_{\lambda}\!\!=\!0, k_{\rho}\!\!=\!0 \rangle$ & $^{2}D_{5/2}$ & $3603^{+17}_{-17}$ & $\dagger\dagger$ & $\dagger$ & $174^{+85}_{-86}$ & $\dagger$ \\ 
$\vert l_{\lambda}\!\!=\!0, l_{\rho}\!\!=\!2, k_{\lambda}\!\!=\!0, k_{\rho}\!\!=\!0 \rangle$ & $^{4}D_{1/2}$ & $3573^{+27}_{-26}$ & $\dagger\dagger$ & $\dagger$ & $218^{+140}_{-139}$ & $\dagger$ \\ 
$\vert l_{\lambda}\!\!=\!0, l_{\rho}\!\!=\!2, k_{\lambda}\!\!=\!0, k_{\rho}\!\!=\!0 \rangle$ & $^{4}D_{3/2}$ & $3600^{+20}_{-20}$ & $\dagger\dagger$ & $\dagger$ & $285^{+144}_{-145}$ & $\dagger$ \\ 
$\vert l_{\lambda}\!\!=\!0, l_{\rho}\!\!=\!2, k_{\lambda}\!\!=\!0, k_{\rho}\!\!=\!0 \rangle$ & $^{4}D_{5/2}$ & $3645^{+15}_{-16}$ & $\dagger\dagger$ & $\dagger$ & $212^{+103}_{-104}$ & $\dagger$ \\ 
$\vert l_{\lambda}\!\!=\!0, l_{\rho}\!\!=\!2, k_{\lambda}\!\!=\!0, k_{\rho}\!\!=\!0 \rangle$ & $^{4}D_{7/2}$ & $3708^{+25}_{-25}$ & $\dagger\dagger$ & $\dagger$ & $383^{+192}_{-194}$ & $\dagger$ \\ 
\hline \hline
\end{tabular}
\caption{$\Omega_c(ssc)$ states. The flavor multiplet is specified with the symbol $\mathcal{F}$. The first column shows the quantum states $\left| l_{\lambda}(l_r),l_{\rho}, k_{\lambda}(k_r),k_{\rho}\right\rangle$ for the three-quark (quark-diquark) model, where $l_{\lambda,\rho}(l_r)$ are the orbital angular momenta and $k_{\lambda,\rho}(k_r)$ the number of nodes  of the $\lambda$($r$) and $\rho$ oscillators. Furthermore, $N=n_\rho+n_\lambda$ $(N=n_r)$ separate the energy bands $N=0,1,2$. The second column contains the spectroscopic notation $^{2S+1}L_J$ for each state and is defined by the total angular momentum ${\bf J}_{\rm tot} =\bf L_{\rm tot} + {\bf S}_{\rm tot} $, where ${\bf S}_{\rm tot} = {\bf S}_{\rm lt}+\frac{1}{2}$, and ${\bf L}_{\rm tot}= {\bf l}_{\rho}+{\bf l}_{\lambda}$ (${\bf L}_{\rm tot}= {\bf l}_{r}$, for the quark-diquark model). The predicted masses, computed within the three-quark model, are shown in the third column, whereas their corresponding total strong decay widths are shown in the sixth column. The predicted masses, computed within the quark-diquark framework, are presented in the fourth column.  Our theoretical results are compared with the experimental masses of the fifth column and the experimental decay widths of the seventh column taken from PDG  \cite{PhysRevD.98.030001} and Ref.  \cite{PhysRevLett.118.182001}. The (*) indicate the experimental mass and decay width values included in the fits. The $\dagger$ indicates that there is no reported experimental mass or decay for that state. The $\dagger\dagger$ indicates that there is no quark-diquark prediction for that state.}
\label{tab:All_mass_Omega}
\end{table*}
 
%%%%%%%%SIGMA_C%%%%%%
\begin{table*}[htbp]
\begin{tabular}{c  c| c c c c c }\hline \hline
            &  & Three-quark &  Quark-diquark    &               &              &  \\ 
$\Sigma_{c}(nnc)$&  & Predicted   &    Predicted   &  Experimental &  Predicted            & Experimental \\ 
$\mathcal{F}={\bf {6}}_{\rm f}$  & $^{2S+1}L_{J}$ & Mass (MeV)  &   Mass (MeV)   &  Mass (MeV)   &  $\Gamma_{tot}$ (MeV) & $\Gamma$ (MeV) \\ \hline
\hline
 $N=0$  &  &  &  &  &  \\ 
$\vert l_{\lambda}\!\!=\!0, l_{\rho}\!\!=\!0, k_{\lambda}\!\!=\!0, k_{\rho}\!\!=\!0 \rangle$ & $^{2}S_{1/2}$ & $2456^{+11}_{-11}$ & $2451^{+11}_{-11}$ & $2453.5\pm 0.9$ (*) & $2^{+1}_{-1}$ & $2.3\pm 0.3\pm 0.3$ (*) \\ 
$\vert l_{\lambda}\!\!=\!0, l_{\rho}\!\!=\!0, k_{\lambda}\!\!=\!0, k_{\rho}\!\!=\!0 \rangle$ & $^{4}S_{3/2}$ & $2525^{+11}_{-11}$ & $2524^{+11}_{-11}$ & $2518.1\pm 2.8$ (*) & $15^{+8}_{-8}$ & $17.2\pm 2.3\pm 3.1$ (*) \\ 
\hline
 $N=1$  &  &  &  &  &  \\ 
$\vert l_{\lambda}\!\!=\!1, l_{\rho}\!\!=\!0, k_{\lambda}\!\!=\!0, k_{\rho}\!\!=\!0 \rangle$ & $^{2}P_{1/2}$ & $2811^{+12}_{-12}$ & $2798^{+14}_{-14}$ & $2800.0\pm 20.0$ (*) &  $21^{+9}_{-8}$ & $75\pm 60$ (*) \\ 
$\vert l_{\lambda}\!\!=\!1, l_{\rho}\!\!=\!0, k_{\lambda}\!\!=\!0, k_{\rho}\!\!=\!0 \rangle$ & $^{4}P_{1/2}$ & $2853^{+17}_{-17}$ & $2845^{+18}_{-18}$ & $\dagger$ & $26^{+12}_{-13}$ & $\dagger$ \\ 
$\vert l_{\lambda}\!\!=\!1, l_{\rho}\!\!=\!0, k_{\lambda}\!\!=\!0, k_{\rho}\!\!=\!0 \rangle$ & $^{2}P_{3/2}$ & $2838^{+12}_{-13}$ & $2823^{+15}_{-15}$ & $\dagger$ &  $86^{+40}_{-37}$ & $\dagger$ \\ 
$\vert l_{\lambda}\!\!=\!1, l_{\rho}\!\!=\!0, k_{\lambda}\!\!=\!0, k_{\rho}\!\!=\!0 \rangle$ & $^{4}P_{3/2}$ & $2880^{+13}_{-13}$ & $2871^{+13}_{-13}$ & $\dagger$ & $60^{+25}_{-19}$ & $\dagger$ \\ 
$\vert l_{\lambda}\!\!=\!1, l_{\rho}\!\!=\!0, k_{\lambda}\!\!=\!0, k_{\rho}\!\!=\!0 \rangle$ & $^{4}P_{5/2}$ & $2925^{+16}_{-16}$ & $2913^{+16}_{-16}$ & $\dagger$ &  $164^{+95}_{-86}$ & $\dagger$ \\ 
$\vert l_{\lambda}\!\!=\!0, l_{\rho}\!\!=\!1, k_{\lambda}\!\!=\!0, k_{\rho}\!\!=\!0 \rangle$ & $^{2}P_{1/2}$ & $2994^{+16}_{-17}$ & $\dagger\dagger$ & $\dagger$ & $125^{+61}_{-60}$ & $\dagger$ \\ 
$\vert l_{\lambda}\!\!=\!0, l_{\rho}\!\!=\!1, k_{\lambda}\!\!=\!0, k_{\rho}\!\!=\!0 \rangle$ & $^{2}P_{3/2}$ & $3021^{+17}_{-17}$ & $\dagger\dagger$ & $\dagger$ & $125^{+63}_{-61}$ & $\dagger$ \\ 
\hline
 $N=2$  &  &  &  &  &  \\ 
$\vert l_{\lambda}\!\!=\!2, l_{\rho}\!\!=\!0, k_{\lambda}\!\!=\!0, k_{\rho}\!\!=\!0 \rangle$ & $^{2}D_{3/2}$ & $3175^{+17}_{-17}$ & $3153^{+21}_{-21}$ & $\dagger$ & $129^{+68}_{-69}$ & $\dagger$ \\ 
$\vert l_{\lambda}\!\!=\!2, l_{\rho}\!\!=\!0, k_{\lambda}\!\!=\!0, k_{\rho}\!\!=\!0 \rangle$ & $^{2}D_{5/2}$ & $3220^{+19}_{-19}$ & $3195^{+23}_{-23}$ & $\dagger$ & $216^{+124}_{-122}$ & $\dagger$ \\ 
$\vert l_{\lambda}\!\!=\!2, l_{\rho}\!\!=\!0, k_{\lambda}\!\!=\!0, k_{\rho}\!\!=\!0 \rangle$ & $^{4}D_{1/2}$ & $3190^{+28}_{-27}$ & $3175^{+28}_{-28}$ & $\dagger$ & $99^{+51}_{-51}$ & $\dagger$ \\ 
$\vert l_{\lambda}\!\!=\!2, l_{\rho}\!\!=\!0, k_{\lambda}\!\!=\!0, k_{\rho}\!\!=\!0 \rangle$ & $^{4}D_{3/2}$ & $3217^{+22}_{-22}$ & $3201^{+23}_{-23}$ & $\dagger$ & $155^{+76}_{-75}$ & $\dagger$ \\ 
$\vert l_{\lambda}\!\!=\!2, l_{\rho}\!\!=\!0, k_{\lambda}\!\!=\!0, k_{\rho}\!\!=\!0 \rangle$ & $^{4}D_{5/2}$ & $3262^{+18}_{-18}$ & $3243^{+19}_{-20}$ & $\dagger$ & $227^{+95}_{-94}$ & $\dagger$ \\ 
$\vert l_{\lambda}\!\!=\!2, l_{\rho}\!\!=\!0, k_{\lambda}\!\!=\!0, k_{\rho}\!\!=\!0 \rangle$ & $^{4}D_{7/2}$ & $3326^{+26}_{-26}$ & $3302^{+28}_{-28}$ & $\dagger$ & $385^{+214}_{-215}$ & $\dagger$ \\ 
$\vert l_{\lambda}\!\!=\!0, l_{\rho}\!\!=\!0, k_{\lambda}\!\!=\!1, k_{\rho}\!\!=\!0 \rangle$ & $^{2}S_{1/2}$ & $3202^{+17}_{-17}$ & $3178^{+21}_{-21}$ & $\dagger$ & $8^{+3}_{-3}$ & $\dagger$ \\ 
$\vert l_{\lambda}\!\!=\!0, l_{\rho}\!\!=\!0, k_{\lambda}\!\!=\!1, k_{\rho}\!\!=\!0 \rangle$ & $^{4}S_{3/2}$ & $3271^{+18}_{-18}$ & $3251^{+20}_{-20}$ & $\dagger$ & $7^{+3}_{-3}$ & $\dagger$ \\ 
$\vert l_{\lambda}\!\!=\!0, l_{\rho}\!\!=\!0, k_{\lambda}\!\!=\!0, k_{\rho}\!\!=\!1 \rangle$ & $^{2}S_{1/2}$ & $3567^{+31}_{-31}$ & $\dagger\dagger$ & $\dagger$ & $19^{+10}_{-9}$ & $\dagger$ \\ 
$\vert l_{\lambda}\!\!=\!0, l_{\rho}\!\!=\!0, k_{\lambda}\!\!=\!0, k_{\rho}\!\!=\!1 \rangle$ & $^{4}S_{3/2}$ & $3637^{+33}_{-33}$ & $\dagger\dagger$ & $\dagger$ & $26^{+13}_{-13}$ & $\dagger$ \\ 
$\vert l_{\lambda}\!\!=\!1, l_{\rho}\!\!=\!1, k_{\lambda}\!\!=\!0, k_{\rho}\!\!=\!0 \rangle$ & $^{2}D_{3/2}$ & $3358^{+22}_{-23}$ & $\dagger\dagger$ & $\dagger$ & $386^{+189}_{-189}$ & $\dagger$ \\ 
$\vert l_{\lambda}\!\!=\!1, l_{\rho}\!\!=\!1, k_{\lambda}\!\!=\!0, k_{\rho}\!\!=\!0 \rangle$ & $^{2}D_{5/2}$ & $3403^{+24}_{-24}$ & $\dagger\dagger$ & $\dagger$ & $334^{+174}_{-172}$ & $\dagger$ \\ 
$\vert l_{\lambda}\!\!=\!1, l_{\rho}\!\!=\!1, k_{\lambda}\!\!=\!0, k_{\rho}\!\!=\!0 \rangle$ & $^{2}P_{1/2}$ & $3367^{+22}_{-22}$ & $\dagger\dagger$ & $\dagger$ & $8^{+6}_{-6}$ & $\dagger$ \\ 
$\vert l_{\lambda}\!\!=\!1, l_{\rho}\!\!=\!1, k_{\lambda}\!\!=\!0, k_{\rho}\!\!=\!0 \rangle$ & $^{2}P_{3/2}$ & $3394^{+23}_{-23}$ & $\dagger\dagger$ & $\dagger$ & $32^{+29}_{-28}$ & $\dagger$ \\ 
$\vert l_{\lambda}\!\!=\!1, l_{\rho}\!\!=\!1, k_{\lambda}\!\!=\!0, k_{\rho}\!\!=\!0 \rangle$ & $^{2}S_{1/2}$ & $3385^{+22}_{-23}$ & $\dagger\dagger$ & $\dagger$ & $100^{+52}_{-52}$ & $\dagger$ \\ 
$\vert l_{\lambda}\!\!=\!0, l_{\rho}\!\!=\!2, k_{\lambda}\!\!=\!0, k_{\rho}\!\!=\!0 \rangle$ & $^{2}D_{3/2}$ & $3540^{+31}_{-31}$ & $\dagger\dagger$ & $\dagger$ & $476^{+229}_{-226}$ & $\dagger$ \\ 
$\vert l_{\lambda}\!\!=\!0, l_{\rho}\!\!=\!2, k_{\lambda}\!\!=\!0, k_{\rho}\!\!=\!0 \rangle$ & $^{2}D_{5/2}$ & $3585^{+32}_{-32}$ & $\dagger\dagger$ & $\dagger$ & $722^{+369}_{-371}$ & $\dagger$ \\ 
$\vert l_{\lambda}\!\!=\!0, l_{\rho}\!\!=\!2, k_{\lambda}\!\!=\!0, k_{\rho}\!\!=\!0 \rangle$ & $^{4}D_{1/2}$ & $3555^{+39}_{-39}$ & $\dagger\dagger$ & $\dagger$ & $1150^{+565}_{-558}$ & $\dagger$ \\ 
$\vert l_{\lambda}\!\!=\!0, l_{\rho}\!\!=\!2, k_{\lambda}\!\!=\!0, k_{\rho}\!\!=\!0 \rangle$ & $^{4}D_{3/2}$ & $3582^{+35}_{-35}$ & $\dagger\dagger$ & $\dagger$ & $652^{+319}_{-313}$ & $\dagger$ \\ 
$\vert l_{\lambda}\!\!=\!0, l_{\rho}\!\!=\!2, k_{\lambda}\!\!=\!0, k_{\rho}\!\!=\!0 \rangle$ & $^{4}D_{5/2}$ & $3627^{+33}_{-33}$ & $\dagger\dagger$ & $\dagger$ & $412^{+207}_{-206}$ & $\dagger$ \\ 
$\vert l_{\lambda}\!\!=\!0, l_{\rho}\!\!=\!2, k_{\lambda}\!\!=\!0, k_{\rho}\!\!=\!0 \rangle$ & $^{4}D_{7/2}$ & $3691^{+38}_{-38}$ & $\dagger\dagger$ & $\dagger$ & $1879^{+977}_{-974}$ & $\dagger$ \\ 
\hline \hline
\end{tabular}
\caption{Same as  Table \ref{tab:All_mass_Omega}, but for  $ \Sigma_c(nnc) $ states.}
\label{tab:All_mass_Sigma}
\end{table*}

%%%%%%XI_c prime states%%%%%%%%%%%%%%%%%%%%%
\begin{table*}[htbp]
\begin{tabular}{c  c| c c c c c }\hline \hline
            &  & Three-quark &  Quark-diquark    &               &              &  \\ 
$\Xi'_{c}(snc)$&  & Predicted   &    Predicted   &  Experimental &  Predicted            & Experimental \\ 
$\mathcal{F}={\bf {6}}_{\rm f}$  & $^{2S+1}L_{J}$ & Mass (MeV)  &   Mass (MeV)   &  Mass (MeV)   &  $\Gamma_{tot}$ (MeV) & $\Gamma$ (MeV) \\ \hline
\hline
 $N=0$  &  &  &  &  &  \\ 
$\vert l_{\lambda}\!\!=\!0, l_{\rho}\!\!=\!0, k_{\lambda}\!\!=\!0, k_{\rho}\!\!=\!0 \rangle$ & $^{2}S_{1/2}$ & $2571^{+8}_{-8}$ & $2577^{+10}_{-10}$ & $2578.0\pm 0.9$ (*) & $0$ & $\dagger$ \\ 
$\vert l_{\lambda}\!\!=\!0, l_{\rho}\!\!=\!0, k_{\lambda}\!\!=\!0, k_{\rho}\!\!=\!0 \rangle$ & $^{4}S_{3/2}$ & $2640^{+7}_{-7}$ & $2650^{+9}_{-9}$ & $2645.9\pm 0.71$ (*) & $0.4^{+0.2}_{-0.2}$ & $2.25\pm 0.41$ (*) \\ 
\hline
 $N=1$  &  &  &  &  &  \\ 
$\vert l_{\lambda}\!\!=\!1, l_{\rho}\!\!=\!0, k_{\lambda}\!\!=\!0, k_{\rho}\!\!=\!0 \rangle$ & $^{2}P_{1/2}$ & $2893^{+9}_{-9}$ & $2893^{+11}_{-11}$ & $\dagger$ & $7^{+4}_{-3}$ & $\dagger$ \\ 
$\vert l_{\lambda}\!\!=\!1, l_{\rho}\!\!=\!0, k_{\lambda}\!\!=\!0, k_{\rho}\!\!=\!0 \rangle$ & $^{4}P_{1/2}$ & $2935^{+14}_{-15}$ & $2941^{+14}_{-14}$ & $2923.0\pm 0.35$ & $5^{+2}_{-3}$ & $7.1\pm 2.0$ \\ 
$\vert l_{\lambda}\!\!=\!1, l_{\rho}\!\!=\!0, k_{\lambda}\!\!=\!0, k_{\rho}\!\!=\!0 \rangle$ & $^{2}P_{3/2}$ & $2920^{+9}_{-9}$ & $2919^{+13}_{-13}$ & $2938.5\pm 0.3$ & $28^{+14}_{-14}$ & $15\pm 9$ \\ 
$\vert l_{\lambda}\!\!=\!1, l_{\rho}\!\!=\!0, k_{\lambda}\!\!=\!0, k_{\rho}\!\!=\!0 \rangle$ & $^{4}P_{3/2}$ & $2962^{+9}_{-9}$ & $2966^{+10}_{-10}$ & $2964.9\pm 0.33$ (*) & $19^{+9}_{-9}$ & $14.1\pm 1.6$ (*)\\ 
$\vert l_{\lambda}\!\!=\!1, l_{\rho}\!\!=\!0, k_{\lambda}\!\!=\!0, k_{\rho}\!\!=\!0 \rangle$ & $^{4}P_{5/2}$ & $3007^{+12}_{-12}$ & $3009^{+14}_{-14}$ & $\dagger$ & $43^{+21}_{-21}$ & $\dagger$ \\ 
$\vert l_{\lambda}\!\!=\!0, l_{\rho}\!\!=\!1, k_{\lambda}\!\!=\!0, k_{\rho}\!\!=\!0 \rangle$ & $^{2}P_{1/2}$ & $3040^{+10}_{-9}$ & $\dagger\dagger$ & $3055.9\pm 0.4$ (*) & $157^{+80}_{-80}$ & $7.8\pm 1.9$ (*) \\ 
$\vert l_{\lambda}\!\!=\!0, l_{\rho}\!\!=\!1, k_{\lambda}\!\!=\!0, k_{\rho}\!\!=\!0 \rangle$ & $^{2}P_{3/2}$ & $3067^{+10}_{-10}$ & $\dagger\dagger$ & $3078.6\pm 2.8$ (*) & $100^{+47}_{-48}$ & $4.6\pm 3.3$ (*)\\ 
\hline
 $N=2$  &  &  &  &  &  \\ 
$\vert l_{\lambda}\!\!=\!2, l_{\rho}\!\!=\!0, k_{\lambda}\!\!=\!0, k_{\rho}\!\!=\!0 \rangle$ & $^{2}D_{3/2}$ & $3223^{+14}_{-14}$ & $3218^{+17}_{-17}$ & $\dagger$ & $20^{+10}_{-10}$ & $\dagger$ \\ 
$\vert l_{\lambda}\!\!=\!2, l_{\rho}\!\!=\!0, k_{\lambda}\!\!=\!0, k_{\rho}\!\!=\!0 \rangle$ & $^{2}D_{5/2}$ & $3268^{+16}_{-16}$ & $3261^{+21}_{-20}$ & $\dagger$ & $64^{+32}_{-33}$ & $\dagger$ \\ 
$\vert l_{\lambda}\!\!=\!2, l_{\rho}\!\!=\!0, k_{\lambda}\!\!=\!0, k_{\rho}\!\!=\!0 \rangle$ & $^{4}D_{1/2}$ & $3238^{+25}_{-25}$ & $3239^{+23}_{-24}$ & $\dagger$ & $29^{+15}_{-15}$ & $\dagger$ \\ 
$\vert l_{\lambda}\!\!=\!2, l_{\rho}\!\!=\!0, k_{\lambda}\!\!=\!0, k_{\rho}\!\!=\!0 \rangle$ & $^{4}D_{3/2}$ & $3265^{+19}_{-19}$ & $3265^{+18}_{-18}$ & $\dagger$ & $53^{+26}_{-26}$ & $\dagger$ \\ 
$\vert l_{\lambda}\!\!=\!2, l_{\rho}\!\!=\!0, k_{\lambda}\!\!=\!0, k_{\rho}\!\!=\!0 \rangle$ & $^{4}D_{5/2}$ & $3310^{+13}_{-13}$ & $3308^{+15}_{-14}$ & $\dagger$ & $97^{+46}_{-47}$ & $\dagger$ \\ 
$\vert l_{\lambda}\!\!=\!2, l_{\rho}\!\!=\!0, k_{\lambda}\!\!=\!0, k_{\rho}\!\!=\!0 \rangle$ & $^{4}D_{7/2}$ & $3373^{+23}_{-23}$ & $3368^{+25}_{-25}$ & $\dagger$ & $161^{+80}_{-80}$ & $\dagger$ \\ 
$\vert l_{\lambda}\!\!=\!0, l_{\rho}\!\!=\!0, k_{\lambda}\!\!=\!1, k_{\rho}\!\!=\!0 \rangle$ & $^{2}S_{1/2}$ & $3250^{+13}_{-14}$ & $3244^{+18}_{-18}$ & $\dagger$ & $2^{+1}_{-1}$ & $\dagger$ \\ 
$\vert l_{\lambda}\!\!=\!0, l_{\rho}\!\!=\!0, k_{\lambda}\!\!=\!1, k_{\rho}\!\!=\!0 \rangle$ & $^{4}S_{3/2}$ & $3319^{+14}_{-14}$ & $3316^{+15}_{-15}$ & $\dagger$ & $5^{+2}_{-2}$ & $\dagger$ \\ 
$\vert l_{\lambda}\!\!=\!0, l_{\rho}\!\!=\!0, k_{\lambda}\!\!=\!0, k_{\rho}\!\!=\!1 \rangle$ & $^{2}S_{1/2}$ & $3544^{+19}_{-19}$ & $\dagger\dagger$ & $\dagger$ & $21^{+10}_{-10}$ & $\dagger$ \\ 
$\vert l_{\lambda}\!\!=\!0, l_{\rho}\!\!=\!0, k_{\lambda}\!\!=\!0, k_{\rho}\!\!=\!1 \rangle$ & $^{4}S_{3/2}$ & $3613^{+20}_{-21}$ & $\dagger\dagger$ & $\dagger$ & $29^{+14}_{-14}$ & $\dagger$ \\ 
$\vert l_{\lambda}\!\!=\!1, l_{\rho}\!\!=\!1, k_{\lambda}\!\!=\!0, k_{\rho}\!\!=\!0 \rangle$ & $^{2}D_{3/2}$ & $3370^{+15}_{-15}$ & $\dagger\dagger$ & $\dagger$ & $229^{+112}_{-111}$ & $\dagger$ \\ 
$\vert l_{\lambda}\!\!=\!1, l_{\rho}\!\!=\!1, k_{\lambda}\!\!=\!0, k_{\rho}\!\!=\!0 \rangle$ & $^{2}D_{5/2}$ & $3415^{+17}_{-17}$ & $\dagger\dagger$ & $\dagger$ & $134^{+67}_{-66}$ & $\dagger$ \\ 
$\vert l_{\lambda}\!\!=\!1, l_{\rho}\!\!=\!1, k_{\lambda}\!\!=\!0, k_{\rho}\!\!=\!0 \rangle$ & $^{2}P_{1/2}$ & $3379^{+14}_{-14}$ & $\dagger\dagger$ & $\dagger$ & $3^{+1}_{-1}$ & $\dagger$ \\ 
$\vert l_{\lambda}\!\!=\!1, l_{\rho}\!\!=\!1, k_{\lambda}\!\!=\!0, k_{\rho}\!\!=\!0 \rangle$ & $^{2}P_{3/2}$ & $3406^{+16}_{-16}$ & $\dagger\dagger$ & $\dagger$ & $3^{+1}_{-1}$ & $\dagger$ \\ 
$\vert l_{\lambda}\!\!=\!1, l_{\rho}\!\!=\!1, k_{\lambda}\!\!=\!0, k_{\rho}\!\!=\!0 \rangle$ & $^{2}S_{1/2}$ & $3397^{+15}_{-15}$ & $\dagger\dagger$ & $\dagger$ & $56^{+28}_{-28}$ & $\dagger$ \\ 
$\vert l_{\lambda}\!\!=\!0, l_{\rho}\!\!=\!2, k_{\lambda}\!\!=\!0, k_{\rho}\!\!=\!0 \rangle$ & $^{2}D_{3/2}$ & $3517^{+19}_{-18}$ & $\dagger\dagger$ & $\dagger$ & $319^{+153}_{-158}$ & $\dagger$ \\ 
$\vert l_{\lambda}\!\!=\!0, l_{\rho}\!\!=\!2, k_{\lambda}\!\!=\!0, k_{\rho}\!\!=\!0 \rangle$ & $^{2}D_{5/2}$ & $3563^{+21}_{-21}$ & $\dagger\dagger$ & $\dagger$ & $232^{+116}_{-115}$ & $\dagger$ \\ 
$\vert l_{\lambda}\!\!=\!0, l_{\rho}\!\!=\!2, k_{\lambda}\!\!=\!0, k_{\rho}\!\!=\!0 \rangle$ & $^{4}D_{1/2}$ & $3532^{+29}_{-29}$ & $\dagger\dagger$ & $\dagger$ & $632^{+332}_{-330}$ & $\dagger$ \\ 
$\vert l_{\lambda}\!\!=\!0, l_{\rho}\!\!=\!2, k_{\lambda}\!\!=\!0, k_{\rho}\!\!=\!0 \rangle$ & $^{4}D_{3/2}$ & $3559^{+23}_{-23}$ & $\dagger\dagger$ & $\dagger$ & $452^{+221}_{-223}$ & $\dagger$ \\ 
$\vert l_{\lambda}\!\!=\!0, l_{\rho}\!\!=\!2, k_{\lambda}\!\!=\!0, k_{\rho}\!\!=\!0 \rangle$ & $^{4}D_{5/2}$ & $3604^{+20}_{-20}$ & $\dagger\dagger$ & $\dagger$ & $208^{+99}_{-99}$ & $\dagger$ \\ 
$\vert l_{\lambda}\!\!=\!0, l_{\rho}\!\!=\!2, k_{\lambda}\!\!=\!0, k_{\rho}\!\!=\!0 \rangle$ & $^{4}D_{7/2}$ & $3668^{+28}_{-28}$ & $\dagger\dagger$ & $\dagger$ & $578^{+293}_{-297}$ & $\dagger$ \\ 
\hline \hline
\end{tabular}

\caption{Same as  Table \ref{tab:All_mass_Omega}, but for  $ \Xi'_c(snc) $ states.}
\label{tab:All_mass_Xiprime}
\end{table*}

%%%%%%XI_c states%%%%%%%%%%%%%%%%%%%%%%%%
\begin{table*}[htbp]
\begin{tabular}{c  c| c c c c c }\hline \hline
            &  & Three-quark &  Quark-diquark    &               &              &  \\ 
$\Xi_{c}(snc)$&  & Predicted   &    Predicted   &  Experimental &  Predicted            & Experimental \\ 
${\bf {\bar{3}}}_{\rm f}$  & $^{2S+1}L_{J}$ & Mass (MeV)  &   Mass (MeV)   &  Mass (MeV)   &  $\Gamma_{tot}$ (MeV) & $\Gamma$ (MeV) \\ \hline
\hline
 $N=0$  &  &  &  &  &  \\ 
$\vert l_{\lambda}\!\!=\!0, l_{\rho}\!\!=\!0, k_{\lambda}\!\!=\!0, k_{\rho}\!\!=\!0 \rangle$ & $^{2}S_{1/2}$ & $2466^{+10}_{-10}$ & $2473^{+10}_{-10}$ & $2469.42\pm 1.77$ (*) & $0$ & $\approx 0$ \\ 
\hline
 $N=1$  &  &  &  &  &  \\ 
$\vert l_{\lambda}\!\!=\!1, l_{\rho}\!\!=\!0, k_{\lambda}\!\!=\!0, k_{\rho}\!\!=\!0 \rangle$ & $^{2}P_{1/2}$ & $2788^{+10}_{-10}$ & $2789^{+9}_{-9}$ & $2793.3\pm 0.28$ (*) & $3^{+2}_{-2}$ & $9.5\pm 2.0$ (*) \\ 
$\vert l_{\lambda}\!\!=\!1, l_{\rho}\!\!=\!0, k_{\lambda}\!\!=\!0, k_{\rho}\!\!=\!0 \rangle$ & $^{2}P_{3/2}$ & $2815^{+10}_{-10}$ & $2814^{+9}_{-9}$ & $2818.49\pm 2.07$ (*) & $5^{+2}_{-2}$ & $2.48\pm 0.5$ (*)\\ 
$\vert l_{\lambda}\!\!=\!0, l_{\rho}\!\!=\!1, k_{\lambda}\!\!=\!0, k_{\rho}\!\!=\!0 \rangle$ & $^{2}P_{1/2}$ & $2935^{+12}_{-12}$ & $\dagger\dagger$ & $\dagger$ & $17^{+9}_{-8}$ & $\dagger$ \\ 
$\vert l_{\lambda}\!\!=\!0, l_{\rho}\!\!=\!1, k_{\lambda}\!\!=\!0, k_{\rho}\!\!=\!0 \rangle$ & $^{4}P_{1/2}$ & $2977^{+20}_{-20}$ & $\dagger\dagger$ & $2968.6\pm 3.3$ & $13^{+6}_{-6}$ & $20\pm 3.5$ \\ 
$\vert l_{\lambda}\!\!=\!0, l_{\rho}\!\!=\!1, k_{\lambda}\!\!=\!0, k_{\rho}\!\!=\!0 \rangle$ & $^{2}P_{3/2}$ & $2962^{+12}_{-12}$ & $\dagger\dagger$ & $\dagger$ & $89^{+45}_{-45}$ & $\dagger$ \\ 
$\vert l_{\lambda}\!\!=\!0, l_{\rho}\!\!=\!1, k_{\lambda}\!\!=\!0, k_{\rho}\!\!=\!0 \rangle$ & $^{4}P_{3/2}$ & $3004^{+17}_{-17}$ & $\dagger\dagger$ & $\dagger$ & $56^{+29}_{-31}$ & $\dagger$ \\ 
$\vert l_{\lambda}\!\!=\!0, l_{\rho}\!\!=\!1, k_{\lambda}\!\!=\!0, k_{\rho}\!\!=\!0 \rangle$ & $^{4}P_{5/2}$ & $3049^{+18}_{-19}$ & $\dagger\dagger$ & $\dagger$ & $122^{+59}_{-60}$ & $\dagger$ \\ 
\hline
 $N=2$  &  &  &  &  &  \\ 
$\vert l_{\lambda}\!\!=\!2, l_{\rho}\!\!=\!0, k_{\lambda}\!\!=\!0, k_{\rho}\!\!=\!0 \rangle$ & $^{2}D_{3/2}$ & $3118^{+14}_{-14}$ & $3113^{+14}_{-14}$ & $3122.9\pm 1.23$ & $50^{+24}_{-25}$ & $4\pm 4$ \\ 
$\vert l_{\lambda}\!\!=\!2, l_{\rho}\!\!=\!0, k_{\lambda}\!\!=\!0, k_{\rho}\!\!=\!0 \rangle$ & $^{2}D_{5/2}$ & $3164^{+16}_{-16}$ & $3156^{+16}_{-16}$ & $\dagger$ & $132^{+63}_{-64}$ & $\dagger$ \\ 
$\vert l_{\lambda}\!\!=\!0, l_{\rho}\!\!=\!0, k_{\lambda}\!\!=\!1, k_{\rho}\!\!=\!0 \rangle$ & $^{2}S_{1/2}$ & $3145^{+14}_{-13}$ & $3139^{+13}_{-14}$ & $\dagger$ & $5^{+2}_{-2}$ & $\dagger$ \\ 
$\vert l_{\lambda}\!\!=\!0, l_{\rho}\!\!=\!0, k_{\lambda}\!\!=\!0, k_{\rho}\!\!=\!1 \rangle$ & $^{2}S_{1/2}$ & $3440^{+20}_{-20}$ & $\dagger\dagger$ & $\dagger$ & $7^{+3}_{-3}$ & $\dagger$ \\ 
$\vert l_{\lambda}\!\!=\!1, l_{\rho}\!\!=\!1, k_{\lambda}\!\!=\!0, k_{\rho}\!\!=\!0 \rangle$ & $^{2}D_{3/2}$ & $3265^{+16}_{-16}$ & $\dagger\dagger$ & $\dagger$ & $54^{+26}_{-27}$ & $\dagger$ \\ 
$\vert l_{\lambda}\!\!=\!1, l_{\rho}\!\!=\!1, k_{\lambda}\!\!=\!0, k_{\rho}\!\!=\!0 \rangle$ & $^{2}D_{5/2}$ & $3311^{+17}_{-18}$ & $\dagger\dagger$ & $\dagger$ & $119^{+57}_{-58}$ & $\dagger$ \\ 
$\vert l_{\lambda}\!\!=\!1, l_{\rho}\!\!=\!1, k_{\lambda}\!\!=\!0, k_{\rho}\!\!=\!0 \rangle$ & $^{4}D_{1/2}$ & $3280^{+29}_{-29}$ & $\dagger\dagger$ & $\dagger$ & $24^{+12}_{-12}$ & $\dagger$ \\ 
$\vert l_{\lambda}\!\!=\!1, l_{\rho}\!\!=\!1, k_{\lambda}\!\!=\!0, k_{\rho}\!\!=\!0 \rangle$ & $^{4}D_{3/2}$ & $3307^{+23}_{-23}$ & $\dagger\dagger$ & $\dagger$ & $92^{+46}_{-46}$ & $\dagger$ \\ 
$\vert l_{\lambda}\!\!=\!1, l_{\rho}\!\!=\!1, k_{\lambda}\!\!=\!0, k_{\rho}\!\!=\!0 \rangle$ & $^{4}D_{5/2}$ & $3353^{+19}_{-19}$ & $\dagger\dagger$ & $\dagger$ & $153^{+75}_{-75}$ & $\dagger$ \\ 
$\vert l_{\lambda}\!\!=\!1, l_{\rho}\!\!=\!1, k_{\lambda}\!\!=\!0, k_{\rho}\!\!=\!0 \rangle$ & $^{4}D_{7/2}$ & $3416^{+26}_{-27}$ & $\dagger\dagger$ & $\dagger$ & $194^{+97}_{-96}$ & $\dagger$ \\ 
$\vert l_{\lambda}\!\!=\!1, l_{\rho}\!\!=\!1, k_{\lambda}\!\!=\!0, k_{\rho}\!\!=\!0 \rangle$ & $^{2}P_{1/2}$ & $3274^{+15}_{-15}$ & $\dagger\dagger$ & $\dagger$ & $0.4^{+0.2}_{-0.2}$ & $\dagger$ \\ 
$\vert l_{\lambda}\!\!=\!1, l_{\rho}\!\!=\!1, k_{\lambda}\!\!=\!0, k_{\rho}\!\!=\!0 \rangle$ & $^{2}P_{3/2}$ & $3302^{+16}_{-16}$ & $\dagger\dagger$ & $\dagger$ & $2^{+1}_{-1}$ & $\dagger$ \\ 
$\vert l_{\lambda}\!\!=\!1, l_{\rho}\!\!=\!1, k_{\lambda}\!\!=\!0, k_{\rho}\!\!=\!0 \rangle$ & $^{4}P_{1/2}$ & $3316^{+22}_{-22}$ & $\dagger\dagger$ & $\dagger$ & $0.3^{+0.1}_{-0.1}$ & $\dagger$ \\ 
$\vert l_{\lambda}\!\!=\!1, l_{\rho}\!\!=\!1, k_{\lambda}\!\!=\!0, k_{\rho}\!\!=\!0 \rangle$ & $^{4}P_{3/2}$ & $3344^{+19}_{-19}$ & $\dagger\dagger$ & $\dagger$ & $1.4^{+0.7}_{-0.7}$ & $\dagger$ \\ 
$\vert l_{\lambda}\!\!=\!1, l_{\rho}\!\!=\!1, k_{\lambda}\!\!=\!0, k_{\rho}\!\!=\!0 \rangle$ & $^{4}P_{5/2}$ & $3389^{+21}_{-22}$ & $\dagger\dagger$ & $\dagger$ & $4^{+2}_{-2}$ & $\dagger$ \\ 
$\vert l_{\lambda}\!\!=\!1, l_{\rho}\!\!=\!1, k_{\lambda}\!\!=\!0, k_{\rho}\!\!=\!0 \rangle$ & $^{4}S_{3/2}$ & $3362^{+19}_{-19}$ & $\dagger\dagger$ & $\dagger$ & $36^{+18}_{-18}$ & $\dagger$ \\ 
$\vert l_{\lambda}\!\!=\!1, l_{\rho}\!\!=\!1, k_{\lambda}\!\!=\!0, k_{\rho}\!\!=\!0 \rangle$ & $^{2}S_{1/2}$ & $3293^{+15}_{-15}$ & $\dagger\dagger$ & $\dagger$ & $30^{+15}_{-15}$ & $\dagger$ \\ 
$\vert l_{\lambda}\!\!=\!0, l_{\rho}\!\!=\!2, k_{\lambda}\!\!=\!0, k_{\rho}\!\!=\!0 \rangle$ & $^{2}D_{3/2}$ & $3413^{+20}_{-20}$ & $\dagger\dagger$ & $\dagger$ & $133^{+64}_{-64}$ & $\dagger$ \\ 
$\vert l_{\lambda}\!\!=\!0, l_{\rho}\!\!=\!2, k_{\lambda}\!\!=\!0, k_{\rho}\!\!=\!0 \rangle$ & $^{2}D_{5/2}$ & $3458^{+22}_{-22}$ & $\dagger\dagger$ & $\dagger$ & $119^{+59}_{-59}$ & $\dagger$ \\ 
\hline \hline
\end{tabular}
\caption{Same as Table \ref{tab:All_mass_Omega}, but for  $ \Xi_c(snc) $ states.}
\label{tab:All_mass_Xi}
\end{table*}

%%%Lambdac states%%%%%%%%%%%%%%%%%%%
\begin{table*}[htbp]
\begin{tabular}{c  c| c c c c c }\hline \hline
            &  & Three-quark &  Quark-diquark    &               &              &  \\ 
$\Lambda_{c}(nnc)$&  & Predicted   &    Predicted   &  Experimental &  Predicted            & Experimental \\ 
${\bf {\bar{3}}}_{\rm f}$   & $^{2S+1}L_{J}$ & Mass (MeV)  &   Mass (MeV)   &  Mass (MeV)   &  $\Gamma_{tot}$ (MeV) & $\Gamma$ (MeV) \\ \hline
\hline
 $N=0$  &  &  &  &  &  \\ 
$\vert l_{\lambda}\!\!=\!0, l_{\rho}\!\!=\!0, k_{\lambda}\!\!=\!0, k_{\rho}\!\!=\!0 \rangle$ & $^{2}S_{1/2}$ & $2261^{+11}_{-11}$ & $2264^{+10}_{-10}$ & $2286.5\pm 0.14$ (*) & $0$ & $\approx 0$ \\ 
\hline
 $N=1$  &  &  &  &  &  \\ 
$\vert l_{\lambda}\!\!=\!1, l_{\rho}\!\!=\!0, k_{\lambda}\!\!=\!0, k_{\rho}\!\!=\!0 \rangle$ & $^{2}P_{1/2}$ & $2616^{+10}_{-10}$ & $2611^{+9}_{-9}$ & $2592.3\pm 0.28$ (*) & $2^{+1}_{-1}$ & $2.6\pm 0.6$ (*)\\ 
$\vert l_{\lambda}\!\!=\!1, l_{\rho}\!\!=\!0, k_{\lambda}\!\!=\!0, k_{\rho}\!\!=\!0 \rangle$ & $^{2}P_{3/2}$ & $2643^{+10}_{-10}$ & $2636^{+9}_{-9}$ & $2625.0\pm 0.18$ (*) & $10^{+5}_{-5}$ & $<0.97$ \\ 
$\vert l_{\lambda}\!\!=\!0, l_{\rho}\!\!=\!1, k_{\lambda}\!\!=\!0, k_{\rho}\!\!=\!0 \rangle$ & $^{2}P_{1/2}$ & $2799^{+15}_{-15}$ & $\dagger\dagger$ & $\dagger$ & $60^{+27}_{-28}$ & $\dagger$ \\ 
$\vert l_{\lambda}\!\!=\!0, l_{\rho}\!\!=\!1, k_{\lambda}\!\!=\!0, k_{\rho}\!\!=\!0 \rangle$ & $^{4}P_{1/2}$ & $2841^{+23}_{-23}$ & $\dagger\dagger$ & $\dagger$ & $33^{+16}_{-16}$ & $\dagger$ \\ 
$\vert l_{\lambda}\!\!=\!0, l_{\rho}\!\!=\!1, k_{\lambda}\!\!=\!0, k_{\rho}\!\!=\!0 \rangle$ & $^{2}P_{3/2}$ & $2826^{+15}_{-15}$ & $\dagger\dagger$ & $\dagger$ & $93^{+45}_{-46}$ & $\dagger$ \\ 
$\vert l_{\lambda}\!\!=\!0, l_{\rho}\!\!=\!1, k_{\lambda}\!\!=\!0, k_{\rho}\!\!=\!0 \rangle$ & $^{4}P_{3/2}$ & $2868^{+20}_{-20}$ & $\dagger\dagger$ & $\dagger$ & $122^{+59}_{-59}$ & $\dagger$ \\ 
$\vert l_{\lambda}\!\!=\!0, l_{\rho}\!\!=\!1, k_{\lambda}\!\!=\!0, k_{\rho}\!\!=\!0 \rangle$ & $^{4}P_{5/2}$ & $2913^{+21}_{-21}$ & $\dagger\dagger$ & $\dagger$ & $108^{+52}_{-53}$ & $\dagger$ \\ 
\hline
 $N=2$  &  &  &  &  &  \\ 
$\vert l_{\lambda}\!\!=\!2, l_{\rho}\!\!=\!0, k_{\lambda}\!\!=\!0, k_{\rho}\!\!=\!0 \rangle$ & $^{2}D_{3/2}$ & $2980^{+14}_{-14}$ & $2967^{+14}_{-14}$ & $2856.1\pm 6$ & $70^{+34}_{-34}$ & $68\pm 22$ \\ 
$\vert l_{\lambda}\!\!=\!2, l_{\rho}\!\!=\!0, k_{\lambda}\!\!=\!0, k_{\rho}\!\!=\!0 \rangle$ & $^{2}D_{5/2}$ & $3025^{+15}_{-15}$ & $3009^{+16}_{-16}$ & $2881.63\pm 0.24$ & $171^{+84}_{-83}$ & $5.6\pm 0.8$ \\ 
$\vert l_{\lambda}\!\!=\!0, l_{\rho}\!\!=\!0, k_{\lambda}\!\!=\!1, k_{\rho}\!\!=\!0 \rangle$ & $^{2}S_{1/2}$ & $3007^{+13}_{-13}$ & $2992^{+14}_{-14}$ & $2766\pm 2.4$ & $5^{+3}_{-3}$ & $50$ \\ 
$\vert l_{\lambda}\!\!=\!0, l_{\rho}\!\!=\!0, k_{\lambda}\!\!=\!0, k_{\rho}\!\!=\!1 \rangle$ & $^{2}S_{1/2}$ & $3372^{+29}_{-29}$ & $\dagger\dagger$ & $\dagger$ & $6^{+3}_{-3}$ & $\dagger$ \\ 
$\vert l_{\lambda}\!\!=\!1, l_{\rho}\!\!=\!1, k_{\lambda}\!\!=\!0, k_{\rho}\!\!=\!0 \rangle$ & $^{2}D_{3/2}$ & $3163^{+20}_{-20}$ & $\dagger\dagger$ & $\dagger$ & $70^{+34}_{-34}$ & $\dagger$ \\ 
$\vert l_{\lambda}\!\!=\!1, l_{\rho}\!\!=\!1, k_{\lambda}\!\!=\!0, k_{\rho}\!\!=\!0 \rangle$ & $^{2}D_{5/2}$ & $3208^{+21}_{-21}$ & $\dagger\dagger$ & $\dagger$ & $133^{+65}_{-65}$ & $\dagger$ \\ 
$\vert l_{\lambda}\!\!=\!1, l_{\rho}\!\!=\!1, k_{\lambda}\!\!=\!0, k_{\rho}\!\!=\!0 \rangle$ & $^{4}D_{1/2}$ & $3178^{+33}_{-32}$ & $\dagger\dagger$ & $\dagger$ & $34^{+17}_{-17}$ & $\dagger$ \\ 
$\vert l_{\lambda}\!\!=\!1, l_{\rho}\!\!=\!1, k_{\lambda}\!\!=\!0, k_{\rho}\!\!=\!0 \rangle$ & $^{4}D_{3/2}$ & $3205^{+28}_{-27}$ & $\dagger\dagger$ & $\dagger$ & $106^{+50}_{-52}$ & $\dagger$ \\ 
$\vert l_{\lambda}\!\!=\!1, l_{\rho}\!\!=\!1, k_{\lambda}\!\!=\!0, k_{\rho}\!\!=\!0 \rangle$ & $^{4}D_{5/2}$ & $3250^{+24}_{-24}$ & $\dagger\dagger$ & $\dagger$ & $163^{+80}_{-79}$ & $\dagger$ \\ 
$\vert l_{\lambda}\!\!=\!1, l_{\rho}\!\!=\!1, k_{\lambda}\!\!=\!0, k_{\rho}\!\!=\!0 \rangle$ & $^{4}D_{7/2}$ & $3313^{+30}_{-30}$ & $\dagger\dagger$ & $\dagger$ & $230^{+118}_{-116}$ & $\dagger$ \\ 
$\vert l_{\lambda}\!\!=\!1, l_{\rho}\!\!=\!1, k_{\lambda}\!\!=\!0, k_{\rho}\!\!=\!0 \rangle$ & $^{2}P_{1/2}$ & $3172^{+20}_{-20}$ & $\dagger\dagger$ & $\dagger$ & $0.5^{+0.3}_{-0.3}$ & $\dagger$ \\ 
$\vert l_{\lambda}\!\!=\!1, l_{\rho}\!\!=\!1, k_{\lambda}\!\!=\!0, k_{\rho}\!\!=\!0 \rangle$ & $^{2}P_{3/2}$ & $3199^{+20}_{-20}$ & $\dagger\dagger$ & $\dagger$ & $2^{+1}_{-1}$ & $\dagger$ \\ 
$\vert l_{\lambda}\!\!=\!1, l_{\rho}\!\!=\!1, k_{\lambda}\!\!=\!0, k_{\rho}\!\!=\!0 \rangle$ & $^{4}P_{1/2}$ & $3214^{+26}_{-26}$ & $\dagger\dagger$ & $\dagger$ & $0.3^{+0.2}_{-0.2}$ & $\dagger$ \\ 
$\vert l_{\lambda}\!\!=\!1, l_{\rho}\!\!=\!1, k_{\lambda}\!\!=\!0, k_{\rho}\!\!=\!0 \rangle$ & $^{4}P_{3/2}$ & $3241^{+24}_{-24}$ & $\dagger\dagger$ & $\dagger$ & $1.5^{+0.7}_{-0.7}$ & $\dagger$ \\ 
$\vert l_{\lambda}\!\!=\!1, l_{\rho}\!\!=\!1, k_{\lambda}\!\!=\!0, k_{\rho}\!\!=\!0 \rangle$ & $^{4}P_{5/2}$ & $3286^{+26}_{-26}$ & $\dagger\dagger$ & $\dagger$ & $4^{+2}_{-2}$ & $\dagger$ \\ 
$\vert l_{\lambda}\!\!=\!1, l_{\rho}\!\!=\!1, k_{\lambda}\!\!=\!0, k_{\rho}\!\!=\!0 \rangle$ & $^{4}S_{3/2}$ & $3259^{+24}_{-24}$ & $\dagger\dagger$ & $\dagger$ & $32^{+16}_{-16}$ & $\dagger$ \\ 
$\vert l_{\lambda}\!\!=\!1, l_{\rho}\!\!=\!1, k_{\lambda}\!\!=\!0, k_{\rho}\!\!=\!0 \rangle$ & $^{2}S_{1/2}$ & $3190^{+20}_{-20}$ & $\dagger\dagger$ & $\dagger$ & $30^{+15}_{-15}$ & $\dagger$ \\ 
$\vert l_{\lambda}\!\!=\!0, l_{\rho}\!\!=\!2, k_{\lambda}\!\!=\!0, k_{\rho}\!\!=\!0 \rangle$ & $^{2}D_{3/2}$ & $3345^{+30}_{-30}$ & $\dagger\dagger$ & $\dagger$ & $154^{+75}_{-74}$ & $\dagger$ \\ 
$\vert l_{\lambda}\!\!=\!0, l_{\rho}\!\!=\!2, k_{\lambda}\!\!=\!0, k_{\rho}\!\!=\!0 \rangle$ & $^{2}D_{5/2}$ & $3390^{+30}_{-30}$ & $\dagger\dagger$ & $\dagger$ & $202^{+102}_{-102}$ & $\dagger$ \\ 
\hline \hline
\end{tabular}
\caption{Same as Table \ref{tab:All_mass_Omega}, but for  $ \Lambda_c(nnc) $ states.}
\label{tab:All_mass_Lambda}
\end{table*}

%%%%%%%%%%%%%%%%%%%%%%%%%%%%%%%%%%%%%%%%%%%
\begin{figure}
    \centering
    \caption{$\Omega_c$ mass spectra and tentative quantum number assignments based on the three-quark model Hamiltonian of Eqs. \ref{MassFormula} and \ref{eq:Hho}. The theoretical predictions and their uncertainties (red lines and bands) are compared with the experimental results (red lines and bands) reported in the PDG \cite{Zyla:2020zbs}. The experimental errors are too small to be evaluated on this energy scale.}
    \includegraphics[width=0.5\textwidth]{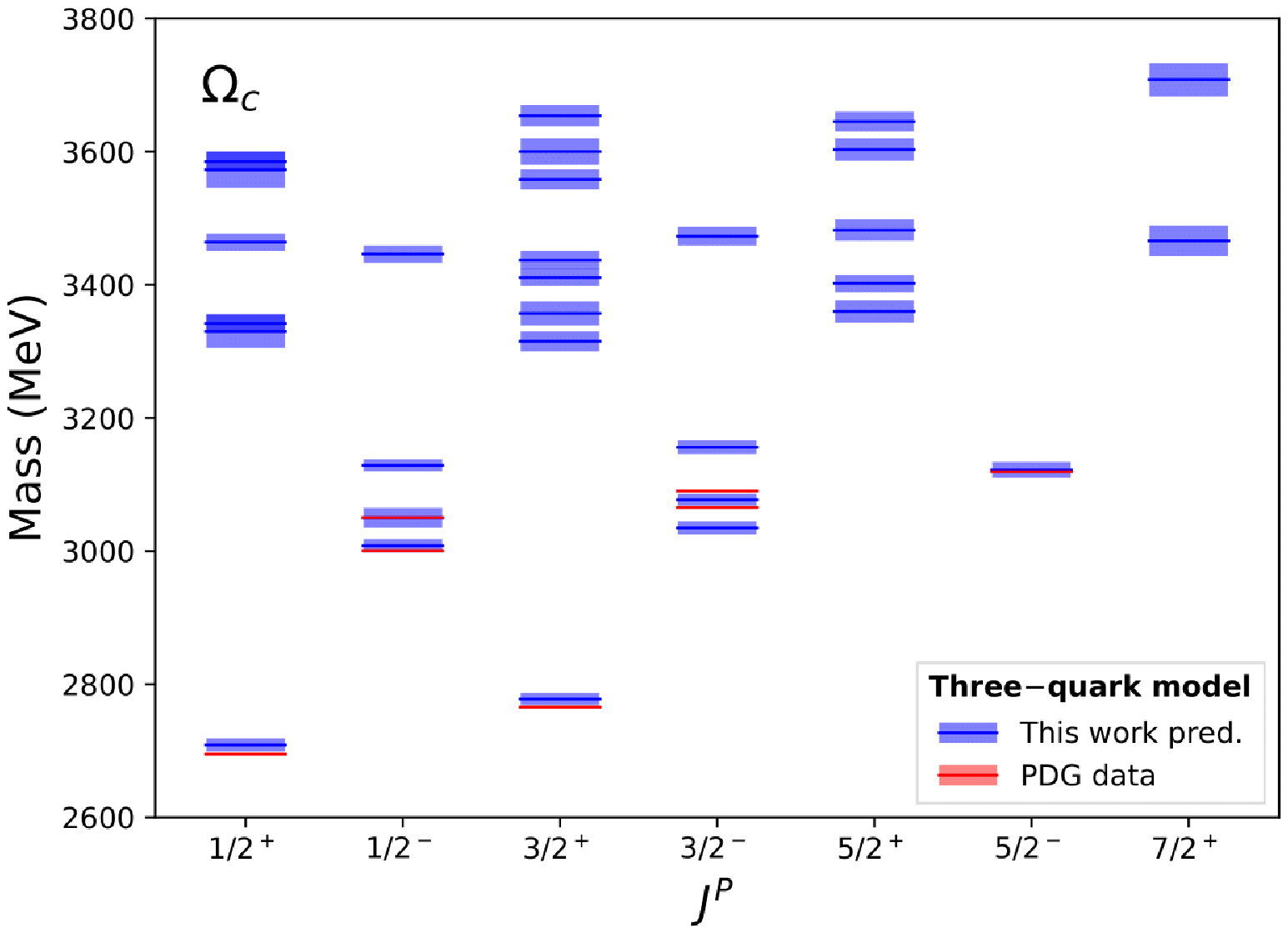}
    
    \label{fig:omegas}
\end{figure}

\begin{figure}
    \centering
    \caption{Same as Figure \ref{fig:omegas}, but for $\Xi'_c$ states.}
    \includegraphics[width=0.5\textwidth]{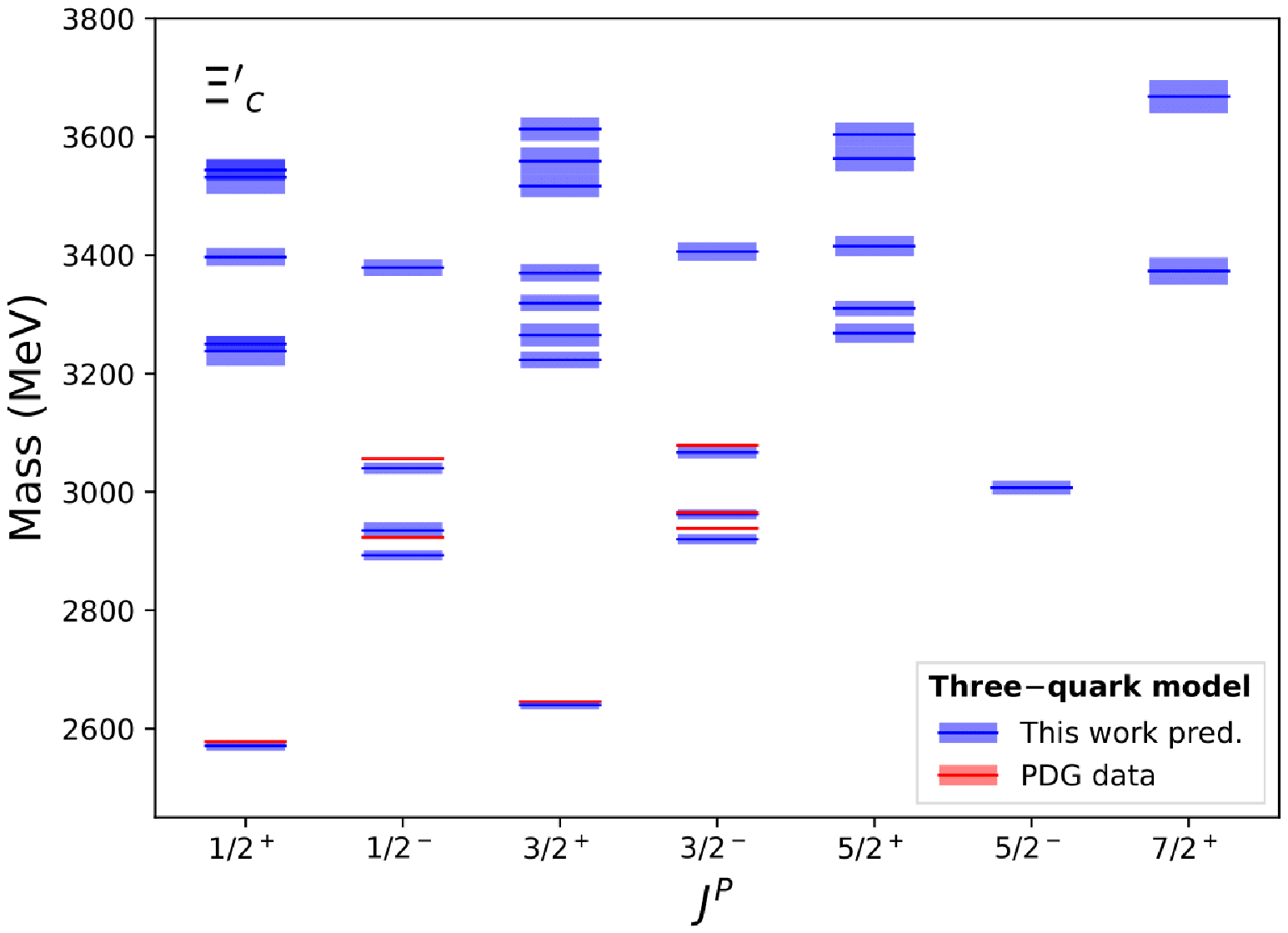}
    \label{fig:cascades}
\end{figure}

\begin{figure}
    \centering
    \caption{Same as  Figure \ref{fig:omegas}, but for $\Sigma_c$ states.}
    \label{fig:sigmas}
    \includegraphics[width=0.5\textwidth]{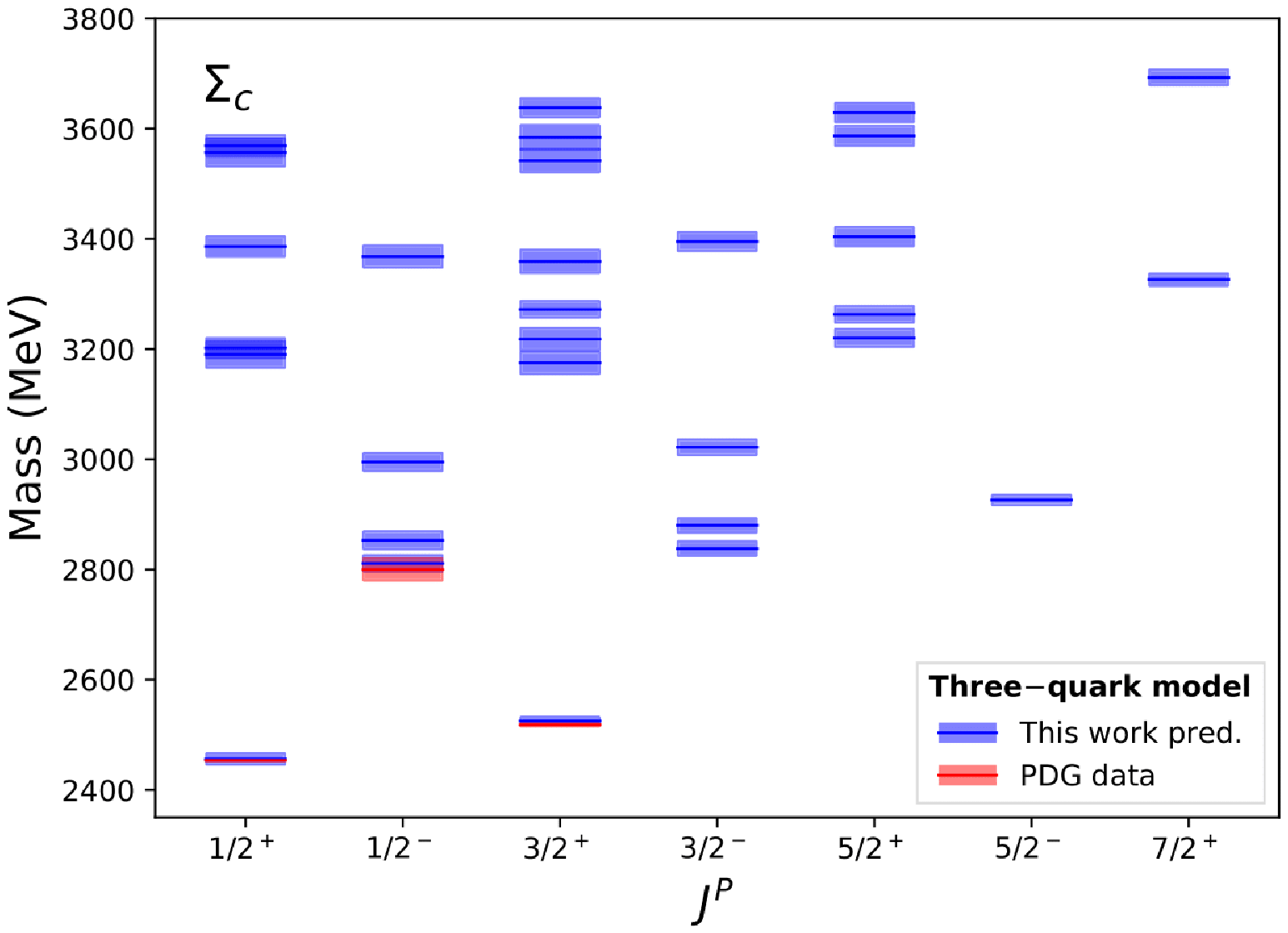}
\end{figure}

\begin{figure}
    \centering
     \caption{Same as  Figure \ref{fig:omegas}, but for $\Lambda_c$ states.}
    \includegraphics[width=0.5\textwidth]{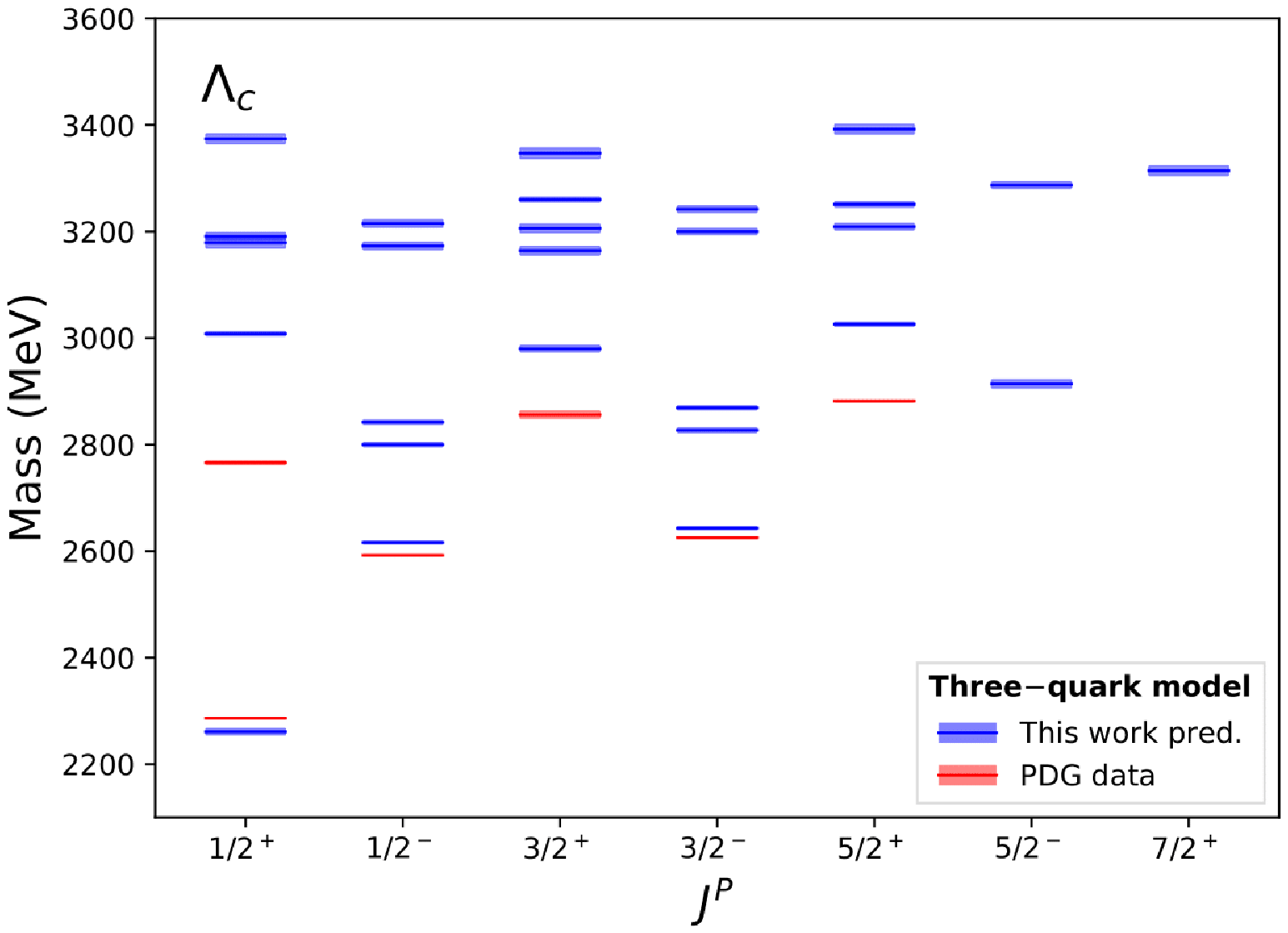}
    \label{fig:lambdas}
\end{figure}

\begin{figure}
    \centering
    \caption{Same as  Figure \ref{fig:omegas}, but for $\Xi_c$ states.}
    \includegraphics[width=0.5\textwidth]{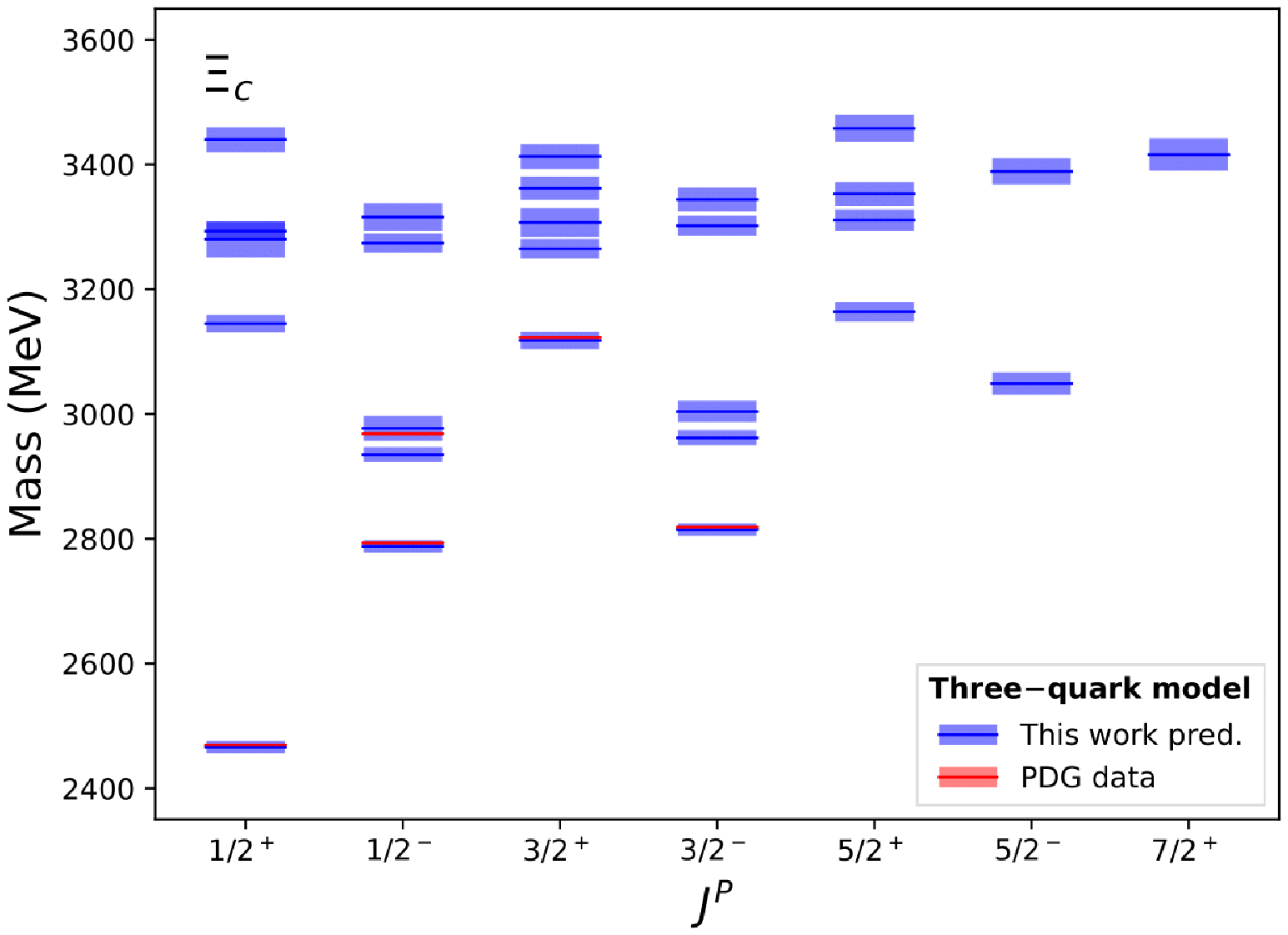}
    \label{fig:cascades_anti3}
\end{figure}

%% Diquark
%%%%%%%%%%%%%%%%%%%%%%%%%%%%%%%%%%%%%%%%%%%
\begin{figure}
    \centering
    \caption{$\Omega_c$ mass spectra and tentative quantum number assignments based on the quark-diquark model Hamiltonian of Eqs. \ref{MassFormula} and \ref{eq:Hhodi}. The theoretical predictions and their uncertainties (red lines and bands) are compared with the experimental results (red lines and bands) reported in the PDG \cite{Zyla:2020zbs}. The experimental errors are too small to be evaluated on this energy scale.}
    \includegraphics[width=0.5\textwidth]{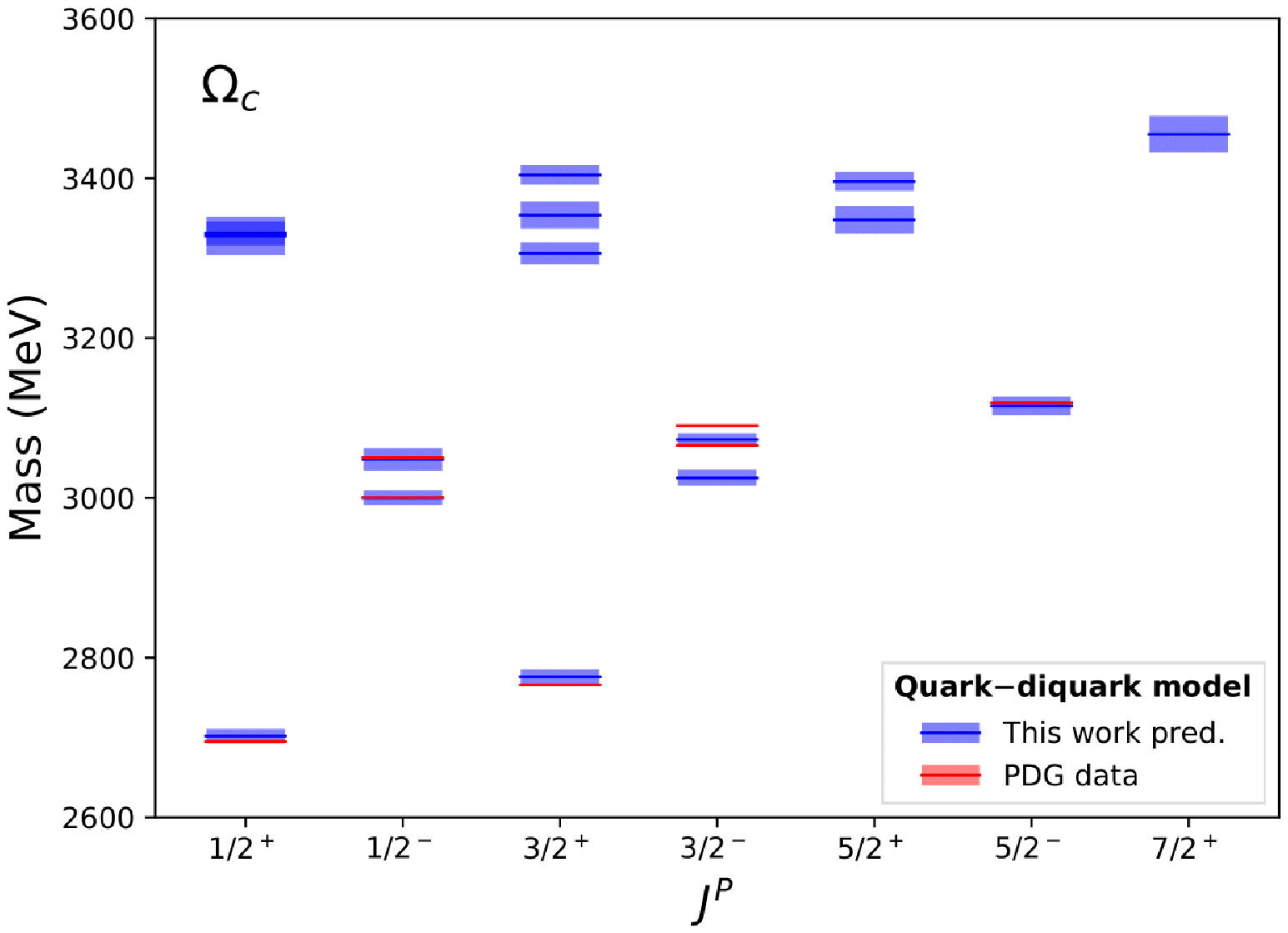}
    
    \label{fig:omegasD}
\end{figure}

\begin{figure}
    \centering
    \caption{Same as  Figure \ref{fig:omegasD}, but for $\Xi'_c$ states.}
    \includegraphics[width=0.5\textwidth]{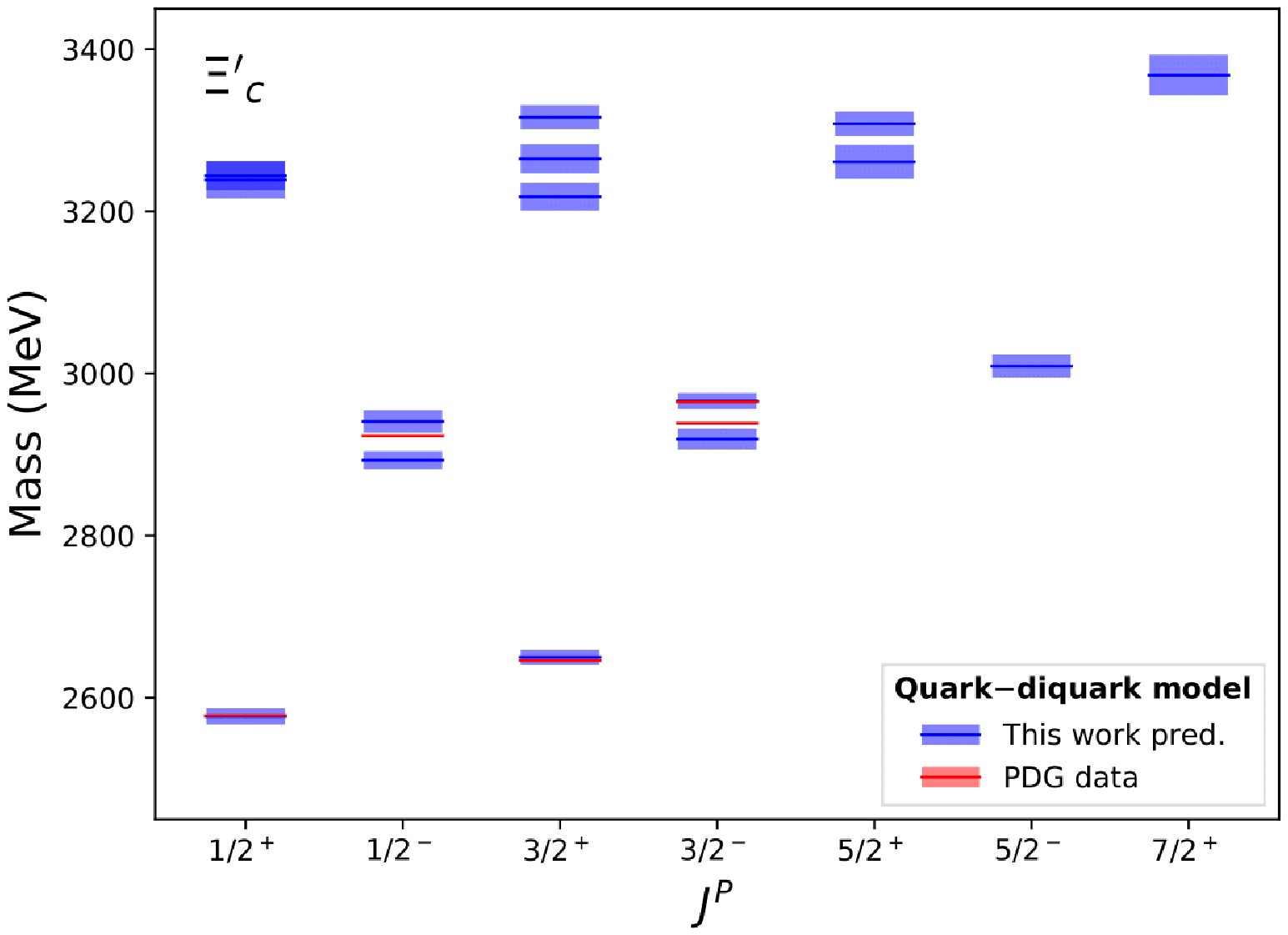}
    \label{fig:cascadesD}
\end{figure}

\begin{figure}
    \centering
    \caption{Same as  Figure \ref{fig:omegasD}, but for $\Sigma_c$ states.}
    \label{fig:sigmasD}
    \includegraphics[width=0.5\textwidth]{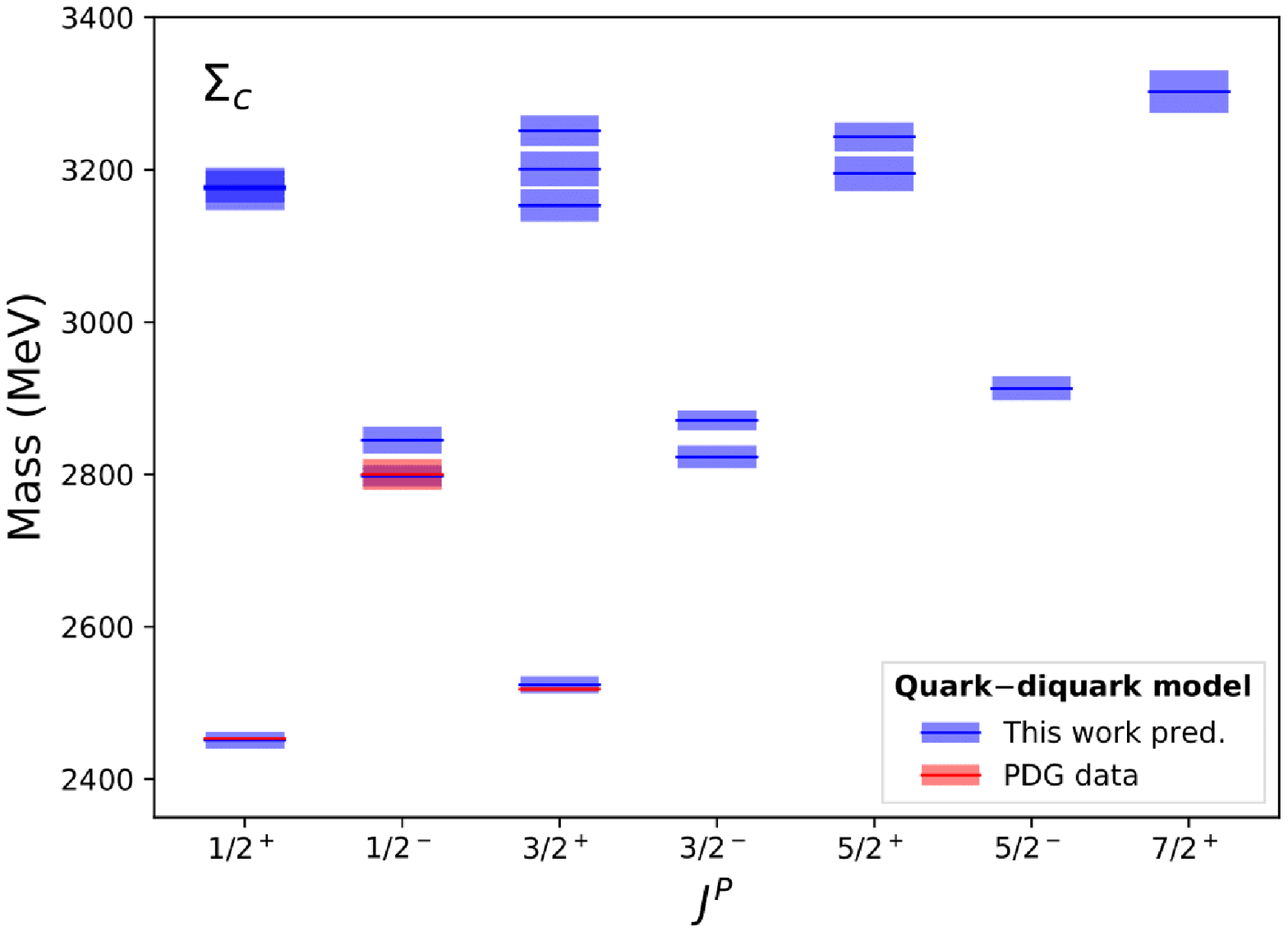}
\end{figure}

\begin{figure}
    \centering
     \caption{Same as  Figure \ref{fig:omegasD}, but for $\Lambda_c$ states.}
    \includegraphics[width=0.5\textwidth]{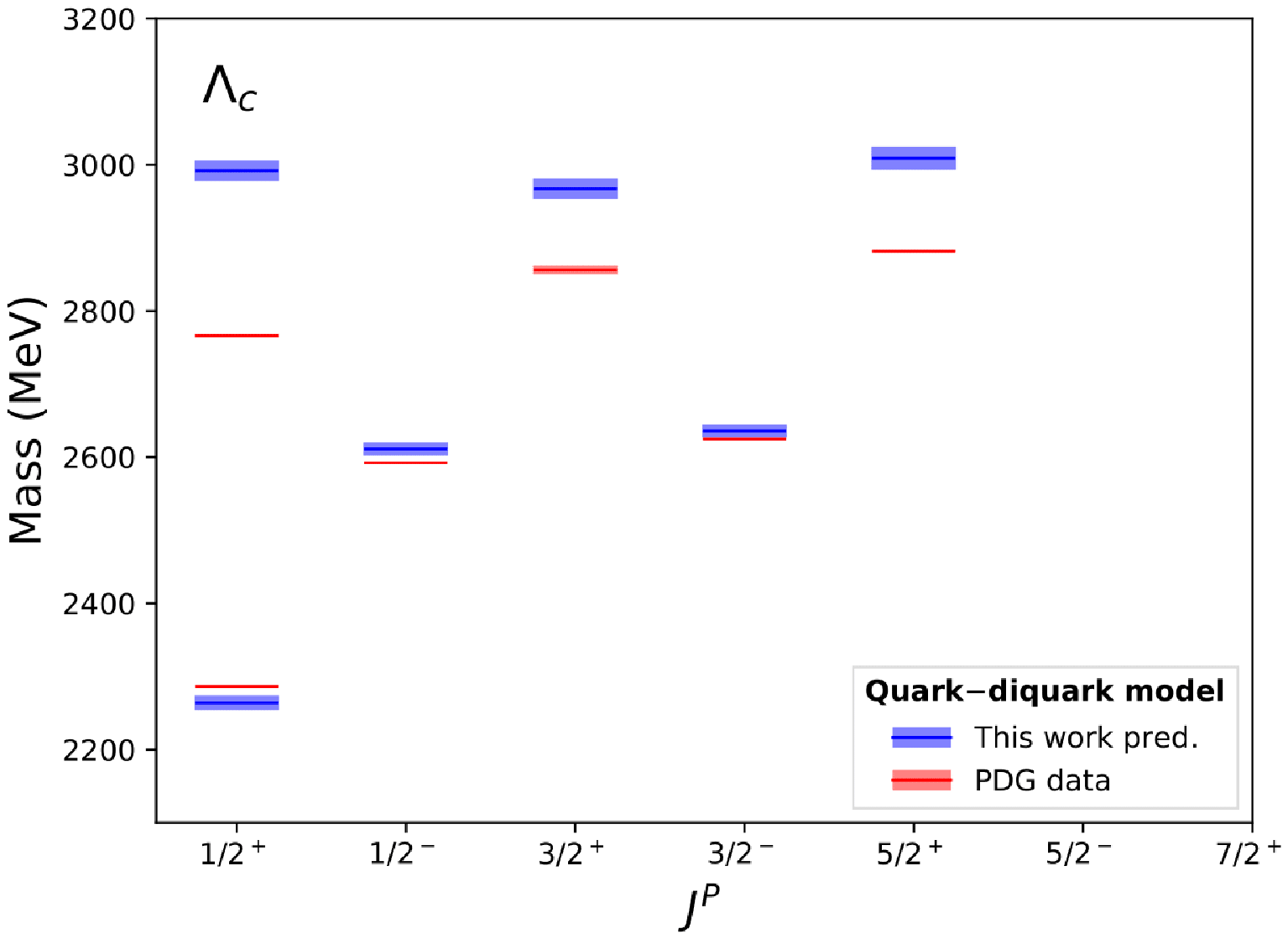}
    \label{fig:lambdasD}
\end{figure}

\begin{figure}
    \centering
    \caption{Same as  Figure \ref{fig:omegasD}, but for $\Xi_c$ states.}
    \includegraphics[width=0.5\textwidth]{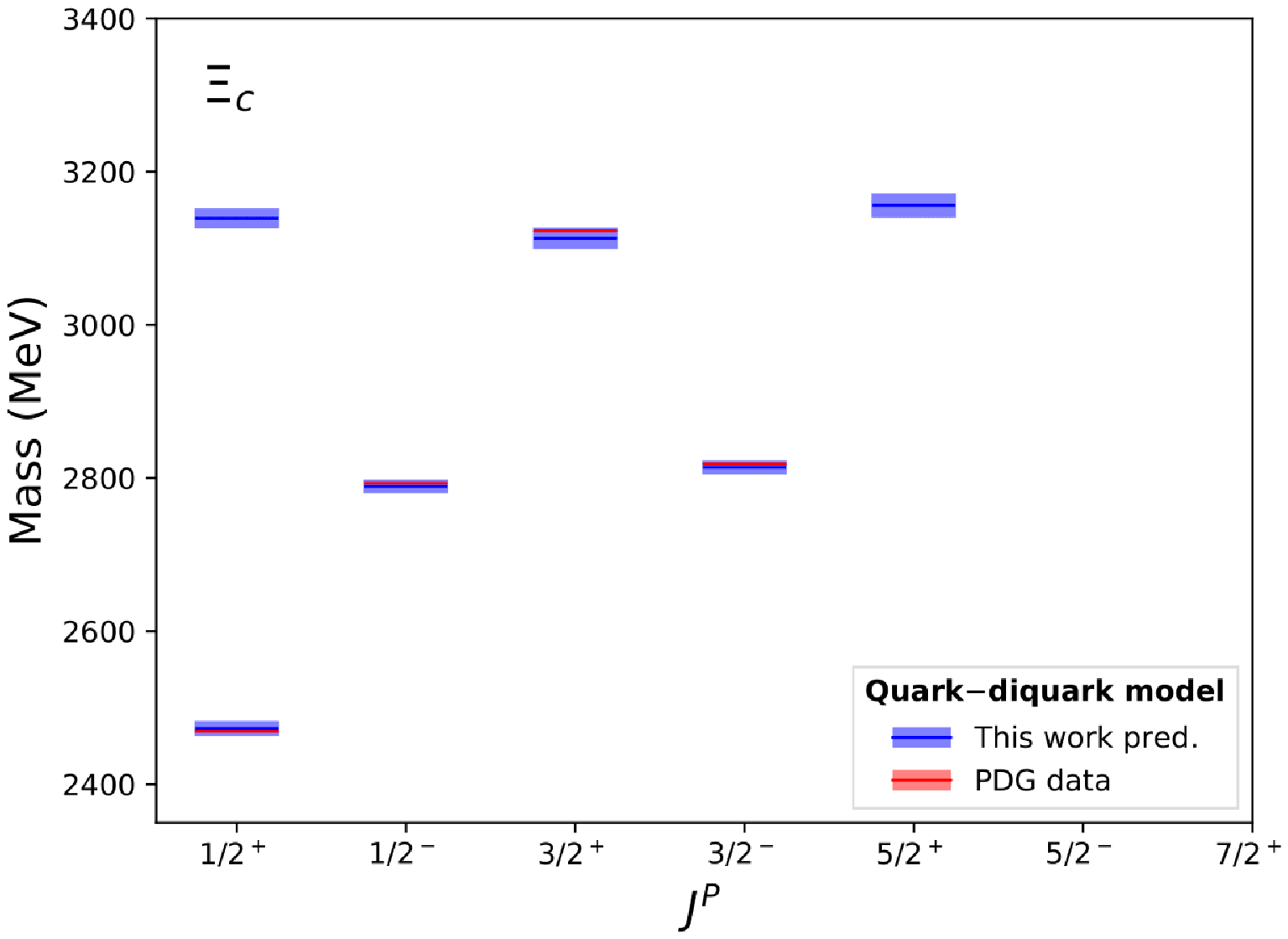}
    \label{fig:cascades_anti3D}
\end{figure}

In this section, we present our results regarding the mass spectra and total decay widths of charmed baryons. The mass spectra are computed via the mass formula of Eq.~\ref{MassFormula}. 
The theoretical masses and their uncertainties are reported in the third column for the three-quark system and in the fourth column for the quark-diquark system in Tables~\ref{tab:All_mass_Omega}-\ref{tab:All_mass_Lambda}. The theoretical decay widths for the three-quark system are computed by using the $^3P_0$ model described in Sec.~\ref{secIIC}. The  masses used in the decay width calculations are the three-quark model theoretical predictions of Tables~\ref{tab:All_mass_Omega}-\ref{tab:All_mass_Lambda}. 
 Each open-flavor channel decay width is obtained via Eq.~\ref{gamma}, and the total decay width is the sum over all the channels. 
 The theoretical total-decay widths and their uncertainties for the three-quark system are reported in the fifth column of Tables~\ref{tab:All_mass_Omega}-\ref{tab:All_mass_Lambda}.  The partial decay widths in the open-flavor channels are  reported  in   Tables~\ref{tab:part_dec_Omega}-\ref{tab:part_dec_lambdas} of Appendix \ref{partialdw}. % In the calculations of the partial decay widths displayed in Tables~\ref{tab:part_dec_Omega}-\ref{tab:part_dec_lambdas}, we use our three-quark model theoretical predictions of Tables~\ref{tab:All_mass_Omega}-\ref{tab:All_mass_Lambda}. 
 In  Tables~\ref{tab:part_dec_Omega}-\ref{tab:part_dec_lambdas}, the partial decay widths  denoted by 0  are forbidden by phase space, while the ones denoted by $-$ are forbidden by selection rules.

Our proposed quantum number assignments for the charmed baryon states are summarized in Figures~\ref{fig:omegas}-\ref{fig:cascades_anti3} within the three-quark model. There is a good agreement between the predicted mass pattern spectrum and the experimental data. Furthermore, we present our charmed baryon spectrum on using  the quark-diquark framework in Figures \ref{fig:omegasD}-\ref{fig:cascades_anti3D}.
In the following subsections, we discuss our assignments to the available data reported in the PDG \cite{Zyla:2020zbs}. 

\subsection{Assignments of charmed baryons} 

First, we discuss our assignments based on our theoretical analyses of the charmed baryons $\Omega_c$, $\Xi'_c$, $\Sigma_c$, $\Xi_c$, and $\Lambda_c$. As a first criterion, we use our mass spectrum to identify the charmed baryon resonances, and the decay width as a secondary criterion. The classification for the quark-diquark model is equivalent to that of the three-quark model when we describe ground states and $\lambda$-type excitations. When states are identified as $\rho$-type excitations in the three-quark model, there are no equivalent states in the quark-diquark model (see Tables~\ref{tab:All_mass_Omega}-\ref{tab:All_mass_Lambda}).
%%%%%%%%%%%%%%%%%%%%%%%%%%%%%%%%%%%%%%%

\subsubsection{$\Omega_c$}
Our results for the $\Omega_c$ resonances are reported in Table~\ref{tab:All_mass_Omega}; they are in good agreement with the experimental masses reported in the PDG. Our results are also consistent with our previous calculations \cite{Santopinto2019}. Here we have extended  our calculations up to $D$-wave states. The $\Omega_c$ and $\Omega_c(2770)$  states are well reproduced \cite{Zyla:2020zbs} in our model. They are identified as the ground states with quantum numbers (QN)  $J^P=1/2^+$ and  $J^P=3/2^+$; note that these QN have not been yet measured however they have been identified  by quark model predictions \cite{Zyla:2020zbs}. The observed $\Omega_c(3000)$ ~\cite{PhysRevLett.118.182001,PhysRevD.97.051102} resonance could be identified as a $P_\lambda$-wave state with $J^P=1/2^-$, where the total-internal-spin is $S=1/2$;  our theoretical width is compatible with the experimental value. Our assignment of $\Omega_c(3000)$ is supported by lattice QCD calculations  \cite{Padmanath:2017lng}, and is also compatible with diquark model interpretations of Ref.~\cite{Karliner:2017kfm},  and predictions of QCD sum rule approaches \cite{Wang:2017zjw}.  The $\Omega_c(3050)$ has an excellent match in our model; the mass is well reproduced, but the width is slightly overestimated; the $\Omega_c(3050)$ is identified as the $J^P=1/2^-$ state, with total-internal-spin $S=3/2$.  Our assignment for $\Omega_c(3050)$  is likewise supported by Ref. \cite{Padmanath:2017lng,Karliner:2017kfm,Wang:2017zjw}, but it also has been identified as a $J^P=3/2^-$ state, see. Ref.  \cite{Wang:2017hej}. In our calculations, the central value deviates $20$ MeV for $\Omega_c(3065)$; however,  the width is overestimated. Hence, we identify the observed $\Omega_c(3065)$ as the state $J^P=3/2^-$ with internal-total-spin $S=1/2$. It should be noted that our state $J^P=3/2^-$ is lighter in mass that the state $J^P=1/2^-$; this may be a numerical consequence of the fit. However, this opens the possibility of interchanging the assignments of the $\Omega_c(3050)$ and $\Omega_c(3065)$ states, with $J^P=3/2^-$ and $J^P=1/2^-$, respectively. The identification of $\Omega_c(3050)$ as a $J^P=3/2^-$ state is supported by Ref. \cite{Wang:2017hej,Padmanath:2017lng,Karliner:2017kfm,Wang:2017zjw}. Only future experiments will confirm the right order and  the assignments. The $\Omega_c(3090)$ is identified as the state $J^P=3/2^-$ with spin 
$S=3/2$; however, its theoretical mass is slightly underestimated, but the theoretical width is in good agreement with the experimental value.  The $\Omega_c(3090)$  assignment is compatible with  the predictions of Ref. \cite{Padmanath:2017lng,Karliner:2017kfm,Wang:2017zjw}, but also has been identified as a $J^P=5/2^-$ state, see Ref.  \cite{Wang:2017hej} . Finally, the mass of the $\Omega_c(3120)$ resonance is well reproduced in our model; it is identified as the state $J^P=5/2^-$ with spin $S=3/2$. This state was not confirmed by Belle~\cite{PhysRevD.97.051102}, other interpretations are therefore possible. Since this state is very narrow, it can be described as a molecular state~\cite{Santopinto2019}. The identification  of $\Omega_c(3120)$ as $J^P=5/2^-$ is supported  by Ref. \cite{Padmanath:2017lng,Karliner:2017kfm,Wang:2017zjw}, but this picture could not be confirmed by the recent LHCb analysis~\cite{LHCb:2021ptx}.   In the present work the identifications of the $\Omega_c(3000)$ and $\Omega_c(3050)$ are selected to have $J^P = 1/2^-$. The $\Omega_c(3066)$ and $\Omega_c(3090)$ both are selected to have $J^P = 3/2^-$, and the $\Omega_c(3119)$ is possibly a $J^P=5/2^-$ state in accordance with Ref. \cite{Padmanath:2017lng,Karliner:2017kfm,Wang:2017zjw}. Nevertheless,
other interpretations were proposed in Refs.~\cite{Wang:2017hej,Wang:2017kfr} based on the constituent quark model: the $\Omega_c(3000)$, $\Omega_c(3050)$, $\Omega_c(3066)$, and $\Omega_c(3090)$ are $\Omega_c(1P)$ states are predicted to have $J^P=1/2^-$, $3/2^-$, $3/2^-$, and $5/2^-$ respectively.  The $\Omega_c(3119)$ may correspond to one of the two $\Omega_c(2S)$ states. 

%%%%%%%%%%%%%%%%%%%%%%%%%%%%%%%%%%
\subsubsection{$\Sigma_c$}
Our results for $\Sigma_c$ states are reported in Table \ref{tab:All_mass_Sigma}. There are only three experimentally observed $\Sigma_c$ states, all of which have masses that are in excellent agreement with our predictions. $\Sigma_c(2455)$ is identified as the ground state $J^P=1/2^+$. The quantum numbers have not yet been measured, and our predicted masses and decay widths are  in good agreement with the experimental data. We find a similar situation in the case of $\Sigma_c(2520)$, which is identified as a ground state with a spin excitation $J^P=3/2^+$. The quantum numbers have not yet been measured, but our theoretical mass is in good agreement with the experimental data, and the decay width is well reproduced. $\Sigma_c(2800)$ is identified as the first $P_\lambda$-wave excitation, with the assignment $J^P=1/2^-$; the theoretical mass and width are compatible with the experimental data.   The lack of data limits the identification of the $\Sigma_c$ states. For instance,  Ref. \cite{Wang:2021bmz} utilizes the chiral quark model to identify the $\Sigma_c(2800)$ as two overlapping $P$-wave $\Sigma_c$ resonances with $J^P=3/2^-$ and $J^P=5/2^-$, respectively. 

%%%%%%%%%%%%%%%%%%%%%%%%%%%%%
\subsubsection{$\Xi'_c$ and  $\Xi_c$}
Our results for $\Xi'_c$ resonances are reported in Table~\ref{tab:All_mass_Xiprime} and those for $\Xi_c$ are reported in Table~\ref{tab:All_mass_Xi}. The $\Xi'_c$ states belong to the sextet configuration and the $\Xi_c$ states belong to the anti-$\bf 3$-plet. Identifying the available data for these states is more complex, since there are several theoretical excited states for $\Xi'_c$ and $\Xi_c$ in the same energy region. 
Additionally, for these states some experimental data are puzzling, as Ref. \cite{Aaij:2020yyt} reports a state with a central mass close to 2965 MeV. Hence, further studies are required in order to establish whether this narrow resonance is a different baryon from the narrow resonance at 2970 MeV found by Belle~\cite{Yelton:2016fqw}. Moreover, the Belle collaboration recently measured the quantum numbers of $\Xi^+_c(2970)$ to be $J^P=1/2^+$ \cite{PhysRevD.103.L111101}, which could indicate that this state is a radial excitation. The  $\Xi'_c$ and $\Xi_c$ ground states are well reproduced in our model, and are identified as $J^P=1/2^+$ of the sextet and anti-$\bf 3$-plet, respectively. $\Xi_c(2645)$ is identified as the $J^P=3/2^+$ member of the sextet. In our model, its mass is well reproduced and the width is underestimated. The  $\Xi_c(2790)$ and $\Xi_c(2815)$ states are identified as $J^P=1/2^-$ and $J^P=3/2^-$ respectively, see Table \ref{tab:All_mass_Xi}; these quantum numbers, which have not yet been measured, refer to the first orbital excitations of the anti-$\bf 3$-plet states. $\Xi_c(2923)$ is identified as the $P_{\lambda}$-wave excitation $J^P=1/2^-$ with spin $S=3/2$ that belongs to the sextet ($\Xi_c^\prime(1P)$ states); the theoretical width is compatible with the experimental value.  The assignment of $\Xi_c(2923)$ as $J^P=1/2^-$  is  consistent with our previous work \cite{Bijker:2020tns} and  supported by the QCD sum rule approaches in Ref. \cite{Yang:2020zjl,Yang:2021lce}; although in Ref. \cite{Wang:2020gkn} it is identified as a $J^P=3/2^-$ state. The $\Xi_c(2930)$ is  identified as a $P_{\lambda}$-wave $J^P=3/2^-$ state with $S=1/2$ that belongs to the sextet; our theoretical mass deviates by 5 MeV, but our theoretical width is in good agreement with the experimental value.  The assignment of $\Xi_c(2930)$ as $J^P=3/2^-$ state is supported by several approaches \cite{Bijker:2020tns,Yang:2020zjl,Yang:2021lce,Wang:2020gkn}. However, there is another possible assignment to this resonance, as the $J^P=1/2^-$ state with $S=1/2$ is close in mass, and belongs to the anti-$\bf 3$-plet; only future experiments will determine the correct assignment.
 
Furthermore, there is a puzzle regarding the state observed by LHCb \cite{Aaij:2020yyt}: it has not been established whether the $\Xi^0_c(2965)$ state is the isospin partner of $\Xi^+_c(2970)$, or a different state. The complexity of identifying $\Xi_c(2965)$ is revealed by the fact that our model predicts two states, which adapt equally well for this state. The first $\Xi_c(2965)$ assignment is $J^P=3/2^-$ with $S=3/2$; this belongs to the sextet, and refers to a $P_\lambda$-wave excitation. The experimental mass and the width are well reproduced.  The identification of $\Xi_c(2965)$  as   $J^P=3/2^-$ is compatible with the  QCD sum rule approach \cite{Yang:2020zjl,Yang:2021lce}, but   in  Ref.~\cite{Wang:2020gkn} the $\Xi_c(2965)$ resonance may correspond to the $\lambda$-mode $\Xi_c^\prime(1P)$ state with $J^P = 5/2^-$. A second identification of $\Xi_c(2965)$ is a $P_\rho$-wave $J^P=1/2^-$ state with $S=3/2$; thus, $\Xi_c(2965)$  would belong to the anti-$\bf 3$-plet, since we obtain a similar mass of $2978\pm6$ MeV and width that is compatible with the experimental value. It is noteworthy that, if the experiments confirm that there is a $\Xi_c$ state at 2965 MeV which it is not a Roper state, it would mean that we could identify this state as a  member of the sextet or anti-$\bf 3$-plet: either as a $P_\lambda$-wave excitation or as a $P_\rho$-wave excitation. The latter would imply that the charmed baryons behave as three-quark systems, instead of quark-diquark systems.  Future experiments will help disentangle this puzzle.    % but of course we can have a superposition of these descriptions. 
There is a similar situation for $\Xi_c(3055)$, where we also have two possible assignments. The first one is its identification as the first $P_\rho$-wave state in the sextet, with $J^P=1/2^-$ and $S=1/2$. Our theoretical mass exhibits a deviation of only 6 MeV, but the width is overestimated. The other possible $\Xi_c(3055)$ assignment is to the $P_\rho$-wave state in the anti-$\bf 3$-plet, with $J^P=5/2^-$ and $S=3/2$. Here, the mass is well reproduced and the width is overestimated. $\Xi_c(3080)$ is identified as the $P_\rho$-wave of the sextet, with $J^P=1/2^-$ and $S=3/2$. While our theoretical mass is well reproduced, our width is overestimated.  Also, there are other possible interpretations of  $\Xi_c(3055)$  and $\Xi_c(3080)$. For instance in Ref.~\cite{Chen:2017aqm} the authors identified $\Xi_c(3055)$ and $\Xi_c(3080)$ together from the $D$-wave charmed baryon doublet of the anti-$\bf 3$-plet. Finally, $\Xi_c(3123)$ is identified as the  first $D_\lambda$-wave excitation $J^P=3/2^+$ with $S=1/2$ of the anti-$\bf 3$-plet. The mass is well reproduced in our model but the width is overestimated.

%%%%%%%%%%%%%%%%%%%%%%%%%%%%%%%%%%%
%\subsubsection{}
\subsubsection{$\Lambda_c$}
Our results for $\Lambda_c$ baryons are reported in Table \ref{tab:All_mass_Lambda}.
  The $\Lambda^+_c$ is identified as the ground state  $J^P=1/2^+$, with $S=1/2$; its mass is well reproduced, with a small deviation of 15 MeV. For the excited states, we can observe a systematic deviation that exhibits the failure of the h.o. potential for these states. Nevertheless, the patterns in the theoretical mass spectrum can describe the experimental one.
  $\Lambda_c(2595)^+$ and $\Lambda_c(2625)^+$ are identified as our two $P_\lambda$-wave excitations  $J^P=1/2^-$ and  $J^P=3/2^-$, respectively, both with $S=1/2$; their masses are reproduced with a deviation of 15 MeV. The theoretical width is compatible with the experimental for $\Lambda_c(2595)^+$ states value but overestimated for $\Lambda_c(2625)^+$.   The identification of  $\Lambda_c(2595)^+$ and $\Lambda_c(2625)^+$ are  in agreement with various quark models~\cite{Ebert:2011kk,Roberts:2007ni,Yoshida:2015tia,Chen:2016iyi}.  If the $\Lambda_c(2765)^+$ or $\Sigma_c(2765)$  state is identified as the $\Lambda_c$ state in our model, there is no resonance within this energy region. Although it is close in energy to our predicted state $\Lambda_c(2800)^+$, $\Lambda_c(2765)^+$ is expected to be a Roper-like resonance. Consequently, we fail to reproduce the $\Lambda_c(2765)^+$ mass in our model, since our first theoretical radial excitation is the $\Lambda_c(3007)^+$ state. The observed $\Lambda_c(2860)^+$ is identified with a $J^P=$ $3/2^+$ state, having a significant predicted mass deviation of 100 MeV, but the  theoretical decay width is well reproduced. Finally, the observed $\Lambda_c(2880)^+$ is identified with  {\color{red} a $J^P=$} $5/2^+$ state. In this case, we also have a predicted mass with a deviation of 100 MeV,  and a large deviation of the decay width.  The identifications of  $\Lambda_c(2860)^+$ and $\Lambda_c(2880)^+$ as $D$-wave states,  $J^P = 3/2^+$  and $J^P = 5/2^+$  are in agreement with quark models~\cite{Ebert:2011kk,Roberts:2007ni,Yoshida:2015tia,Chen:2016iyi}.

%%%%%%%%%%%%%%%%%%%%%%%%%%%%%%%%%%%%%%%%%%%%%%%%%%%%%%
\section{Comparison between the three-quark and  quark-diquark structures}
\label{comparison}
In the light baryon sector, the successful constituent quark model reproduces the baryon mass spectra by assuming that  the  constituent quarks have roughly the same mass. This implies that the two oscillators, $\rho$ and $\lambda$, have  the same frequency, $\omega_{\rho}\simeq \omega_{\lambda}$, 
meaning that the $\lambda$ and $\rho$ modes are degenerate in the mass spectrum.  In the charm sector,  we have a mass splitting between the  $\lambda$ and $\rho$  modes, which is given by  $\omega_\lambda -\omega_\rho\simeq122$ MeV for $\Omega_c$ baryons, 
 by $\omega_\lambda -\omega_\rho\simeq147$ MeV for $\Xi_c$ and $\Xi_c'$ baryons, and  by $\omega_\lambda -\omega_\rho\simeq183$ MeV for $\Sigma_c$ and $\Lambda_c$ baryons. Consequently, we may expect to find the $\rho$-mode excited states in future experiments. However, given that  the $\rho$ states have not been observed yet, it seems that the charmed baryons can have a special internal structure which corresponds  to the quark-diquark configuration. The reduction of the effective degrees of freedom in the quark-diquark picture means fewer predicted states. We notice that in the case of $\Lambda_c$ and $\Xi_c$ baryons, the number of states decreases drastically in the quark-diquark model, see Tables \ref{tab:All_mass_Lambda} and \ref{tab:All_mass_Xi} respectively. 
The lack of experimental data prevents us from reaching a decisive conclusion about which description is better. 

For instance, for the $\Omega_c$ baryons,
we have identified all the five $P_\lambda$-wave excited states  with the experimental ones.  We also expect to observe the two $P_\rho$-wave  excited states, $\Omega_c (3129)$ with $J^P=1/2^-$, and $\Omega_c (3156)$ with  ${\color{red}J^P}=3/2^-$. The existence of these states may indicate that the charmed baryons are not quark-diquark systems.

\section{Conclusions}
\label{conclusion}
We have calculated the mass spectra, the strong partial decay widths and the total decay widths of {\color{red}the} charmed baryons up to the $D$-wave.
All charmed baryons are simultaneously described by a global fit in which the same set of model parameters predicts the charmed baryon masses and strong partial decay widths in all the possible decay channels up to the $D$-wave.  Moreover, the  charmed baryon mass spectra are given in both the three-quark and the quark-diquark schemes.  
Propagation of the parameter uncertainties via a Monte Carlo bootstrap method is also included. This is often missing in theoretical papers on this subject. Nevertheless, it is necessary in order to guarantee a rigorous treatment of the uncertainties in the predicted mass spectra and decay widths.
Our mass and strong partial decay width predictions are in good agreement with the available experimental data, and show the ability to guide future experimental  searches by LHCb, Belle and Belle II.
Moreover, for all the possible decay channels, we provide the flavor coupling coefficients, which are relevant to further theoretical investigations on charmed baryon strong decay widths.
To the best of our knowledge, considering that the calculations of the strong decay widths are barely sensitive to the specific model used, our strong partial decay width predictions constitute the most complete calculation in the charmed baryon sector up to date.

\section*{Acknowledgments}
This work was supported in part by INFN, sezione di Genova.
The authors acknowledge financial support from CONACyT, M\'exico (postdoctoral fellowship for H. Garc\'ia-Tecocoatzi); National Research Foundation of Korea (grants
2020R1I1A1A01066423,\\ 2019R1I1A3A01058933,  2018R1A6A1A06024970).

%start appendix
\appendix
%\section{Decay model}
\section{Baryon wave functions}
\label{BWF}
\label{app}
\subsection{The harmonic oscillator wave functions} % used in our
%calculation, in terms of $\omega_{\rho}$ and $\omega_{\lambda}$}

In the heavy-light sector, the the $\rho$- and $\lambda$-modes decouple; therefore, they can be distinguished through an analysis of the heavy-light baryon mass spectra.
This is because there is a difference in frequency between the $\rho$- and $\lambda$-modes, 
\begin{eqnarray}
\omega_{\rho}=\sqrt{\frac{3 K_Q}{m_{\rho}}} \;\; \text{and} \;\;\omega_{\lambda}=\sqrt{\frac{3 K_Q}{m_{\lambda}}},
\end{eqnarray}
where $m_{\rho}$ and $m_{\lambda}$ are defined in Section \ref{secIIA} .  We write the baryon wave functions in terms of $\omega_{\rho}$ and $\omega_{\lambda}$ by using the relation $\alpha^2_{\rho,\lambda}=\omega_{\rho,\lambda}m_{\rho,\lambda}$ . 

Also, we use the standard Jacobi coordinates:   
\begin{eqnarray}
\bf p_{\rho} &=& \frac{1}{2} (\bf p_1 - p_2)  \mbox{ },
\nonumber\\
\bf p_{\lambda} &=& \frac{1}{3} (\bf p_1 +  p_2 - 2  p_3)  \mbox{ },
\nonumber\\
\bf P_{cm} &=& \bf p_1+  p_2 +  p_3  \mbox{ },
\label{momRSB}
\end{eqnarray}
for the baryon, and
\begin{eqnarray}
\bf q_c &=& \frac{1}{2} (\bf p_3 -  p_5)  \mbox{ },
\end{eqnarray}
 for the meson. In this coordinate system, $\bf p_3$ refers to the charm quark and $\bf p_{1,2}$ to the light quarks. Finally, $\bf p_5$ is the anti-quark momentum.

\noindent
For the $S$-wave charmed baryon, we have,
\begin{eqnarray}
\psi(0,0,0,0)&=&3^{3/4}\;(\frac{1}{\pi
\omega_{\rho}m_{\rho}})^{\frac{3}{4}}\,(\frac{1}{\pi
\omega_{\lambda}m_{\lambda}})^{\frac{3}{4}}\nonumber \\ &\times&\,\exp\Big[{-\frac{
{\mathbf{p}}^2_{\rho}}{2\omega_{\rho}m_{\rho}} -\frac{
{\mathbf{p}}^2_{\lambda}}{2\omega_{\lambda}m_{\lambda}}}\Big].\nonumber\\&&
\end{eqnarray}

\noindent
For the $P$-wave charmed baryon, we have,
\begin{eqnarray}
\psi(1,m,0,0)&=&-i\;3^{3/4}\,\Big(\frac{8}{3\sqrt{\pi}}\Big)^{{1}/{2}}\Big(\frac{1}{
\omega_{\rho}m_{\rho}}\Big)^{ {5}/{4}}\,{\cal
Y}^m_1({\mathbf{p}}_{\rho})\nonumber\\&\times&\Big(\frac{1}{\pi
\omega_{\lambda}m_{\lambda}}\Big)^{{3}/{4}}\,\exp\Big[{-\frac{{\mathbf{p}}^2_{\rho}}{2\omega_{\rho}m_{\rho}}
-\frac{{\mathbf{p}}^2_{\lambda}}{2\omega_{\lambda}m_{\lambda}}}\Big],\nonumber\\&& \\
\psi(0,0,1,m)&=&-i\;3^{3/4}\,\Big(\frac{8}{3\sqrt{\pi}}\Big)^{{1}/{2}}\Big(\frac{1}{
\omega_{\lambda}m_{\lambda}}\Big)^{{5}/{4}}\,{\cal
Y}^m_1({\mathbf{p}}_{\lambda})\nonumber\\&\times& \Big(\frac{1}{\pi
\omega_{\rho}m_{\rho}}\Big)^{{3}/{4}}\,\exp\Big[{-\frac{{\mathbf{p}}^2_{\rho}}{2\omega_{\rho}m_{\rho}}
-\frac{{\mathbf{p}}^2_{\lambda}}{2\omega_{\lambda}m_{\lambda}}}\Big].\nonumber\\&&
\end{eqnarray}

\noindent
For the $D$-wave charmed baryon, we have,
\begin{eqnarray}
\psi
(2,m,0,0)&=&3^{3/4}\;\Big(\frac{16}{15\sqrt{\pi}}\Big)^{{1}/{2}}\Big(\frac{1}{
\omega_{\rho}m_{\rho}}\Big)^{{7}/{4}}\,{\cal
Y}^m_2({\mathbf{p}}_{\rho})\nonumber\\&\times&\Big(\frac{1}{\pi
\omega_{\lambda}m_{\lambda}}\Big)^{{3}/{4}}\,\exp\Big[{-\frac{{\mathbf{p}}^2_{\rho}}{2\omega_{\rho}m_{\rho}}
-\frac{{\mathbf{p}}^2_{\lambda}}{2\omega_{\lambda}m_{\lambda}}}\Big], \nonumber\\&&\\
\psi(0,0,2,m)&=&3^{3/4}\;\Big(\frac{16}{15\sqrt{\pi}}\Big)^{{1}/{2}}\Big(\frac{1}{
\omega_{\lambda}m_{\lambda}}\Big)^{{7}/{4}}\,{\cal
Y}^m_2({\mathbf{p}}_{\lambda})\nonumber\\&\times&\Big(\frac{1}{\pi
\omega_{\rho}m_{\rho}}\Big)^{{3}/{4}}\,\exp\Big[{-\frac{{\mathbf{p}}^2_{\rho}}{2\omega_{\rho}m_{\rho}}
-\frac{ {\mathbf{p}}^2_{\lambda}}{2\omega_{\lambda}m_{\lambda}}}\Big],\nonumber\\&&\\
\psi(1,m,1,m')&=&-3^{3/4}\;\Big(\frac{8}{3\sqrt{\pi}}\Big)^{{1}/{2}}\Big(\frac{1}{
\omega_{\rho}m_{\rho}}\Big)^{{5}/{4}}\,{\cal Y}^m_1(
{\mathbf{p}}_{\rho})\nonumber\\&\times&\Big(\frac{8}{3\sqrt{\pi}}\Big)^{{1}/{2}}\Big(\frac{1}{
\omega_{\lambda}m_{\lambda}}\Big)^{{5}/{4}}{\cal
Y}^{m'}_1({\mathbf{p}}_{\lambda})\nonumber\\&\times&
\exp\Big[{-\frac{{\mathbf{p}}^2_{\rho}}{2\omega_{\rho}m_{\rho}}
-\frac{{\mathbf{p}}^2_{\lambda}}{2\omega_{\lambda}m_{\lambda}}}\Big].
\end{eqnarray}
\noindent
Here $\mathcal{Y}_{l}^{m}(\mathbf{p})$ is the solid harmonic.
The wave functions of the first radially excited charmed baryons $\psi(k_\lambda,k_\rho)$ are
\begin{eqnarray}
\psi(1,0)&=&3^{3/4}\;\Big(\frac{2}{3}\Big)^{{1}/{2}}\Big(\frac{1}{\pi^4
\omega_{\lambda}m_{\lambda} \omega_{\rho}m_{\rho}}\Big)^{{3}/{4}} \Big[ \frac{3}{2}-\frac{ \mathbf{p}^2_{\lambda}}{\omega_{\lambda}m_{\lambda}} \Big]\nonumber\\&\times&
\exp\Big[{-\frac{{\mathbf{p}}^2_{\rho}}{2\omega_{\rho}m_{\rho}}
-\frac{{\mathbf{p}}^2_{\lambda}}{2\omega_{\lambda}m_{\lambda}}}\Big],
\end{eqnarray}
\begin{eqnarray}
\psi(0,1)&=&3^{3/4}\;\Big(\frac{2}{3}\Big)^{{1}/{2}}\Big(\frac{1}{\pi^4
\omega_{\lambda}m_{\lambda} \omega_{\rho}m_{\rho}}\Big)^{{3}/{4}} \Big[ \frac{3}{2}-\frac{ \mathbf{p}^2_{\rho}}{\omega_{\rho}m_{\rho}} \Big]\nonumber\\&\times&
\exp\Big[{-\frac{{\mathbf{p}}^2_{\rho}}{2\omega_{\rho}m_{\rho}}
-\frac{{\mathbf{p}}^2_{\lambda}}{2\omega_{\lambda}m_{\lambda}}}\Big].
\end{eqnarray}

\noindent
The ground state wave function of the meson is
\begin{eqnarray}
\psi(0,0)=\Big(\frac{{R}^2}{\pi}\Big)^{{3}/{4}} \exp
\Big[-\frac{R^2({\mathbf{p}}_3-{\mathbf{p}}_5)^2}{8}\Big].
\end{eqnarray}
\section{Charmed-baryon flavor wave functions }
In the charm sector, we consider the  {\bf 6-plet} and the {\bf $\bar 3$-plet}  representation of the flavor wave functions. \label{flavorcb}
In the following subsections, we give the flavor wave functions of a charmed baryon $A_c$ and its isospin quantum numbers  $|A_c,I,M_I\rangle$.
\subsubsection{\bf 6-plet}

\begin{eqnarray}
|\Omega_c,0,0 \rangle:&=&|ssc\rangle\\
|\Xi^{\prime0}_c,1/2,-1/2 \rangle:&=&\frac{1}{\sqrt{2}}(|dsc\rangle+|sdc\rangle)\\
|\Xi^{\prime+}_c,1/2,1/2 \rangle:&=&\frac{1}{\sqrt{2}}(|usc\rangle+|suc\rangle)\\
|\Sigma^{++}_c,1,1 \rangle:&=&|uuc\rangle\\
|\Sigma^0_c,1,-1 \rangle:&=&|ddc\rangle\\
|\Sigma^+_c,1, 0\rangle:&=&\frac{1}{\sqrt{2}}(|udc\rangle+|duc\rangle)
\end{eqnarray}

\subsubsection{\bf $\bar 3$-plet}
\begin{eqnarray}
|\Xi^{0}_c,1/2,-1/2 \rangle:&=&\frac{1}{\sqrt{2}}(|dsc\rangle-|sdc\rangle)\\
|\Xi^{+}_c,1/2,1/2 \rangle:&=&\frac{1}{\sqrt{2}}(|usc\rangle-|suc\rangle)\\
|\Lambda^+_c,0,0\rangle:&=&\frac{1}{\sqrt{2}}(|udc\rangle-|duc\rangle)
\end{eqnarray}

%%%%%%%%%%%%%%%%%%%%%%%%%%%%%%%%%%%%%%%%%

\section{Light-baryon wave functions}
\setcounter{equation}{0}
Whenever our final states contained a light baryon, we used the following conventions, considering 
the $S_3$ invariant space-spin-flavor ($\Psi=\psi \chi \phi$). Thus, the light-baryon wave functions are given by 
\ba
^{2}8[56,L^P] &:& \psi_S ( \chi_{\rho} \phi_{\rho} 
+ \chi_{\lambda} \phi_{\lambda})/\sqrt{2} ~,
\nonumber\\
^{2}8[70,L^P] &:& [ \psi_{\rho} ( \chi_{\rho} \phi_{\lambda} 
+ \chi_{\lambda} \phi_{\rho}) + \psi_{\lambda} ( \chi_{\rho} \phi_{\rho} 
- \chi_{\lambda} \phi_{\lambda} ) ]/2 ~, 
\nonumber\\
^{4}8[70,L^P] &:& ( \psi_{\rho} \phi_{\rho} 
+ \psi_{\lambda} \phi_{\lambda} ) \chi_S/\sqrt{2} ~,
\nonumber\\
^{2}8[20,L^P] &:& \psi_A ( \chi_{\rho} \phi_{\lambda} 
- \chi_{\lambda} \phi_{\rho} )/\sqrt{2} ~,
\nonumber\\
^{4}10[56,L^P] &:& \psi_S \chi_S \phi_S ~,
\nonumber\\
^{2}10[70,L^P] &:& ( \psi_{\rho} \chi_{\rho} 
+ \psi_{\lambda} \chi_{\lambda} ) \phi_S/\sqrt{2} ~,
\nonumber\\
^{2}1[70,L^P] &:& ( \psi_{\rho} \chi_{\lambda} 
- \psi_{\lambda} \chi_{\rho} ) \phi_A/\sqrt{2} ~,
\nonumber\\
^{4}1[20,L^P] &:& \psi_A \chi_S \phi_A~. 
\ea
The quark orbital angular momentum $\bf{L}$ is coupled with the spin 
$\bf{S}$ to yield the total angular momentum $\bf{J}$ of the baryon.

\subsection{Light-baryon flavor wave functions}
\label{flavorlb}
For the flavor wave functions $|(p,q),I,M_I,Y \rangle$ we adopt the 
convention of Ref.~\cite{deSwart:1963pdg} with $(p,q)=(g_1-g_2,g_2)$.\\
\begin{itemize}
\item {\bf The octet baryons} 
\begin{equation}
	\begin{array}{rcccl}
	|(1,1),\frac{1}{2},\frac{1}{2},1 \rangle & : & \phi_{\rho}(p) & = & \frac{1}{\sqrt 2} [ |udu \rangle - |duu \rangle ] \\
	& : & \phi_{\lambda}(p) & = & \frac{1}{\sqrt 6} [ 2|uud \rangle - |udu \rangle \\ 
	& & & & - |duu \rangle ]
	\end{array}
\end{equation}
\begin{equation}
	\begin{array}{rcccl}
	|(1,1),1,1,0 \rangle & : & \phi_{\rho}(\Sigma^+) & = & \frac{1}{\sqrt 2} [ |suu \rangle - |usu \rangle ]  \\
	& : & \phi_{\lambda}(\Sigma^+) & = &  \frac{1}{\sqrt 6} [ |suu \rangle + |usu \rangle \\
	& & & & - 2|uus \rangle ] 
	\end{array}
\end{equation}
\begin{equation}
	\begin{array}{rcccl}
	|(1,1),0,0,0 \rangle & : & \phi_{\rho}(\Lambda) & = & \frac{1}{\sqrt{12}} [ 2|uds \rangle - 2|dus \rangle \\
	& & & & - |dsu \rangle + |sdu \rangle  \\
	& & & & - |sud \rangle + |usd \rangle ] \\
	& : & \phi_{\lambda}(\Lambda) & = & \frac{1}{2} [- |dsu \rangle - |sdu \rangle \\
	& & & & + |sud \rangle + |usd \rangle ]
	\end{array}
\end{equation}
\begin{equation}
	\begin{array}{rcccl}
	|(1,1),\frac{1}{2},\frac{1}{2},-1 \rangle & : & \phi_{\rho}(\Xi^0) & = & \frac{1}{\sqrt 2} [ |sus \rangle - |uss \rangle ]  \\
	& : & \phi_{\lambda}(\Xi^0) & = & \frac{1}{\sqrt 6} [ 2|ssu \rangle \\ 
	& & & & - |sus \rangle - |uss \rangle ] 
	\end{array}
\end{equation}
\item {\bf The decuplet baryons}
\begin{equation}
	\begin{array}{rcccl}
	|(3,0),\frac{3}{2},\frac{3}{2},1 \rangle & : & \phi_S(\Delta^{++}) & = & |uuu \rangle
	\end{array}
\end{equation}
\begin{equation}
	\begin{array}{rcccl}
	|(3,0),1,1,0 \rangle & : & \phi_S(\Sigma^{+}) & = & \frac{1}{\sqrt 3} [ |suu \rangle + |usu \rangle \\
	& & & & + |uus \rangle ]
	\end{array}
\end{equation}
\begin{equation}
	\begin{array}{rcccl}
	|(3,0),\frac{1}{2},\frac{1}{2},-1 \rangle & : & \phi_S(\Xi^{0}) & = & \frac{1}{\sqrt 3} [ |ssu \rangle + |sus \rangle \\
	& & & & + |uss \rangle ]
	\end{array}
\end{equation}
\begin{equation}
	\begin{array}{rcccl}
	|(3,0),0,0,-2 \rangle & : & \phi_S(\Omega^{-}) & = &  |sss \rangle 
	\end{array}
\end{equation}
\item {\bf The singlet baryon}
\begin{equation}
	\begin{array}{rcccl}
	|(0,0),0,0,0 \rangle & : & \phi_A(\Lambda) & = & \frac{1}{\sqrt 6} [ |uds \rangle - |dus \rangle \\
	& & & & + |dsu \rangle - |sdu \rangle \\
	& & & & + |sud \rangle - |usd \rangle ]
	\end{array}
\end{equation}
\end{itemize}

%%%%%%%%%%%%%%%%%%%%%%%%%%%%%%%%%%%%%%%%%%%%%%%%%%%
\section{Meson flavor wave functions}

In the following subsections, we give the flavor wave functions of a $C$  meson  and its isospin quantum numbers  $|C,I,M_I\rangle$.
\label{appme}
\subsection{Pseudoscalar mesons}
Since the mixing angle $\theta_{\eta \eta'}$ between $\eta$ and $\eta'$ is small, we set $\theta_{\eta \eta'} =0$. Thus, we identify $\eta=\eta_8$ and $\eta'=\eta_1$.
\begin{itemize}
 \item {\bf The octet mesons}
 \begin{eqnarray}
|\pi^+,1,1\rangle&=& -|u\bar{d}\rangle \nonumber\\
|\pi^0,1,0\rangle&=&\frac{1}{\sqrt{2}}[|u\bar{u}\rangle-|d\bar{d}\rangle]\nonumber\\ \nonumber
|\pi^-,1,-1\rangle&=&|d\bar{u}\rangle \\\nonumber
|K^+,1/2,1/2\rangle &=& -|u\bar{s}\rangle\\
|K^-,1/2,-1/2\rangle &=& |s\bar{u}\rangle\\\nonumber
|K^0,1/2,-1/2\rangle&=&- |d\bar{s}\rangle\\\nonumber
|\bar{K}^0,1/2,1/2\rangle&=& -|s\bar{d}\rangle\\\nonumber
|\eta ,0,0\rangle&=& \frac{1}{\sqrt{6}} [ |u\bar{u}\rangle +|d\bar{d}\rangle- 2|s\bar{s}\rangle ]
\end{eqnarray} 
\item {\bf The singlet meson}
 \begin{eqnarray}
  |\eta',0,0 \rangle&=& \frac{1}{\sqrt{3}}[ |u\bar{u}\rangle +|d\bar{d}\rangle+|s\bar{s}\rangle ]
 \end{eqnarray}
\end{itemize}

\subsection{Vector mesons}
We consider that the $\phi$ meson is a pure $s\bar s$ state; thus, we have the following wave functions:
\begin{itemize}
 \item {\bf The octet mesons}
 \begin{eqnarray}
|\rho^+,1,1\rangle&=& -|u\bar{d}\rangle \nonumber\\
|\rho^0,1,0\rangle&=&\frac{1}{\sqrt{2}}[|u\bar{u}\rangle-|d\bar{d}\rangle]\nonumber\\ \nonumber
|\rho^-,1,-1\rangle&=&|d\bar{u}\rangle \\\nonumber
|K^{*+},1/2,1/2\rangle &=& -|u\bar{s}\rangle\\
|K^{*-},1/2,-1/2\rangle &=& |s\bar{u}\rangle\\\nonumber
|K^{*0},1/2,-1/2\rangle&=&- |d\bar{s}\rangle\\\nonumber
|\bar{K}^{*0},1/2,1/2\rangle&=& -|s\bar{d}\rangle\\\nonumber
|\omega,0,0 \rangle&=& \frac{1}{\sqrt{2}} [ |u\bar{u}\rangle +|d\bar{d}\rangle ]
\end{eqnarray} 
\item {\bf The singlet meson}
 \begin{eqnarray}
  |\phi,0,0 \rangle&=&  |s\bar{s}\rangle 
 \end{eqnarray}
\end{itemize}

\subsection{Charmed mesons}

In the case of charmed-$D$ mesons, the flavor wave functions are the same for the pseudoscalar  and  vector states. We use the following flavor wave functions:
\ba
\begin{array}{lllcc}
|D^{+ *}_s,0,0\rangle&=&|D^{+ }_s,0,0\rangle&=&|c\bar{s}\rangle\\
 |D^{+ *},1/2,1/2\rangle&=& |D^{+ },1/2,1/2\rangle&=& |c\bar{d}\rangle\\
|D^{0 *},1/2,-1/2\rangle&=& |D^{0 },1/2,-1/2\rangle&=& |c\bar{u}\rangle
\end{array}
\ea

\section{Flavor coupling}
\label{flavor}
In the following subsections, we give the flavor coefficients $\mathcal{F}_{A\rightarrow BC}$ used to calculate the transition amplitudes. % In this appendix we denote the decay as $A\rightarrow BC$ as opposed in Sec.~\ref{secIIC}, that was used $X\rightarrow YZ$ to avoid confusion with the parameters of Eq.~\ref{eq:mass}. 
We compute 
$\mathcal{F}_{A\rightarrow BC}=\langle\phi_B \phi_C|\phi_0 \phi_A  \rangle$ where $\phi_{(A,B,C)}$ refers to the initial flavor wave function of a charmed baryon $\phi_{A_c}$, final baryon $\phi_B$, and  final meson $\phi_C$, respectively;  $\phi^{45}_{0}=(u\bar u +d\bar d +s \bar s)/\sqrt3$ is the flavor singlet-wave function of $SU(3)$. In addition, we compute the flavor decay coefficients of the isospin channels, since we assume that the isospin symmetry holds even though it is slightly broken. The corresponding charge channels are obtained by multiplying our $\mathcal{F}_{A\rightarrow BC}$ by the corresponding Clebsch-Gordan coefficient in the isospin space, using the convention of the isospin quantum numbers of the baryon and meson  flavor wave functions found in \ref{flavorcb}, \ref{flavorlb}, and \ref{appme}. Thus, the flavor charge channel  for a specific projection  $(I,M_I)$ in the isospin space is obtained   as follows: 

\ba
\mathcal{F}_{A (I_A, M_{I_A})\rightarrow B(I_B, M_{I_B})C(I_C, M_{I_C})} =\nonumber\\ \langle\phi_B,I_B, M_{I_B}, \phi_C,I_C, M_{I_C}|\phi_0,0,0,  \phi_A ,I_A, M_{I_A}\rangle_F \nonumber \\= \langle I_B, M_{I_B},I_C, M_{I_C}|I_A, M_{I_A}\rangle\mathcal{F}_{A\rightarrow BC},
\ea
where $\langle I_B, M_{I_B},I_C, M_{I_C}|I_A, M_{I_A}\rangle$ is a Clebsch-Gordan coefficient and the flavor functions $\phi_i$ of each baryon and meson have a specific isospin projection $M_{I}$.

\subsection{Charmed baryons and pseudoscalar mesons}
We give the squared flavor-coupling coefficients, $\mathcal{F}^2_{A\rightarrow BC}$, when the final states have a pseudoscalar light  meson. Here, $A$ and $B$ are charmed baryons, and  the subindexes ${\bf 6}_{\rm f}$ and  $ \mathbf{\bar 3}_{\rm f} $ refer to the sextet and the anti-triplet baryon multiplets. The $C$ is a pseudoscalar meson and the subindexes $\bf 8 $ and $\bf 1$ refer to the octet and singlet meson multiplets, respectively. 
\begin{itemize}
    \item $ A_{\mathbf{6}_{\rm f}} \rightarrow  B_{\mathbf{6}_{\rm f}}+C_{\bf 8}$ 

\begin{eqnarray}
\left( \begin{array}{c} \Omega_c \\ \\ \Sigma_c \\ \\\Xi'_c \end{array} \right) &\rightarrow &
\left( \begin{array}{cccc} &\Xi'_c K  & \Omega_c \eta\\ \\
\Xi'_c K & \Sigma_c \pi & \Sigma_c \eta \\ \\
\Sigma_c K & \Xi'_c \pi & \Xi'_c \eta  %\\ \Sigma^* \bar{K} & \Xi^* \pi & \Xi^* \eta_8 & \Omega K 
\end{array} \right) 
\nonumber\\
&=& \left( \begin{array}{cccc} & \frac{1}{3} & \frac{2}{9}\\  \\
\frac{1 }{6} & \frac{1}{3} & \frac{1}{18} \\ \\
\frac{1 }{4} & \frac{1}{8} & \frac{1}{72} \\  %\\-\frac{1}{3} & \frac{1}{3} & \frac{1+2\frac{m_n}{m_s} }{9} & -\frac{\sqrt 2}{3}\frac{m_n}{m_s}
\end{array} \right) 
\end{eqnarray}

\item $ A_{\mathbf{6}_{\rm f}} \rightarrow  B_{\mathbf{6}_{\rm f}}+C_{\bf 1}$ 

\begin{eqnarray}
\left( \begin{array}{c} \Omega_c \\ \\ \Sigma_c \\ \\\Xi'_c \end{array} \right) \rightarrow 
\left( \begin{array}{c} \Omega_c \eta' \\ \\ \Sigma_c \eta' \\ \\ \Xi'_c \eta' \\  \end{array} \right) = 
 \left( \begin{array}{c} \frac{1}{3} \\ \\ \frac{1}{9} \\  \\ \frac{1}{9}  \end{array} \right) 
\label{eqn:10101_1}
\end{eqnarray}

  \item $ A_{\mathbf{6}_{\rm f}} \rightarrow  B_{ \mathbf{\bar3}_{\rm f}}+C_{\bf 8}$ 
  
\begin{eqnarray}
\left( \begin{array}{c} \Omega_c \\ \\ \Sigma_c \\ \\\Xi'_c \end{array} \right) &\rightarrow &
\left( \begin{array}{cccc} &\Xi_c K  & \\ \\
\Xi_c K & \Lambda_c \pi &  \\ \\
\Lambda_c K & \Xi_c \pi & \Xi_c \eta  %\\ \Sigma^* \bar{K} & \Xi^* \pi & \Xi^* \eta_8 & \Omega K 
\end{array} \right) 
\nonumber\\
&=& \left( \begin{array}{cccc} & \frac{1}{3} & \\  \\
\frac{1 }{6} & \frac{1}{2} &  \\ \\
\frac{1 }{12} & \frac{1}{8} & \frac{1}{72} \\  %\\-\frac{1}{3} & \frac{1}{3} & \frac{1+2\frac{m_n}{m_s} }{9} & -\frac{\sqrt 2}{3}\frac{m_n}{m_s}
\end{array} \right) 
\end{eqnarray}

\item $ A_{\mathbf{6}_{\rm f}} \rightarrow  B_{\mathbf{\bar3}_{\rm f}}+C_{\bf 1}$ 

\begin{eqnarray}
\left(   \Xi'_c  \right) \rightarrow 
\left(    \Xi_c \eta'  \right) = 
 \left(  \frac{1}{9}  \right) 
\label{eqn:10101_2}
\end{eqnarray}

\item $ A_{\mathbf{\bar 3}_{\rm f}} \rightarrow  B_{\mathbf{6}_{\rm f}}+C_{\bf 8}$ 

\begin{eqnarray}
\left( \begin{array}{c}  \Lambda_c \\ \\\Xi_c \end{array} \right) &\rightarrow &
\left( \begin{array}{cccc} 
\Xi'_c K & \Sigma_c \pi & \\ \\
\Sigma_c K & \Xi'_c \pi & \Xi'_c \eta  %\\ \Sigma^* \bar{K} & \Xi^* \pi & \Xi^* \eta_8 & \Omega K 
\end{array} \right) 
\nonumber\\
&=& \left( \begin{array}{cccc} 
\frac{1 }{6} & \frac{1}{2} &  \\ \\
\frac{1 }{4} & \frac{1}{8} & \frac{1}{72} \\  %\\-\frac{1}{3} & \frac{1}{3} & \frac{1+2\frac{m_n}{m_s} }{9} & -\frac{\sqrt 2}{3}\frac{m_n}{m_s}
\end{array} \right) 
\end{eqnarray}

\item $ A_{\mathbf{\bar 3}_{\rm f}} \rightarrow  B_{\mathbf{6}_{\rm f}}+C_{\bf 1}$ 

\begin{eqnarray}
\left( \begin{array}{c}   \Xi_c \end{array} \right) &\rightarrow &
\left( \begin{array}{c} 
 \Xi'_c \eta'  %\\ \Sigma^* \bar{K} & \Xi^* \pi & \Xi^* \eta_8 & \Omega K 
\end{array} \right) 
=\Big( 
 \frac{1}{9} \Big)
\end{eqnarray}

\item $ A_{\mathbf{\bar 3}_{\rm f}} \rightarrow  B_{\mathbf{\bar 3}_{\rm f}}+C_{\bf 8}$ 

\begin{eqnarray}
\left( \begin{array}{c}  \Lambda_c \\ \\\Xi_c \end{array} \right) &\rightarrow &
\left( \begin{array}{cccc} 
\Xi_c K & \Lambda_c \eta & \\ \\
\Lambda_c K & \Xi_c \pi & \Xi_c \eta  %\\ \Sigma^* \bar{K} & \Xi^* \pi & \Xi^* \eta_8 & \Omega K 
\end{array} \right) 
\nonumber\\
&=& \left( \begin{array}{cccc} 
\frac{1 }{6} & \frac{1}{18} &  \\ \\
\frac{1 }{12} & \frac{1}{8} & \frac{1}{72} \\  %\\-\frac{1}{3} & \frac{1}{3} & \frac{1+2\frac{m_n}{m_s} }{9} & -\frac{\sqrt 2}{3}\frac{m_n}{m_s}
\end{array} \right) 
\end{eqnarray}

\item $ A_{\mathbf{\bar 3}_{\rm f}} \rightarrow  B_{\mathbf{\bar 3}_{\rm f}}+C_{\bf 1}$ 

\begin{eqnarray}
\left( \begin{array}{c}  \Lambda_c \\ \\\Xi_c \end{array} \right) &\rightarrow &
\left( \begin{array}{c} 
 \Lambda_c \eta' \\ \\
 \Xi_c \eta'  %\\ \Sigma^* \bar{K} & \Xi^* \pi & \Xi^* \eta_8 & \Omega K 
\end{array} \right) 
=\left( \begin{array}{c} 
\frac{1 }{9}  \\ \\
\frac{1 }{9}   %\\-\frac{1}{3} & \frac{1}{3} & \frac{1+2\frac{m_n}{m_s} }{9} & -\frac{\sqrt 2}{3}\frac{m_n}{m_s}
\end{array} \right) 
\end{eqnarray}

\end{itemize}
%%%%%%%%%%%%%%%%%%%%%%%%%%%%%%%%%%%%%%%%%%%%%%%%%%%%%%%%%%%%%%%%%%%%%%%%%%%%%%%%%%%%
\subsection{Charmed baryons and vector mesons}
We give the squared flavor-coupling coefficients, $\mathcal{F}^2_{A\rightarrow BC}$, when the final states have a vector-light  meson. Here $A$ and $B$ are charmed baryons, and  the subindexes ${\bf 6}_{\rm f}$ and  $ \mathbf{\bar 3}_{\rm f} $ refer to the sextet and the anti-triplet baryon multiplets. The $C$ is a vector meson  and the subindexes $\bf 8 $ and $\bf 1$ refer to the octet and singlet meson multiplets respectively. 
\begin{itemize}
    \item $ A_{\mathbf{6}_{\rm f}} \rightarrow  B_{\mathbf{6}_{\rm f}}+C_{\bf 8}$ 

\begin{eqnarray}
\left( \begin{array}{c} \Omega_c \\ \\ \Sigma_c \\ \\\Xi'_c \end{array} \right) &\rightarrow &
\left( \begin{array}{cccc} &\Xi'_c K^*  & \\ \\
\Xi'_c K^* & \Sigma_c \rho & \Sigma_c \omega \\ \\
\Sigma_c K^* & \Xi'_c \rho & \Xi'_c \omega %\\ \Sigma^* \bar{K} & \Xi^* \pi & \Xi^* \eta_8 & \Omega K 
\end{array} \right) 
\nonumber\\
&=& \left( \begin{array}{cccc} & \frac{1}{3} & \\  \\
\frac{1 }{6} & \frac{1}{3} & \frac{1}{6} \\ \\
\frac{1 }{4} & \frac{1}{8} & \frac{1}{24} \\  %\\-\frac{1}{3} & \frac{1}{3} & \frac{1+2\frac{m_n}{m_s} }{9} & -\frac{\sqrt 2}{3}\frac{m_n}{m_s}
\end{array} \right) 
\end{eqnarray}

\item $ A_{\mathbf{6}_{\rm f}} \rightarrow  B_{\mathbf{6}_{\rm f}}+C_{\bf 1}$ 

\begin{eqnarray}
\left( \begin{array}{c} \Omega_c \\ \\ \Sigma_c \\ \\\Xi'_c \end{array} \right) \rightarrow 
\left( \begin{array}{c} \Omega_c \phi \\ \\ \Sigma_c \phi \\ \\ \Xi'_c \phi \\  \end{array} \right) = 
 \left( \begin{array}{c} \frac{1}{9} \\ \\ 0\\  \\ \frac{1}{2}  \end{array} \right) 
\label{eqn:10101_3}
\end{eqnarray}

  \item $ A_{\mathbf{6}_{\rm f}} \rightarrow  B_{ \mathbf{\bar3}_{\rm f}}+C_{\bf 8}$ 
  
\begin{eqnarray}
\left( \begin{array}{c} \Omega_c \\ \\ \Sigma_c \\ \\\Xi'_c \end{array} \right) &\rightarrow &
\left( \begin{array}{cccc} &\Xi_c K ^* & \\ \\
\Xi_c K^* & \Lambda_c \rho &  \\ \\
\Lambda_c K^* & \Xi_c \rho & \Xi_c \omega %\\ \Sigma^* \bar{K} & \Xi^* \pi & \Xi^* \eta_8 & \Omega K 
\end{array} \right) 
\nonumber\\
&=& \left( \begin{array}{cccc} & \frac{1}{3} & \\  \\
\frac{1 }{6} & \frac{1}{3} &  \\ \\
\frac{1 }{12} & \frac{1}{8} & \frac{1}{24} \\  %\\-\frac{1}{3} & \frac{1}{3} & \frac{1+2\frac{m_n}{m_s} }{9} & -\frac{\sqrt 2}{3}\frac{m_n}{m_s}
\end{array} \right) 
\end{eqnarray}

\item $ A_{\mathbf{6}_{\rm f}} \rightarrow  B_{\mathbf{\bar3}_{\rm f}}+C_{\bf 1}$ 

\begin{eqnarray}
\left(   \Xi'_c  \right) \rightarrow 
\left(    \Xi_c \phi  \right) = 
 \left(  \frac{1}{12}  \right) 
\label{eqn:10101_4}
\end{eqnarray}

\item $ A_{\mathbf{\bar 3}_{\rm f}} \rightarrow  B_{\mathbf{6}_{\rm f}}+C_{\bf 8}$ 

\begin{eqnarray}
\left( \begin{array}{c}  \Lambda_c \\ \\\Xi_c \end{array} \right) &\rightarrow &
\left( \begin{array}{cccc} 
\Xi'_c K^* & \Sigma_c \rho & \\ \\
\Sigma_c K^* & \Xi'_c \rho & \Xi'_c \omega  %\\ \Sigma^* \bar{K} & \Xi^* \pi & \Xi^* \eta_8 & \Omega K 
\end{array} \right) 
\nonumber\\
&=& \left( \begin{array}{cccc} 
\frac{1 }{6} & \frac{1}{2} &  \\ \\
\frac{1 }{4} & \frac{1}{8} & \frac{1}{24} \\  %\\-\frac{1}{3} & \frac{1}{3} & \frac{1+2\frac{m_n}{m_s} }{9} & -\frac{\sqrt 2}{3}\frac{m_n}{m_s}
\end{array} \right) 
\end{eqnarray}

\item $ A_{\mathbf{\bar 3}_{\rm f}} \rightarrow  B_{\mathbf{6}_{\rm f}}+C_{\bf 1}$ 

\begin{eqnarray}
\left( \begin{array}{c}   \Xi_c \end{array} \right) &\rightarrow &
\left( \begin{array}{c} 
 \Xi'_c \phi  %\\ \Sigma^* \bar{K} & \Xi^* \pi & \Xi^* \eta_8 & \Omega K 
\end{array} \right) 
=\Big( 
 \frac{1}{12} \Big)
\end{eqnarray}

\item $ A_{\mathbf{\bar 3}_{\rm f}} \rightarrow  B_{\mathbf{\bar 3}_{\rm f}}+C_{\bf 8}$ 

\begin{eqnarray}
\left( \begin{array}{c}  \Lambda_c \\ \\\Xi_c \end{array} \right) &\rightarrow &
\left( \begin{array}{cccc} 
\Xi_c K^* & \Lambda_c \omega & \\ \\
\Lambda_c K^* & \Xi_c \rho & \Xi_c \omega %\\ \Sigma^* \bar{K} & \Xi^* \pi & \Xi^* \eta_8 & \Omega K 
\end{array} \right) 
\nonumber\\
&=& \left( \begin{array}{cccc} 
\frac{1 }{6} & \frac{1}{6} &  \\ \\
\frac{1 }{12} & \frac{1}{8} & \frac{1}{24} \\  %\\-\frac{1}{3} & \frac{1}{3} & \frac{1+2\frac{m_n}{m_s} }{9} & -\frac{\sqrt 2}{3}\frac{m_n}{m_s}
\end{array} \right) 
\end{eqnarray}

\item $ A_{\mathbf{\bar 3}_{\rm f}} \rightarrow  B_{\mathbf{\bar 3}_{\rm f}}+C_{\bf 1}$ 

\begin{eqnarray}
\left( \begin{array}{c}  \Lambda_c \\ \\\Xi_c \end{array} \right) &\rightarrow &
\left( \begin{array}{c} 
 \Lambda_c \phi \\ \\
 \Xi_c \phi  %\\ \Sigma^* \bar{K} & \Xi^* \pi & \Xi^* \eta_8 & \Omega K 
\end{array} \right) 
=\left( \begin{array}{c} 
0  \\ \\
\frac{1 }{12}   %\\-\frac{1}{3} & \frac{1}{3} & \frac{1+2\frac{m_n}{m_s} }{9} & -\frac{\sqrt 2}{3}\frac{m_n}{m_s}
\end{array} \right) 
\end{eqnarray}

\end{itemize}

\subsection{Light baryons and charm-(pseudoscalar/vector) mesons}
We give the $\mathcal{F}^2_{A\rightarrow BC}$ when the final states have a light baryon and a charm-(pseudoscalar/vector) meson. Since the mesons $D^0$ and $D^+$ form an isospin doublet, both are treated as $D$ in the tables; whereas $D_s$ is separated by the strangeness content. The subindexes $\mathbf{6}_{\rm f}$ and  $\mathbf{\bar 3}_{\rm f} $ refer to the sextet and the anti-triplet baryon multiplets for the initial charmed baryon $A$, whereas the final $B$ baryons can have subindexes $\bf 8$ or $\bf 10$, according to whether the final light baryon belongs to the octet or decuplet baryon multiplets. 
Additionally, owing to the symmetry of the wave functions of the octet-light baryons, see \ref{flavorlb}, we can have only $\rho$ or $\lambda$ contributions in the final states, as indicated by a superindex.

\begin{itemize}
    \item $A_{\mathbf{6}_{\rm f}} \rightarrow  B_{\bf 10}+C$ 
    
    \begin{eqnarray}
\left( \begin{array}{c} \Omega_c \\ \\ \Sigma_c \\ \\ \Xi'_c \end{array} \right) &\rightarrow &
\left( \begin{array}{cccc} \Xi^*_{10}D&\Omega_{10} D_s  \\ \\
\Delta D & \Sigma^*_{10}D_s  \\ \\
\Sigma^*_{10} D & \Xi^*_{10} D_s %\\ \Sigma^* \bar{K} & \Xi^* \pi & \Xi^* \eta_8 & \Omega K 
\end{array} \right) 
= \left( \begin{array}{cccc} \frac{2}{9}& \frac{1}{3}  \\  \\
\frac{4 }{9} & \frac{1}{9} \\ \\
\frac{1 }{3} & \frac{2}{9}  \\  %\\-\frac{1}{3} & \frac{1}{3} & \frac{1+2\frac{m_n}{m_s} }{9} & -\frac{\sqrt 2}{3}\frac{m_n}{m_s}
\end{array} \right) 
\end{eqnarray}
    
       \item $A_{\mathbf{6}_{\rm f}} \rightarrow  B_{\bf 8}+C$ 
    
    \begin{eqnarray}
\left( \begin{array}{c} \Omega_c \\ \\ \Sigma_c \\ \\ \Xi'_c \end{array} \right) &\rightarrow &
\left( \begin{array}{cccc} &\Xi^\lambda_8 D \\ \\
N^\lambda D & \Sigma^\lambda_8D_s  \\ \\
\Sigma^\lambda_8 D & \Xi^\lambda_8 D_s %\\ \Sigma^* \bar{K} & \Xi^* \pi & \Xi^* \eta_8 & \Omega K 
\end{array} \right) 
= \left( \begin{array}{cccc} & \frac{4}{9} & \\  \\
\frac{2 }{9} & \frac{2}{9} & \\ \\
\frac{1 }{6} & \frac{1}{9} & \\  %\\-\frac{1}{3} & \frac{1}{3} & \frac{1+2\frac{m_n}{m_s} }{9} & -\frac{\sqrt 2}{3}\frac{m_n}{m_s}
\end{array} \right) 
\end{eqnarray}
    
     \item $A_{\mathbf{\bar 3}_{\rm f}} \rightarrow  B_{\bf 8}+C$ 
    
    \begin{eqnarray}\left( \begin{array}{c}
 \Lambda_c \\ \\ \Xi_c \end{array} \right) &\rightarrow &
\left( \begin{array}{cccc} 
N^\rho D & \Lambda^\rho_8D_s &  \\ \\
\Sigma^\rho_8 D & \Xi^\rho_8 D_s & \Lambda^\rho_8 D%\\ \Sigma^* \bar{K} & \Xi^* \pi & \Xi^* \eta_8 & \Omega K 
\end{array} \right) 
\nonumber\\
&=& \left( \begin{array}{cccc} 
\frac{2 }{3} & \frac{2}{9} & \\ \\
\frac{1 }{2} & \frac{1}{3} & \frac{1}{18} \\  %\\-\frac{1}{3} & \frac{1}{3} & \frac{1+2\frac{m_n}{m_s} }{9} & -\frac{\sqrt 2}{3}\frac{m_n}{m_s}
\end{array} \right) 
\end{eqnarray}
\end{itemize}

\section{Partial decay widths}
\label{partialdw}
The partial decay widths, $\Gamma_{A \rightarrow BC} $, of an initial baryon $A$ decaying to a final baryon $B$ plus a meson $C$, in all the open-flavor channels, are shown in Tables~\ref{tab:part_dec_Omega}-\ref{tab:part_dec_lambdas}. Here, we give the contribution of the isospin channels. The charge channel width for the $A$ baryon  with isospin projection $|A,I,M_{I_A}\rangle$ can be obtain as follows 
\ba
\Gamma_{A (I_A, M_{I_A})\rightarrow B(I_B, M_{I_B})C(I_C, M_{I_C})} =\nonumber\\ \langle I_B, M_{I_B},I_C, M_{I_C}|I_A, M_{I_A}\rangle^2\Gamma_{A\rightarrow BC},
\ea
where  $\langle I_B, M_{I_B},I_C, M_{I_C}|I_A, M_{I_A}\rangle$ is a Clebsch-Gordan coefficient, and the partial decay width  $\Gamma_{A\rightarrow BC}$ can be extracted from Tables~\ref{tab:part_dec_Omega}-\ref{tab:part_dec_lambdas}. 

%% Tables begin
%%%%%%%%%%%%%%%%%%%%%%%%%%%%%%%%%%%%%%%%%%%%%%%%%%
%%%%%%%%%%%%%%%%%%%%%%%%%%%%%%%%%%%%%%%%%%%%%%%%%%\
%hook%
\begin{turnpage}
\begin{table*}[htbp]
{\scriptsize

\begin{tabular}{c |  p{0.58cm}  p{0.58cm}  p{0.58cm}  p{0.58cm}  p{0.58cm}  p{0.58cm}  p{0.58cm}  p{0.58cm}  p{0.58cm}  p{0.58cm}  p{0.58cm}  p{0.58cm}  p{0.58cm}  p{0.58cm}p{0.75cm}} \hline \hline
$\Omega_c(ssc)$  & $\Xi_{c} K$  & $\Xi'_{c} K$  & $\Xi^{*}_{c} K$  & $\Xi_{c} K^{*}$  & $\Xi'_{c}K^{*}$  & $\Xi^{*}_{c} K^{*}$  & $\Omega_{c} \eta$  & $\Omega^{*}_{c} \eta$  & $\Omega_{c} \phi$  & $\Omega^{*}_{c} \phi$  & $\Omega_{c} \eta'$  & $\Omega^{*}_{c} \eta'$  & $\Xi_{8} D$  & $\Xi_{10} D$  & Tot $\Gamma$  \\ 
$\mathcal{F}={\bf {6}}_{\rm f}$ &&&&&&&&&&&&&&\\ \hline
$\Omega_c(2709)$ $^{2}S_{1/2}$&$0$   &$0$   &$0$   &$0$   &$0$   &$0$   &$0$   &$0$   &$0$   &$0$   &$0$   &$0$   &$0$   &$0$   &$0$  \\
$\Omega_c(2778)$ $^{4}S_{3/2}$&$0$   &$0$   &$0$   &$0$   &$0$   &$0$   &$0$   &$0$   &$0$   &$0$   &$0$   &$0$   &$0$   &$0$   &$0$  \\
$\Omega_c(3008)$ $^{2}P_{1/2}$&4.1   &$0$   &$0$   &$0$   &$0$   &$0$   &$0$   &$0$   &$0$   &$0$   &$0$   &$0$   &$0$   &$0$   &4.1  \\
$\Omega_c(3050)$ $^{4}P_{1/2}$&7.5   &0.1   &$0$   &$0$   &$0$   &$0$   &$0$   &$0$   &$0$   &$0$   &$0$   &$0$   &$0$   &$0$   &7.6  \\
$\Omega_c(3035)$ $^{2}P_{3/2}$&26.3   &$0$   &$0$   &$0$   &$0$   &$0$   &$0$   &$0$   &$0$   &$0$   &$0$   &$0$   &$0$   &$0$   &26.3  \\
$\Omega_c(3077)$ $^{4}P_{3/2}$&6.3   &0.4   &$0$   &$0$   &$0$   &$0$   &$0$   &$0$   &$0$   &$0$   &$0$   &$0$   &$0$   &$0$   &6.7  \\
$\Omega_c(3122)$ $^{4}P_{5/2}$&40.9   &8.9   &0.3   &$0$   &$0$   &$0$   &$0$   &$0$   &$0$   &$0$   &$0$   &$0$   &$0$   &$0$   &50.1  \\
$\Omega_c(3129)$ $^{2}P_{1/2}$&-   &8.9   &5.5   &$0$   &$0$   &$0$   &$0$   &$0$   &$0$   &$0$   &$0$   &$0$   &$0$   &$0$   &14.4  \\
$\Omega_c(3156)$ $^{2}P_{3/2}$&-   &61.1   &10.5   &$0$   &$0$   &$0$   &$0$   &$0$   &$0$   &$0$   &$0$   &$0$   &$0$   &$0$   &71.6  \\
$\Omega_c(3315)$ $^{2}D_{3/2}$&1.9   &1.8   &2.3   &$0$   &$0$   &$0$   &0.3   &-   &$0$   &$0$   &$0$   &$0$   &4.3   &$0$   &10.6  \\
$\Omega_c(3360)$ $^{2}D_{5/2}$&5.4   &5.1   &0.5   &$0$   &$0$   &$0$   &1.2   &-   &$0$   &$0$   &$0$   &$0$   &12.2   &$0$   &24.4  \\
$\Omega_c(3330)$ $^{4}D_{1/2}$&0.2   &0.2   &3.3   &$0$   &$0$   &$0$   &0.1   &0.1   &$0$   &$0$   &$0$   &$0$   &12.3   &$0$   &16.2  \\
$\Omega_c(3357)$ $^{4}D_{3/2}$&2.0   &0.5   &5.2   &0.2   &$0$   &$0$   &0.2   &0.6   &$0$   &$0$   &$0$   &$0$   &21.7   &$0$   &30.4  \\
$\Omega_c(3402)$ $^{4}D_{5/2}$&5.0   &1.2   &5.0   &1.6   &$0$   &$0$   &0.3   &1.2   &$0$   &$0$   &$0$   &$0$   &46.9   &1.1   &62.3  \\
$\Omega_c(3466)$ $^{4}D_{7/2}$&7.8   &2.0   &5.0   &2.6   &$0$   &$0$   &0.8   &0.9   &$0$   &$0$   &$0$   &$0$   &83.2   &20.9   &123.2  \\
$\Omega_c(3342)$ $^{2}S_{1/2}$&0.2   &0.3   &0.1   &$0$   &$0$   &$0$   &0.1   &-   &$0$   &$0$   &$0$   &$0$   &0.5   &$0$   &1.2  \\
$\Omega_c(3411)$ $^{4}S_{3/2}$&0.2   &0.1   &0.4   &0.2   &$0$   &$0$   &-   &0.1   &$0$   &$0$   &$0$   &$0$   &2.1   &0.2   &3.3  \\
$\Omega_c(3585)$ $^{2}S_{1/2}$&0.3   &1.0   &0.7   &3.0   &11.6   &0.1   &1.1   &0.5   &$0$   &$0$   &$0$   &$0$   &-   &-   &18.3  \\
$\Omega_c(3654)$ $^{4}S_{3/2}$&0.1   &0.1   &1.2   &2.8   &1.0   &17.2   &0.2   &1.4   &$0$   &$0$   &-   &$0$   &-   &-   &24.0  \\
$\Omega_c(3437)$ $^{2}D_{3/2}$&-   &6.5   &107.0   &53.5   &$0$   &$0$   &4.0   &27.0   &$0$   &$0$   &$0$   &$0$   &-   &-   &198.0  \\
$\Omega_c(3482)$ $^{2}D_{5/2}$&-   &56.4   &16.8   &17.2   &0.4   &$0$   &20.9   &3.3   &$0$   &$0$   &$0$   &$0$   &-   &-   &115.0  \\
$\Omega_c(3446)$ $^{2}P_{1/2}$&-   &-   &1.4   &0.4   &$0$   &$0$   &-   &0.3   &$0$   &$0$   &$0$   &$0$   &-   &-   &2.1  \\
$\Omega_c(3473)$ $^{2}P_{3/2}$&-   &1.1   &0.8   &0.7   &$0$   &$0$   &0.3   &0.2   &$0$   &$0$   &$0$   &$0$   &-   &-   &3.1  \\
$\Omega_c(3464)$ $^{2}S_{1/2}$&-   &7.6   &22.3   &36.1   &0.2   &$0$   &8.6   &13.5   &$0$   &$0$   &$0$   &$0$   &-   &-   &88.3  \\
$\Omega_c(3558)$ $^{2}D_{3/2}$&18.4   &16.8   &18.8   &28.7   &115.1   &-   &8.0   &11.2   &$0$   &$0$   &$0$   &$0$   &-   &-   &217.0  \\
$\Omega_c(3603)$ $^{2}D_{5/2}$&48.3   &49.7   &14.5   &3.6   &30.8   &0.7   &23.3   &3.4   &$0$   &$0$   &$0$   &$0$   &-   &-   &174.3  \\
$\Omega_c(3573)$ $^{4}D_{1/2}$&9.4   &0.4   &33.1   &38.8   &7.0   &111.6   &0.3   &17.0   &$0$   &$0$   &$0$   &$0$   &-   &-   &217.6  \\
$\Omega_c(3600)$ $^{4}D_{3/2}$&22.1   &4.9   &28.2   &77.3   &16.8   &111.4   &2.2   &21.9   &$0$   &$0$   &$0$   &$0$   &-   &-   &284.8  \\
$\Omega_c(3645)$ $^{4}D_{5/2}$&38.6   &10.8   &32.8   &47.8   &13.2   &43.7   &5.4   &19.7   &$0$   &$0$   &$0$   &$0$   &-   &-   &212.0  \\
$\Omega_c(3708)$ $^{4}D_{7/2}$&72.1   &18.0   &88.1   &107.8   &18.0   &38.9   &8.4   &29.2   &0.1   &$0$   &2.3   &0.1   &-   &-   &383.0  \\
\hline \hline
\end{tabular}

}
\caption{$\Omega_c(ssc)$ state partial decay widths (in MeV). The order of the states is the same as in Table \ref{tab:All_mass_Omega}. The predicted masses, reported in Table \ref{tab:All_mass_Omega}, are obtained by the three-quark model Hamiltonian of Eqs. \ref{MassFormula} and \ref{eq:Hho}. 
The partial decay widths  are computed by means of Eq. \ref{gamma}. For each state is also reported the spectroscopic notation  $^{2S+1}L_J$, where ${\bf J} ={\bf L}_{\rm tot} + {\bf S}_{\rm tot} $ is the total angular momentum,  ${\bf S}_{\rm tot} = {\bf S}_{\rho}+\frac{1}{2}$, and ${\bf L}_{\rm tot}= {\bf l}_{\rho}+{\bf l}_{\lambda}$.
The partial decay widths  denoted by 0 and $-$ are forbidden by phase space and selection rules, respectively.}
\label{tab:part_dec_Omega}
\end{table*}
\end{turnpage}

\begin{turnpage}
\begin{table*}[htbp]
{\scriptsize
%\hspace{0cm}

\begin{tabular}{c |  p{0.58cm}  p{0.58cm}  p{0.58cm}  p{0.58cm}  p{0.58cm}  p{0.58cm}  p{0.58cm}  p{0.58cm}  p{0.58cm}  p{0.58cm}  p{0.58cm}  p{0.58cm}  p{0.58cm}  p{0.58cm}  p{0.58cm}  p{0.58cm}  p{0.58cm}  p{0.58cm}  p{0.58cm}  p{0.58cm}  p{0.58cm}  p{0.58cm}  p{0.58cm}  p{0.58cm}  p{0.58cm}  p{0.58cm}p{0.75cm}} \hline \hline
$\Sigma_c(nnc)$  & $\Sigma_{c} \pi$  & $\Sigma^{*}_{c} \pi$  & $\Lambda_{c} \pi$  & $\Sigma_{c} \eta$  & $\Xi_{c} K$  & $\Sigma_{c}\rho$  & $\Sigma^{*}_{c}\rho$  & $\Lambda_{c}\rho$  & $\Sigma^{*}_{c}\eta$  & $\Sigma_{c}\eta'$  & $\Sigma^{*}_{c}\eta'$  & $\Xi'_{c}K$  & $\Xi^{*}_{c}K$  & $\Xi_c K^{*}$  & $\Xi'_c K^{*}$  & $\Xi^{*}_{c} K^{*}$  & $\Sigma_c\omega$  & $\Sigma^{*}_{c}\omega$  & $N D$  & $\Sigma_{8} D_{s}$  & $N D^{*}$  & $\Delta D$  & $N^{*}_{1} D$  & $N^{*}_{2} D$  & $N^{*}_{3} D$  & $N^{*}_{4} D$  & Tot $\Gamma$  \\ 
$\mathcal{F}={\bf {6}}_{\rm f}$ &&&&&&&&&&&&&&&&&&&&&&&&&&\\ \hline
$\Sigma_c(2456)$ $^{2}S_{1/2}$&$0$   &$0$   &1.7   &$0$   &$0$   &$0$   &$0$   &$0$   &$0$   &$0$   &$0$   &$0$   &$0$   &$0$   &$0$   &$0$   &$0$   &$0$   &$0$   &$0$   &$0$   &$0$   &$0$   &$0$   &$0$   &$0$   &1.7  \\
$\Sigma_c(2525)$ $^{4}S_{3/2}$&$0$   &$0$   &14.9   &$0$   &$0$   &$0$   &$0$   &$0$   &$0$   &$0$   &$0$   &$0$   &$0$   &$0$   &$0$   &$0$   &$0$   &$0$   &$0$   &$0$   &$0$   &$0$   &$0$   &$0$   &$0$   &$0$   &14.9  \\
$\Sigma_c(2811)$ $^{2}P_{1/2}$&2.2   &17.4   &0.1   &$0$   &$0$   &$0$   &$0$   &$0$   &$0$   &$0$   &$0$   &$0$   &$0$   &$0$   &$0$   &$0$   &$0$   &$0$   &0.8   &$0$   & $0$   &$0$   &$0$   &$0$   &$0$   &$0$   &20.5  \\
$\Sigma_c(2853)$ $^{4}P_{1/2}$&0.7   &9.9   &1.7   &$0$   &$0$   &$0$   &$0$   &$0$   &$0$   &$0$   &$0$   &$0$   &$0$   &$0$   &$0$   &$0$   &$0$   &$0$   &13.3   &$0$   & $0$   &$0$   &$0$   &$0$   &$0$   &$0$   & 25.6  \\
$\Sigma_c(2838)$ $^{2}P_{3/2}$&28.7   &2.8   &45.2   &$0$   &$0$   &$0$   &$0$   &$0$   &$0$   &$0$   &$0$   &$0$   &$0$   &$0$   &$0$   &$0$   &$0$   &$0$   &9.0   &$0$   & $0$   &$0$   &$0$   &$0$   &$0$   &$0$   & 85.7  \\
$\Sigma_c(2880)$ $^{4}P_{3/2}$&1.6   &39.0   &9.8   &$0$   &$0$   &$0$   &$0$   &$0$   &$0$   &$0$   &$0$   &$0$   &$0$   &$0$   &$0$   &$0$   &$0$   &$0$   &9.8   &$0$   & $0$   &$0$   &$0$   &$0$   &$0$   &$0$   & 60.2  \\
$\Sigma_c(2925)$ $^{4}P_{5/2}$&10.5   &22.8   &64.8   &$0$   &$0$   &$0$   &$0$   &$0$   &$0$   &$0$   &$0$   &$0$   &$0$   &$0$   &$0$   &$0$   &$0$   &$0$   &65.4   &$0$   & $0$   &$0$   &$0$   &$0$   &$0$   &$0$   & 163.5 \\
$\Sigma_c(2994)$ $^{2}P_{1/2}$&0.5   &123.8   &-   &0.3   &-   &$0$   &$0$   &$0$   &$0$   &$0$   &$0$   &$0$   &$0$   &$0$   &$0$   &$0$   &$0$   &$0$   &-   &$0$   &-   &$0$   &$0$   &$0$   &$0$   &$0$   &124.6  \\
$\Sigma_c(3021)$ $^{2}P_{3/2}$&64.0   &55.4   &-   &5.4   &-   &$0$   &$0$   &0.2   &$0$   &$0$   &$0$   &$0$   &$0$   &$0$   &$0$   &$0$   &$0$   &$0$   &-   &$0$   &-   &$0$   &$0$   &$0$   &$0$   &$0$   &125.0  \\
$\Sigma_c(3175)$ $^{2}D_{3/2}$&2.5   &3.3   &4.9   &0.3   &0.7   &$0$   &$0$   &4.3   &0.3   &$0$   &$0$   &0.4   &0.2   &$0$   &$0$   &$0$   &$0$   &$0$   &3.8   &1.4   & 30.0   &77.0   &$0$   &$0$   &$0$   &$0$   & 129.1  \\
$\Sigma_c(3220)$ $^{2}D_{5/2}$&7.6   &1.8   &12.8   &0.8   &1.9   &0.1   &$0$   &0.3   &0.1   &$0$   &$0$   &1.3   &0.1   &$0$   &$0$   &$0$   &$0$   &$0$   &11.3   &4.5   &77.4   &96.1   &$0$   &$0$   &$0$   &$0$   & 216.1  \\
$\Sigma_c(3190)$ $^{4}D_{1/2}$&-   &5.4   &2.4   &-   &0.3   &$0$   &$0$   &5.5   &0.4   &$0$   &$0$   &0.1   &0.4   &$0$   &$0$   &$0$   &$0$   &$0$   &0.4   &5.1   & 42.8   &36.3   &$0$   &$0$   &$0$   &$0$   & 99.1  \\
$\Sigma_c(3217)$ $^{4}D_{3/2}$&0.7   &5.4   &5.8   &0.1   &0.8   &0.1   &$0$   &12.2   &0.8   &$0$   &$0$   &0.2   &0.9   &$0$   &$0$   &$0$   &$0$   &$0$   &17.1   &8.4   & 46.1   &56.9   &$0$   &$0$   &$0$   &$0$   & 155.5  \\
$\Sigma_c(3262)$ $^{4}D_{5/2}$&1.7   &5.4   &10.3   &0.2   &1.8   &0.5   &$0$   &8.6   &0.8   &$0$   &$0$   &0.3   &1.4   &$0$   &$0$   &$0$   &0.2   &$0$   &40.5   &18.3   & 49.0   &87.5   &$0$   &$0$   &$0$   &$0$   & 227.4 \\
$\Sigma_c(3326)$ $^{4}D_{7/2}$&2.7   &12.2   &19.0   &0.3   &3.0   &0.8   &0.3   &13.7   &0.8   &$0$   &$0$   &0.7   &1.0   &$0$   &$0$   &$0$   &0.4   &0.1   &64.1   &35.0   & 77.4   &153.1   &$0$   &$0$   &$0$   &$0$   & 385.1  \\
$\Sigma_c(3202)$ $^{2}S_{1/2}$&0.2   &0.1   &-   &-   &0.1   &$0$   &$0$   &0.4   &-   &$0$   &$0$   &0.1   &-   &$0$   &$0$   &$0$   &$0$   &$0$   &0.1   &0.2   &  1.8   &4.8   &$0$   &$0$   &$0$   &$0$   & 7.8  \\
$\Sigma_c(3271)$ $^{4}S_{3/2}$&-   &0.2   &-   &-   &0.1   &-   &$0$   &0.5   &0.1   &$0$   &$0$   &-   &0.1   &$0$   &$0$   &$0$   &-   &$0$   &0.1   &0.8   & 1.0   & 3.9   &$0$   &$0$   &$0$   &$0$   & 6.7  \\
$\Sigma_c(3567)$ $^{2}S_{1/2}$&0.3   &0.1   &1.7   &-   &-   &6.6   &0.1   &0.1   &-   &0.5   &0.2   &0.2   &0.2   &1.0   &4.0   &-   &3.5   &0.1   &-   &-   &-   &-   &-   &-   &-   &$0$   &18.6  \\
$\Sigma_c(3637)$ $^{4}S_{3/2}$&0.2   &0.4   &2.6   &-   &-   &0.2   &9.2   &0.2   &-   &0.1   &0.6   &-   &0.2   &0.8   &0.3   &6.0   &0.1   &4.8   &-   &-   &-   &-   &-   &-   &-   &-   &25.7  \\
$\Sigma_c(3358)$ $^{2}D_{3/2}$&7.7   &94.8   &-   &0.9   &-   &37.7   &0.4   &166.5   &17.1   &$0$   &$0$   &3.0   &36.8   &1.6   &$0$   &$0$   &18.0   &0.2   &-   &-   &-   &-   &1.7   &0.1   &$0$   &$0$   &386.5  \\
$\Sigma_c(3403)$ $^{2}D_{5/2}$&77.0   &78.6   &-   &9.3   &-   &3.8   &6.1   &84.8   &4.2   &0.3   &$0$   &20.6   &5.5   &2.4   &$0$   &$0$   &1.9   &2.9   &-   &-   &-   &-   &20.6   &15.6   &$0$   &$0$   &333.6  \\
$\Sigma_c(3367)$ $^{2}P_{1/2}$&-   &1.7   &-   &-   &-   &0.2   &-   &1.6   &0.2   &$0$   &$0$   &-   &0.3   &-   &$0$   &$0$   &0.1   &-   &-   &-   &-   &-   &3.2   &0.7   &$0$   &$0$   &8.0  \\
$\Sigma_c(3394)$ $^{2}P_{3/2}$&1.0   &0.9   &-   &0.1   &-   &0.1   &-   &1.7   &0.1   &$0$   &$0$   &0.3   &0.2   &-   &$0$   &$0$   &-   &-   &-   &-   &-   &-   &24.3   &3.5   &$0$   &$0$   &32.2  \\
$\Sigma_c(3385)$ $^{2}S_{1/2}$&8.6   &4.4   &-   &0.4   &-   &23.7   &4.9   &7.5   &1.9   &$0$   &$0$   &3.5   &9.4   &3.7   &$0$   &$0$   &11.9   &2.3   &-   &-   &-   &-   &16.7   &0.9   &$0$   &$0$   &99.8  \\
$\Sigma_c(3540)$ $^{2}D_{3/2}$&43.6   &4.5   &84.1   &4.2   &9.9   &159.7   &0.4   &3.1   &2.0   &3.3   &2.9   &8.4   &7.7   &12.4   &46.8   &$0$   &81.8   &0.2   &-   &-   &-   &-   &-   &-   &0.8   &0.1   &475.9  \\
$\Sigma_c(3585)$ $^{2}D_{5/2}$&80.7   &46.1   &126.5   &10.2   &23.4   &200.6   &1.1   &68.6   &4.5   &9.7   &0.8   &23.6   &7.9   &2.3   &14.2   &0.2   &96.6   &0.6   &-   &-   &-   &-   &-   &-   &2.8   &1.7   &722.1  \\
$\Sigma_c(3555)$ $^{4}D_{1/2}$&14.5   &16.2   &137.8   &0.8   &8.1   &10.3   &564.6   &14.2   &4.4   &0.4   &4.8   &0.6   &14.0   &17.0   &3.0   &43.9   &5.2   &284.8   &-   &-   &-   &-   &-   &-   &-   &5.5   &1150.1  \\
$\Sigma_c(3582)$ $^{4}D_{3/2}$&12.8   &4.4   &94.4   &1.3   &11.9   &19.3   &226.5   &33.4   &2.1   &1.0   &8.7   &2.5   &10.4   &33.5   &7.2   &42.4   &9.8   &115.9   &-   &-   &-   &-   &-   &-   &4.6   &9.5   &651.6  \\
$\Sigma_c(3627)$ $^{4}D_{5/2}$&12.5   &68.8   &64.8   &1.9   &17.3   &13.8   &21.5   &62.7   &6.9   &2.4   &9.3   &4.9   &15.2   &20.8   &5.8   &17.8   &6.9   &10.8   &-   &-   &-   &-   &-   &-   &32.0   &16.0   &412.1  \\
$\Sigma_c(3691)$ $^{4}D_{7/2}$&30.7   &173.4   &186.5   &3.9   &35.8   &56.6   &529.9   &275.3   &21.7   &3.9   &11.2   &8.8   &45.4   &54.9   &9.0   &22.1   &27.8   &255.6   &-   &-   &-   &-   &-   &-   &126.8   &-   &1879.3  \\
\hline \hline
\end{tabular}

}
\caption{Same as \ref{tab:part_dec_Omega}, but for $\Sigma_c(nnc)$ states. The order of the states is the same as in Table \ref{tab:All_mass_Sigma}. The predicted masses, reported in Table \ref{tab:All_mass_Sigma}, are obtained by the three-quark model Hamiltonian of Eqs. \ref{MassFormula} and \ref{eq:Hho}.  $N^*_1$, $N^*_2$, $N^*_3$, and $N^*_4$ represent $N(1520)$, $N(1535)$, $N(1680)$, and $N(1720)$, respectively.}
\label{tab:part_dec_Sigma}
\end{table*}
\end{turnpage}

\begin{turnpage}
%\begin{sideways}
\begin{table*}[htbp]
{\scriptsize
%\hspace{0cm}
%\input{tables/decay_indi_cascades_paper}
\begin{tabular}{c |  p{0.58cm}  p{0.58cm}  p{0.58cm}  p{0.58cm}  p{0.58cm}  p{0.58cm}  p{0.58cm}  p{0.58cm}  p{0.58cm}  p{0.58cm}  p{0.58cm}  p{0.58cm}  p{0.58cm}  p{0.58cm}  p{0.58cm}  p{0.58cm}  p{0.58cm}  p{0.58cm}  p{0.58cm}  p{0.58cm}  p{0.58cm}  p{0.58cm}  p{0.58cm}  p{0.58cm}  p{0.58cm}  p{0.58cm}  p{0.58cm}  p{0.58cm}p{0.75cm}} \hline \hline
$\Xi'_c(snc)$  & $\Lambda_{c} K$  & $\Xi_{c} \pi$  & $\Xi'_{c} \pi$  & $\Xi^{*}_{c} \pi$  & $\Sigma_{c} K$  & $\Sigma^{*}_{c} K$  & $\Xi_{c} \eta$  & $\Lambda_{c} K^{*}$  & $\Xi_{c} \rho$  & $\Xi'_{c} \rho$  & $\Xi^{*}_{c} \rho$  & $\Sigma_{c} K^{*}$  & $\Sigma^{*}_{c} K^{*}$  & $\Xi'_{c} \eta$  & $\Xi^{*}_{c} \eta$  & $\Xi_{c} \eta'$  & $\Xi'_{c} \eta'$  & $\Xi^{*}_{c} \eta'$  & $\Xi_{c} \omega$  & $\Xi'_{c} \omega$  & $\Xi^{*}_{c} \omega$  & $\Xi_{c} \phi$  & $\Xi'_{c} \phi$  & $\Xi^{*}_{c} \phi$  & $\Sigma_{8} D$  & $\Xi_{8} D_{s}$  & $\Sigma_{8} D^{*}$  & $\Sigma_{10} D$  & Tot $\Gamma$  \\ 
$\mathcal{F}={\bf {6}}_{\rm f}$ &&&&&&&&&&&&&&&&&&&&&&&&&&&&\\ \hline
$\Xi'_c(2571)$ $^{2}S_{1/2}$&$0$   &$0$   &$0$   &$0$   &$0$   &$0$   &$0$   &$0$   &$0$   &$0$   &$0$   &$0$   &$0$   &$0$   &$0$   &$0$   &$0$   &$0$   &$0$   &$0$   &$0$   &$0$   &$0$   &$0$   &$0$   &$0$   &$0$   &$0$   &$0$  \\
$\Xi'_c(2640)$ $^{4}S_{3/2}$&$0$   &0.4   &$0$   &$0$   &$0$   &$0$   &$0$   &$0$   &$0$   &$0$   &$0$   &$0$   &$0$   &$0$   &$0$   &$0$   &$0$   &$0$   &$0$   &$0$   &$0$   &$0$   &$0$   &$0$   &$0$   &$0$   &$0$   &$0$   &0.4  \\
$\Xi'_c(2893)$ $^{2}P_{1/2}$&0.7   &0.4   &1.1   &5.1   &$0$   &$0$   &$0$   &$0$   &$0$   &$0$   &$0$   &$0$   &$0$   &$0$   &$0$   &$0$   &$0$   &$0$   &$0$   &$0$   &$0$   &$0$   &$0$   &$0$   &$0$   &$0$   &$0$   &$0$   &7.3  \\
$\Xi'_c(2935)$ $^{4}P_{1/2}$&0.8   &0.4   &0.5   &3.2   &0.2   &$0$   &$0$   &$0$   &$0$   &$0$   &$0$   &$0$   &$0$   &$0$   &$0$   &$0$   &$0$   &$0$   &$0$   &$0$   &$0$   &$0$   &$0$   &$0$   &$0$   &$0$   &$0$   &$0$   &5.1  \\
$\Xi'_c(2920)$ $^{2}P_{3/2}$&8.3   &9.5   &9.4   &0.7   &$0$   &$0$   &$0$   &$0$   &$0$   &$0$   &$0$   &$0$   &$0$   &$0$   &$0$   &$0$   &$0$   &$0$   &$0$   &$0$   &$0$   &$0$   &$0$   &$0$   &$0$   &$0$   &$0$   &$0$   &27.9  \\
$\Xi'_c(2962)$ $^{4}P_{3/2}$&1.8   &2.1   &0.5   &13.9   &0.6   &$0$   &$0$   &$0$   &$0$   &$0$   &$0$   &$0$   &$0$   &$0$   &$0$   &$0$   &$0$   &$0$   &$0$   &$0$   &$0$   &$0$   &$0$   &$0$   &$0$   &$0$   &$0$   &$0$   &18.9  \\
$\Xi'_c(3007)$ $^{4}P_{5/2}$&11.3   &13.4   &3.6   &6.2   &7.4   &1.1   &0.1   &$0$   &$0$   &$0$   &$0$   &$0$   &$0$   &$0$   &$0$   &$0$   &$0$   &$0$   &$0$   &$0$   &$0$   &$0$   &$0$   &$0$   &-   &$0$   &$0$   &$0$   &43.1  \\
$\Xi'_c(3040)$ $^{2}P_{1/2}$&-   &-   &0.4   &52.2   &5.5   &98.5   &-   &$0$   &$0$   &$0$   &$0$   &$0$   &$0$   &$0$   &$0$   &$0$   &$0$   &$0$   &$0$   &$0$   &$0$   &$0$   &$0$   &$0$   &-   &$0$   &$0$   &$0$   &156.6  \\
$\Xi'_c(3067)$ $^{2}P_{3/2}$&-   &-   &21.7   &11.9   &50.4   &15.9   &-   &$0$   &$0$   &$0$   &$0$   &$0$   &$0$   &$0$   &$0$   &$0$   &$0$   &$0$   &$0$   &$0$   &$0$   &$0$   &$0$   &$0$   &-   &$0$   &$0$   &$0$   &99.9  \\
$\Xi'_c(3223)$ $^{2}D_{3/2}$&0.6   &0.7   &0.7   &1.1   &1.6   &2.2   &0.1   &0.4   &$0$   &$0$   &$0$   &$0$   &$0$   &-   &-   &$0$   &$0$   &$0$   &$0$   &$0$   &$0$   &$0$   &$0$   &$0$   &2.3   &$0$   &10.7   &0.1   &20.5  \\
$\Xi'_c(3268)$ $^{2}D_{5/2}$&1.8   &2.2   &2.2   &0.3   &4.6   &0.6   &0.2   &0.1   &-   &$0$   &$0$   &$0$   &$0$   &0.1   &-   &$0$   &$0$   &$0$   &-   &$0$   &$0$   &$0$   &$0$   &$0$   &6.6   &$0$   &38.0   &7.8   &64.5  \\
$\Xi'_c(3238)$ $^{4}D_{1/2}$&-   &-   &-   &1.6   &0.1   &3.3   &-   &0.6   &0.1   &$0$   &$0$   &$0$   &$0$   &-   &-   &$0$   &$0$   &$0$   &$0$   &$0$   &$0$   &$0$   &$0$   &$0$   &2.6   &$0$   &19.7   &0.9   &28.9  \\
$\Xi'_c(3265)$ $^{4}D_{3/2}$&0.7   &0.8   &0.2   &2.1   &0.5   &4.6   &0.1   &1.8   &0.4   &$0$   &$0$   &$0$   &$0$   &-   &0.1   &$0$   &$0$   &$0$   &0.1   &$0$   &$0$   &$0$   &$0$   &$0$   &10.2   &$0$   &27.1   &4.3   &53.0  \\
$\Xi'_c(3310)$ $^{4}D_{5/2}$&1.6   &2.0   &0.5   &1.9   &1.1   &4.3   &0.2   &1.7   &1.1   &$0$   &$0$   &$0$   &$0$   &-   &0.1   &$0$   &$0$   &$0$   &0.3   &$0$   &$0$   &$0$   &$0$   &$0$   &24.2   &0.8   &32.8   &24.6   &97.2  \\
$\Xi'_c(3373)$ $^{4}D_{7/2}$&2.5   &3.1   &0.8   &2.7   &1.7   &5.2   &0.3   &2.0   &1.3   &-   &$0$   &0.1   &$0$   &0.1   &0.1   &$0$   &$0$   &$0$   &0.4   &-   &$0$   &$0$   &$0$   &$0$   &37.4   &6.9   &32.7   &63.2   &160.5  \\
$\Xi'_c(3250)$ $^{2}S_{1/2}$&-   &0.1   &0.1   &-   &0.2   &0.1   &-   &0.1   &-   &$0$   &$0$   &$0$   &$0$   &-   &-   &$0$   &$0$   &$0$   &$0$   &$0$   &$0$   &$0$   &$0$   &$0$   &0.2   &$0$   &1.5   &0.1   &2.4  \\
$\Xi'_c(3319)$ $^{4}S_{3/2}$&-   &-   &-   &0.1   &0.1   &0.3   &-   &0.1   &0.1   &$0$   &$0$   &$0$   &$0$   &-   &-   &$0$   &$0$   &$0$   &-   &$0$   &$0$   &$0$   &$0$   &$0$   &0.5   &0.1   &1.3   &1.8   &4.4  \\
$\Xi'_c(3544)$ $^{2}S_{1/2}$&-   &-   &0.1   &0.1   &0.2   &0.2   &-   &0.4   &0.8   &5.0   &0.1   &10.9   &0.1   &-   &-   &0.4   &-   &$0$   &0.3   &1.7   &-   &0.3   &$0$   &$0$   &-   &-   &-   &-   &20.6  \\
$\Xi'_c(3613)$ $^{4}S_{3/2}$&0.1   &-   &-   &0.1   &-   &0.2   &-   &0.2   &0.6   &0.3   &7.3   &0.6   &15.9   &-   &0.1   &0.4   &0.1   &0.1   &0.2   &0.1   &2.4   &0.6   &-   &$0$   &-   &-   &-   &-   &29.3  \\
$\Xi'_c(3370)$ $^{2}D_{3/2}$&-   &-   &1.7   &40.4   &4.0   &90.1   &-   &39.1   &34.0   &1.2   &$0$   &4.0   &$0$   &0.3   &2.8   &$0$   &$0$   &$0$   &10.7   &0.3   &$0$   &$0$   &$0$   &$0$   &-   &-   &-   &-   &228.6  \\
$\Xi'_c(3415)$ $^{2}D_{5/2}$&-   &-   &22.7   &11.8   &48.4   &22.3   &-   &11.3   &9.1   &0.8   &$0$   &2.1   &0.2   &1.7   &0.3   &$0$   &$0$   &$0$   &2.9   &0.2   &$0$   &$0$   &$0$   &$0$   &-   &-   &-   &-   &133.8  \\
$\Xi'_c(3379)$ $^{2}P_{1/2}$&-   &-   &-   &0.6   &-   &1.2   &-   &0.3   &0.2   &-   &$0$   &-   &$0$   &-   &-   &$0$   &$0$   &$0$   &0.1   &-   &$0$   &$0$   &$0$   &$0$   &-   &-   &-   &-   &2.4  \\
$\Xi'_c(3406)$ $^{2}P_{3/2}$&-   &-   &0.4   &0.3   &0.8   &0.7   &-   &0.4   &0.3   &-   &$0$   &-   &$0$   &-   &-   &$0$   &$0$   &$0$   &0.1   &-   &$0$   &$0$   &$0$   &$0$   &-   &-   &-   &-   &3.0  \\
$\Xi'_c(3397)$ $^{2}S_{1/2}$&-   &-   &0.3   &2.5   &1.4   &8.4   &-   &9.7   &15.1   &3.1   &$0$   &8.6   &$0$   &0.5   &1.0   &$0$   &$0$   &$0$   &5.0   &0.8   &$0$   &$0$   &$0$   &$0$   &-   &-   &-   &-   &56.4  \\
$\Xi'_c(3517)$ $^{2}D_{3/2}$&8.4   &9.4   &8.1   &4.5   &17.2   &10.6   &0.6   &4.2   &8.9   &66.7   &-   &151.6   &0.1   &0.6   &0.8   &1.8   &$0$   &$0$   &3.0   &21.8   &-   &1.4   &$0$   &$0$   &-   &-   &-   &-   &319.7  \\
$\Xi'_c(3563)$ $^{2}D_{5/2}$&16.8   &20.4   &21.0   &8.8   &45.3   &18.7   &1.7   &6.2   &3.4   &20.1   &0.4   &51.4   &1.0   &1.8   &0.4   &5.7   &1.2   &$0$   &1.1   &6.4   &0.1   &0.3   &$0$   &$0$   &-   &-   &-   &-   &232.2  \\
$\Xi'_c(3532)$ $^{4}D_{1/2}$&10.4   &9.7   &1.2   &10.0   &2.3   &22.6   &0.2   &8.5   &13.8   &3.9   &139.6   &8.8   &345.6   &-   &1.3   &1.4   &0.1   &$0$   &4.6   &1.3   &43.9   &2.6   &$0$   &$0$   &-   &-   &-   &-   &631.8  \\
$\Xi'_c(3559)$ $^{4}D_{3/2}$&10.1   &11.3   &2.5   &5.4   &5.2   &12.6   &0.7   &16.0   &26.3   &8.0   &85.1   &17.8   &200.3   &0.2   &1.2   &2.5   &0.1   &$0$   &8.8   &2.6   &27.4   &8.1   &$0$   &$0$   &-   &-   &-   &-   &452.2  \\
$\Xi'_c(3604)$ $^{4}D_{5/2}$&10.9   &14.3   &4.1   &13.5   &9.0   &29.3   &1.4   &11.9   &16.2   &5.2   &16.9   &11.2   &34.4   &0.4   &1.2   &5.8   &0.6   &0.4   &5.4   &1.7   &5.7   &8.2   &0.2   &$0$   &-   &-   &-   &-   &207.9  \\
$\Xi'_c(3668)$ $^{4}D_{7/2}$&25.7   &31.2   &7.9   &43.1   &17.1   &93.8   &2.6   &51.1   &51.9   &9.9   &40.0   &24.5   &117.6   &0.6   &2.9   &10.5   &2.1   &1.5   &16.9   &3.2   &12.4   &10.7   &0.8   &0.2   &-   &-   &-   &-   &578.2  \\
\hline \hline
\end{tabular}

}
\caption{Same as \ref{tab:part_dec_Omega}, but for $\Xi'_c(snc)$ states.  The order of the states is the same as in Table \ref{tab:All_mass_Xiprime}. The predicted masses, reported in Table \ref{tab:All_mass_Xiprime},  are obtained by the three-quark model Hamiltonian of Eqs. \ref{MassFormula} and \ref{eq:Hho}. 
}
\label{tab:part_dec_cascades}
\end{table*}
\end{turnpage}

\begin{turnpage}
\begin{table*}[htbp]
{\scriptsize
%\hspace{-7cm}0.58cm
\begin{tabular}{c |  p{0.58cm}  p{0.58cm}  p{0.58cm}  p{0.58cm}  p{0.58cm}  p{0.58cm}  p{0.58cm}  p{0.58cm}  p{0.58cm}  p{0.58cm}  p{0.58cm}  p{0.58cm}  p{0.58cm}  p{0.58cm}  p{0.58cm}  p{0.58cm}  p{0.58cm}  p{0.58cm}  p{0.58cm}  p{0.58cm}  p{0.58cm}  p{0.58cm}  p{0.58cm}  p{0.58cm}  p{0.58cm}  p{0.58cm}  p{0.58cm}  p{0.58cm}p{0.75cm}} \hline \hline
$\Xi_c(snc)$  & $\Lambda_{c} K$  & $\Xi_{c} \pi$  & $\Xi'_{c} \pi$  & $\Xi^{*}_{c} \pi$  & $\Sigma_{c} K$  & $\Sigma^{*}_{c} K$  & $\Xi_{c} \eta$  & $\Lambda_{c} K^{*}$  & $\Xi_{c} \rho$  & $\Xi'_{c} \rho$  & $\Xi^{*}_{c} \rho$  & $\Sigma_{c} K^{*}$  & $\Sigma^{*}_{c} K^{*}$  & $\Xi'_{c} \eta$  & $\Xi^{*}_{c} \eta$  & $\Xi_{c} \eta'$  & $\Xi'_{c} \eta'$  & $\Xi^{*}_{c} \eta'$  & $\Xi_{c} \omega$  & $\Xi'_{c} \omega$  & $\Xi^{*}_{c} \omega$  & $\Xi_{c} \phi$  & $\Xi'_{c} \phi$  & $\Xi^{*}_{c} \phi$  & $\Lambda_{8} D$  & $\Lambda_{8} D^{*}$  & $\Sigma_{8} D$  & $\Lambda_{8}^{*} D$  & Tot $\Gamma$  \\ 
$\mathcal{F}={\bf {\bar{3}}}_{\rm f}$&&&&&&&&&&&&&&&&&&&&&&&&&&&&\\ \hline
$\Xi_c(2466)$ $^{2}S_{1/2}$&$0$   &$0$   &$0$   &$0$   &$0$   &$0$   &$0$   &$0$   &$0$   &$0$   &$0$   &$0$   &$0$   &$0$   &$0$   &$0$   &$0$   &$0$   &$0$   &$0$   &$0$   &$0$   &$0$   &$0$   &$0$   &$0$   &$0$   &$0$   &$0$  \\
$\Xi_c(2788)$ $^{2}P_{1/2}$&-   &-   &0.6   &2.0   &$0$   &$0$   &$0$   &$0$   &$0$   &$0$   &$0$   &$0$   &$0$   &$0$   &$0$   &$0$   &$0$   &$0$   &$0$   &$0$   &$0$   &$0$   &$0$   &$0$   &$0$   &$0$   &$0$   &$0$   &2.6  \\
$\Xi_c(2815)$ $^{2}P_{3/2}$&-   &-   &3.9   &0.6   &$0$   &$0$   &$0$   &$0$   &$0$   &$0$   &$0$   &$0$   &$0$   &$0$   &$0$   &$0$   &$0$   &$0$   &$0$   &$0$   &$0$   &$0$   &$0$   &$0$   &$0$   &$0$   &$0$   &$0$   &4.5  \\
$\Xi_c(2935)$ $^{2}P_{1/2}$&0.8   &0.4   &2.0   &13.3   &0.5   &$0$   &$0$   &$0$   &$0$   &$0$   &$0$   &$0$   &$0$   &$0$   &$0$   &$0$   &$0$   &$0$   &$0$   &$0$   &$0$   &$0$   &$0$   &$0$   &$0$   &$0$   &$0$   &$0$   &17.0  \\
$\Xi_c(2977)$ $^{4}P_{1/2}$&0.6   &0.2   &0.7   &7.7   &3.4   &0.3   &$0$   &$0$   &$0$   &$0$   &$0$   &$0$   &$0$   &$0$   &$0$   &$0$   &$0$   &$0$   &$0$   &$0$   &$0$   &$0$   &$0$   &$0$   &$0$   &$0$   &$0$   &$0$   &12.9  \\
$\Xi_c(2962)$ $^{2}P_{3/2}$&18.7   &21.8   &22.7   &2.0   &23.9   &$0$   &$0$   &$0$   &$0$   &$0$   &$0$   &$0$   &$0$   &$0$   &$0$   &$0$   &$0$   &$0$   &$0$   &$0$   &$0$   &$0$   &$0$   &$0$   &$0$   &$0$   &$0$   &$0$   &89.1  \\
$\Xi_c(3004)$ $^{4}P_{3/2}$&4.0   &4.7   &1.3   &31.5   &2.5   &12.2   &$0$   &$0$   &$0$   &$0$   &$0$   &$0$   &$0$   &$0$   &$0$   &$0$   &$0$   &$0$   &$0$   &$0$   &$0$   &$0$   &$0$   &$0$   &-   &$0$   &-   &$0$   &56.2  \\
$\Xi_c(3049)$ $^{4}P_{5/2}$&25.5   &30.8   &8.4   &17.0   &19.2   &19.0   &2.2   &$0$   &$0$   &$0$   &$0$   &$0$   &$0$   &$0$   &$0$   &$0$   &$0$   &$0$   &$0$   &$0$   &$0$   &$0$   &$0$   &$0$   &-   &$0$   &-   &$0$   &122.1  \\
$\Xi_c(3118)$ $^{2}D_{3/2}$&-   &-   &0.4   &2.2   &0.8   &2.9   &-   &$0$   &$0$   &$0$   &$0$   &$0$   &$0$   &$0$   &$0$   &$0$   &$0$   &$0$   &$0$   &$0$   &$0$   &$0$   &$0$   &$0$   &4.5   &0.5   &39.2   &$0$   &50.5  \\
$\Xi_c(3164)$ $^{2}D_{5/2}$&-   &-   &1.2   &0.5   &2.2   &0.6   &-   &$0$   &$0$   &$0$   &$0$   &$0$   &$0$   &-   &$0$   &$0$   &$0$   &$0$   &$0$   &$0$   &$0$   &$0$   &$0$   &$0$   &12.7   &2.1   &112.5   &$0$   &131.8  \\
$\Xi_c(3145)$ $^{2}S_{1/2}$&-   &-   &0.1   &0.1   &0.1   &0.2   &-   &$0$   &$0$   &$0$   &$0$   &$0$   &$0$   &-   &$0$   &$0$   &$0$   &$0$   &$0$   &$0$   &$0$   &$0$   &$0$   &$0$   &0.5   &0.2   &4.0   &$0$   &5.2  \\
$\Xi_c(3440)$ $^{2}S_{1/2}$&-   &-   &0.2   &0.5   &0.4   &1.2   &-   &0.9   &1.3   &0.5   &-   &1.1   &0.1   &-   &0.1   &-   &$0$   &$0$   &0.4   &0.1   &-   &$0$   &$0$   &$0$   &-   &-   &-   &-   &6.8  \\
$\Xi_c(3265)$ $^{2}D_{3/2}$&1.3   &1.7   &2.5   &10.6   &6.3   &21.8   &0.3   &6.9   &1.6   &$0$   &$0$   &$0$   &$0$   &0.2   &0.3   &$0$   &$0$   &$0$   &0.4   &$0$   &$0$   &$0$   &$0$   &$0$   &-   &-   &-   &$0$   &53.9  \\
$\Xi_c(3311)$ $^{2}D_{5/2}$&18.3   &22.3   &22.1   &1.8   &46.0   &3.1   &1.6   &1.1   &0.8   &$0$   &$0$   &$0$   &$0$   &1.2   &-   &$0$   &$0$   &$0$   &0.3   &$0$   &$0$   &$0$   &$0$   &$0$   &-   &-   &-   &$0$   &118.6  \\
$\Xi_c(3280)$ $^{4}D_{1/2}$&0.2   &0.4   &0.5   &4.6   &1.7   &9.8   &0.4   &3.7   &1.5   &$0$   &$0$   &$0$   &$0$   &0.1   &0.2   &$0$   &$0$   &$0$   &0.4   &$0$   &$0$   &$0$   &$0$   &$0$   &-   &-   &-   &$0$   &23.5  \\
$\Xi_c(3307)$ $^{4}D_{3/2}$&1.4   &1.7   &0.6   &17.4   &1.5   &39.0   &0.3   &16.8   &9.5   &$0$   &$0$   &$0$   &$0$   &0.1   &0.9   &$0$   &$0$   &$0$   &2.7   &$0$   &$0$   &$0$   &$0$   &$0$   &-   &-   &-   &$0$   &91.9  \\
$\Xi_c(3353)$ $^{4}D_{5/2}$&8.1   &9.8   &2.5   &23.8   &5.2   &52.7   &0.7   &23.0   &18.4   &0.2   &$0$   &0.8   &$0$   &0.2   &1.5   &$0$   &$0$   &$0$   &5.7   &$0$   &$0$   &$0$   &$0$   &$0$   &-   &-   &-   &$0$   &152.6  \\
$\Xi_c(3416)$ $^{4}D_{7/2}$&30.6   &37.1   &9.4   &20.1   &20.1   &39.4   &3.0   &15.4   &10.2   &0.6   &0.3   &1.8   &1.0   &0.7   &0.7   &0.3   &$0$   &$0$   &3.2   &0.2   &0.1   &$0$   &$0$   &$0$   &-   &-   &-   &-   &194.2  \\
$\Xi_c(3274)$ $^{2}P_{1/2}$&-   &-   &-   &0.1   &-   &0.2   &-   &-   &-   &$0$   &$0$   &$0$   &$0$   &-   &-   &$0$   &$0$   &$0$   &-   &$0$   &$0$   &$0$   &$0$   &$0$   &-   &-   &-   &$0$   &0.3  \\
$\Xi_c(3302)$ $^{2}P_{3/2}$&0.3   &0.4   &0.4   &0.1   &0.7   &0.1   &-   &-   &-   &$0$   &$0$   &$0$   &$0$   &-   &-   &$0$   &$0$   &$0$   &-   &$0$   &$0$   &$0$   &$0$   &$0$   &-   &-   &-   &$0$   &2.0  \\
$\Xi_c(3316)$ $^{4}P_{1/2}$&-   &-   &-   &0.1   &-   &0.1   &-   &-   &-   &$0$   &$0$   &$0$   &$0$   &-   &-   &$0$   &$0$   &$0$   &-   &$0$   &$0$   &$0$   &$0$   &$0$   &-   &-   &-   &$0$   &0.2  \\
$\Xi_c(3344)$ $^{4}P_{3/2}$&0.1   &0.1   &-   &0.3   &-   &0.5   &-   &0.2   &0.1   &$0$   &$0$   &$0$   &$0$   &-   &-   &$0$   &$0$   &$0$   &-   &$0$   &$0$   &$0$   &$0$   &$0$   &-   &-   &-   &$0$   &1.3  \\
$\Xi_c(3389)$ $^{4}P_{5/2}$&0.5   &0.6   &0.1   &0.4   &0.3   &0.9   &-   &0.4   &0.3   &-   &$0$   &-   &$0$   &-   &-   &$0$   &$0$   &$0$   &0.1   &-   &$0$   &$0$   &$0$   &$0$   &-   &-   &-   &-   &3.6  \\
$\Xi_c(3362)$ $^{4}S_{3/2}$&0.9   &0.3   &0.3   &3.1   &1.0   &9.7   &0.3   &7.2   &8.7   &0.2   &$0$   &0.8   &$0$   &0.2   &0.7   &$0$   &$0$   &$0$   &2.8   &0.1   &$0$   &$0$   &$0$   &$0$   &-   &-   &-   &$0$   &36.3  \\
$\Xi_c(3293)$ $^{2}S_{1/2}$&0.1   &0.3   &2.5   &1.9   &8.0   &5.3   &0.4   &6.2   &3.4   &$0$   &$0$   &$0$   &$0$   &0.6   &0.2   &$0$   &$0$   &$0$   &0.9   &$0$   &$0$   &$0$   &$0$   &$0$   &-   &-   &-   &$0$   &29.8  \\
$\Xi_c(3413)$ $^{2}D_{3/2}$&-   &-   &4.1   &16.4   &8.7   &38.9   &-   &19.5   &23.1   &3.3   &$0$   &8.7   &$0$   &0.3   &1.9   &$0$   &$0$   &$0$   &7.5   &0.9   &$0$   &$0$   &$0$   &$0$   &-   &-   &-   &-   &133.3  \\
$\Xi_c(3458)$ $^{2}D_{5/2}$&-   &-   &12.0   &15.2   &25.7   &30.5   &-   &14.7   &12.2   &0.4   &0.3   &0.8   &1.0   &1.0   &0.5   &-   &$0$   &$0$   &3.9   &0.1   &0.1   &$0$   &$0$   &$0$   &-   &-   &-   &-   &118.4  \\
\hline \hline
\end{tabular}

}
\caption{Same as \ref{tab:part_dec_Omega}, but for $\Xi_c(snc)$ states. The order of the states is the same as in Table \ref{tab:All_mass_Xi}. The predicted masses, reported in Table \ref{tab:All_mass_Xi}, are obtained by the three-quark model Hamiltonian of Eqs. \ref{MassFormula} and \ref{eq:Hho}. 
}
\label{tab:part_dec_cascades_anti3}
\end{table*}
\end{turnpage}

\begin{turnpage}
\begin{table*}[htbp]
{\scriptsize
\begin{tabular}{c |  p{0.58cm}  p{0.58cm}  p{0.58cm}  p{0.58cm}  p{0.58cm}  p{0.58cm}  p{0.58cm}  p{0.58cm}  p{0.58cm}  p{0.58cm}  p{0.58cm}  p{0.58cm}  p{0.58cm}  p{0.58cm}p{0.75cm}} \hline \hline
$\Lambda_c(nnc)$  & $\Sigma_{c} \pi$  & $\Sigma^{*}_{c} \pi$  & $\Lambda_{c} \eta$  & $\Sigma_{c}\rho$  & $\Sigma^{*}\rho$  & $\Lambda_{c}\eta'$  & $\Lambda_{c}\omega$  & $\Xi_{c} K$  & $\Xi'_{c} K$  & $\Xi^{*}_{c} K$  & $\Xi_{c} K^{*}$  & $\Xi'_{c} K^{*}$  & $\Xi^{*}_{c} K^{*}$  & $N D$  & Tot $\Gamma$  \\ 
$\mathcal{F}={\bf {\bar{3}}}_{\rm f}$&&&&&&&&&&&&&&\\ \hline
$\Lambda_c(2261)$ $^{2}S_{1/2}$&$0$   &$0$   &$0$   &$0$   &$0$   &$0$   &$0$   &$0$   &$0$   &$0$   &$0$   &$0$   &$0$   &$0$   &$0$  \\
$\Lambda_c(2616)$ $^{2}P_{1/2}$&1.4   &$0$   &$0$   &$0$   &$0$   &$0$   &$0$   &$0$   &$0$   &$0$   &$0$   &$0$   &$0$   &$0$   &1.4  \\
$\Lambda_c(2643)$ $^{2}P_{3/2}$&9.7   &0.1   &$0$   &$0$   &$0$   &$0$   &$0$   &$0$   &$0$   &$0$   &$0$   &$0$   &$0$   &$0$   &9.8  \\
$\Lambda_c(2799)$ $^{2}P_{1/2}$&7.3   &49.6   &$0$   &$0$   &$0$   &$0$   &$0$   &$0$   &$0$   &$0$   &$0$   &$0$   &$0$   &$0$   &56.9  \\
$\Lambda_c(2841)$ $^{4}P_{1/2}$&2.6   &28.5   &1.4   &$0$   &$0$   &$0$   &$0$   &$0$   &$0$   &$0$   &$0$   &$0$   &$0$   &-   &32.5  \\
$\Lambda_c(2826)$ $^{2}P_{3/2}$&83.9   &7.7   &1.1   &$0$   &$0$   &$0$   &$0$   &$0$   &$0$   &$0$   &$0$   &$0$   &$0$   &-   &92.7  \\
$\Lambda_c(2868)$ $^{4}P_{3/2}$&4.7   &115.8   &1.4   &$0$   &$0$   &$0$   &$0$   &$0$   &$0$   &$0$   &$0$   &$0$   &$0$   &-   &121.9  \\
$\Lambda_c(2913)$ $^{4}P_{5/2}$&31.2   &64.6   &12.2   &$0$   &$0$   &$0$   &$0$   &0.1   &$0$   &$0$   &$0$   &$0$   &$0$   &-   &108.1  \\
$\Lambda_c(2980)$ $^{2}D_{3/2}$&1.7   &8.8   &-   &$0$   &$0$   &$0$   &$0$   &-   &$0$   &$0$   &$0$   &$0$   &$0$   &59.1   &69.6  \\
$\Lambda_c(3025)$ $^{2}D_{5/2}$&4.7   &2.0   &-   &$0$   &$0$   &$0$   &$0$   &-   &$0$   &$0$   &$0$   &$0$   &$0$   &164.5   &171.2  \\
$\Lambda_c(3007)$ $^{2}S_{1/2}$&0.2   &0.4   &-   &$0$   &$0$   &$0$   &$0$   &-   &$0$   &$0$   &$0$   &$0$   &$0$   &4.7   &5.3  \\
$\Lambda_c(3372)$ $^{2}S_{1/2}$&0.2   &0.7   &-   &2.5   &0.4   &-   &1.1   &-   &0.4   &0.8   &0.2   &$0$   &$0$   &-   &6.3  \\
$\Lambda_c(3163)$ $^{2}D_{3/2}$&9.0   &42.0   &1.0   &$0$   &$0$   &$0$   &13.3   &2.7   &1.6   &0.7   &$0$   &$0$   &$0$   &-   &70.3  \\
$\Lambda_c(3208)$ $^{2}D_{5/2}$&90.0   &10.1   &8.3   &0.4   &$0$   &$0$   &1.7   &13.9   &8.4   &0.2   &$0$   &$0$   &$0$   &-   &133.0  \\
$\Lambda_c(3178)$ $^{4}D_{1/2}$&1.0   &19.9   &0.6   &$0$   &$0$   &$0$   &7.5   &3.8   &0.8   &0.7   &$0$   &$0$   &$0$   &-   &34.3  \\
$\Lambda_c(3205)$ $^{4}D_{3/2}$&2.2   &63.0   &0.9   &0.4   &$0$   &$0$   &31.5   &2.9   &0.6   &4.5   &$0$   &$0$   &$0$   &-   &106.0  \\
$\Lambda_c(3250)$ $^{4}D_{5/2}$&10.3   &86.6   &3.8   &3.9   &0.1   &0.2   &39.9   &6.7   &1.2   &10.0   &$0$   &$0$   &$0$   &-   &162.7  \\
$\Lambda_c(3313)$ $^{4}D_{7/2}$&37.3   &99.2   &13.9   &4.5   &2.8   &5.7   &29.8   &26.8   &5.7   &4.4   &$0$   &$0$   &$0$   &-   &230.1  \\
$\Lambda_c(3172)$ $^{2}P_{1/2}$&-   &0.4   &-   &$0$   &$0$   &$0$   &0.1   &-   &-   &-   &$0$   &$0$   &$0$   &-   &0.5  \\
$\Lambda_c(3199)$ $^{2}P_{3/2}$&1.2   &0.3   &0.1   &$0$   &$0$   &$0$   &-   &0.2   &0.1   &-   &$0$   &$0$   &$0$   &-   &1.9  \\
$\Lambda_c(3214)$ $^{4}P_{1/2}$&-   &0.3   &-   &$0$   &$0$   &$0$   &-   &-   &-   &-   &$0$   &$0$   &$0$   &-   &0.3  \\
$\Lambda_c(3241)$ $^{4}P_{3/2}$&0.1   &0.9   &-   &-   &$0$   &$0$   &0.3   &-   &-   &-   &$0$   &$0$   &$0$   &-   &1.3  \\
$\Lambda_c(3286)$ $^{4}P_{5/2}$&0.5   &1.4   &0.2   &0.1   &$0$   &-   &0.6   &0.3   &0.1   &0.1   &$0$   &$0$   &$0$   &-   &3.3  \\
$\Lambda_c(3259)$ $^{4}S_{3/2}$&0.3   &4.7   &0.2   &2.9   &1.2   &0.7   &11.2   &3.9   &1.6   &5.6   &$0$   &$0$   &$0$   &-   &32.3  \\
$\Lambda_c(3190)$ $^{2}S_{1/2}$&4.1   &4.4   &0.7   &0.1   &$0$   &$0$   &10.8   &4.8   &4.4   &0.8   &$0$   &$0$   &$0$   &-   &30.1  \\
$\Lambda_c(3345)$ $^{2}D_{3/2}$&21.0   &49.6   &-   &26.4   &0.3   &-   &33.3   &-   &3.5   &18.8   &0.7   &$0$   &$0$   &-   &153.6  \\
$\Lambda_c(3390)$ $^{2}D_{5/2}$&54.3   &90.1   &-   &2.1   &3.8   &-   &34.5   &-   &10.0   &5.4   &1.5   &$0$   &$0$   &-   &201.7  \\
\hline \hline
\end{tabular}
}
\caption{Same as \ref{tab:part_dec_Omega}, but for $\Lambda_c(nnc)$ states. The order of the states is the same as in Table \ref{tab:All_mass_Lambda}. The predicted masses, reported in Table \ref{tab:All_mass_Lambda}, are obtained by the three-quark model Hamiltonian of Eqs. \ref{MassFormula} and \ref{eq:Hho}.
}
\label{tab:part_dec_lambdas}
\end{table*}
\end{turnpage}

%%%%%%%%%%%%%%%%%%%%%%%%%%%%%%%%%%%%%
\clearpage
\section{DECAY PRODUCTS}
\label{app2}
\begin{table}[h!]%tbp]
%decay product meson mass table
\begin{tabular}{c | l }\hline \hline
                 & Mass in GeV  \\ \hline
$m_{\pi}$       & $0.13725 \pm 0.00295$ \\
$m_{K}$          & $0.49564 \pm 0.00279$ \\
$m_{\eta}$       & $0.54786 \pm 0.00002$ \\
$m_{\eta'}$       & $0.95778 \pm 0.00006$ \\
$m_{\rho}$       & $0.77518 \pm 0.00045$ \\
$m_{K^*}$          & $0.89555 \pm 0.00100$ \\
$m_{\omega}$       & $0.78266 \pm 0.00002$ \\
$m_{\phi}$       & $1.01946 \pm 0.00002$ \\
$m_{D}$          & $1.86672 \pm 0.00193$ \\
$m_{D_s}$          & $1.96835 \pm 0.00007$ \\
$m_{D^*}$          & $2.00855 \pm 0.00180$ \\
$m_{N}$          & $0.93891 \pm 0.00091$ \\
$m_{N(1520)}$    & $ 1.51500\pm 0.00500$ \\
$m_{N(1535)}$        & $1.53000 \pm 0.01500$ \\
$m_{N(1680)}$        & $1.68500\pm 0.00500$ \\
$m_{N(1720)}$        & $1.72000 \pm 0.03500$ \\
$m_{\Delta}$       & $1.23200 \pm 0.00200$ \\
$m_{\Lambda}$    & $1.11568 \pm 0.00001$ \\
$m_{\Lambda(1520)}$    & $1.51900 \pm 0.00010$ \\
$m_{\Xi_8}$        & $1.31820\pm 0.00360$ \\
$m_{\Xi_{10}}$       & $1.53370 \pm 0.00250$ \\
 $m_{\Sigma_8}$     & $1.11932 \pm 0.00340$ \\
$m_{\Sigma_{10}}$ & $1.38460 \pm 0.00460$ \\
$m_{\Lambda_c}$    & $2.28646 \pm 0.00014$ \\
$m_{\Xi_c}$        & $2.46908 \pm 0.00158$ \\
$m_{\Xi'_c}$      & $2.57850 \pm 0.00100$ \\
$m_{\Xi^*_c}$      & $2.64563 \pm 0.00100$ \\
$m_{\Sigma_c}$     & $2.45350 \pm 0.00090$ \\
$m_{\Sigma^*_c}$ & $2.51813 \pm 0.00280$ \\
$m_{\Omega_c}$ & $2.69520 \pm 0.00170$ \\
$m_{\Omega^*_c}$ & $2.76590 \pm 0.00200$ \\
%$m_{\eta}$(GeV)       & $0.548 \pm 0.01$ \\
\hline\hline
\end{tabular}
\caption{Masses of final state baryons and mesons used  in the calculation of the decay widths as from PDG \cite{Zyla:2020zbs}.}
\label{tab:exp_dec}
\end{table}

%\end{linenumbers}

\end{document}